\documentclass[aps,prd,amsmath,floats,floatfix, twocolumn,
superscriptaddress,nofootinbib,showpacs]{revtex4-1}

\pdfoutput=1

\usepackage[colorlinks, pdfborder={0 0 0}, plainpages=false]{hyperref}
\usepackage{graphicx}
\usepackage{xspace}
\usepackage[usenames,dvipsnames]{color}
\usepackage{amssymb}
\usepackage{mathrsfs}

\newcommand{\xten}[1]{\times 10^{#1}}
\newcommand{\sub}[1]{_{\rm #1}}
\newcommand{\map}[1]{\ensuremath{\mathcal{M}_{\rm #1}}}
\newcommand{\distorted}[1]{\hat{#1}}
\newcommand{\inertial}[1]{\bar{#1}}

\newcommand{\grid}[1]{#1}

\newcommand{\roughly}{\mathchar"5218\relax} 

\newcommand{\spin}[1]{\ensuremath{S^{++}_{0.{#1}}}\xspace}
\newcommand{\generic}{\ensuremath{S^{0.99}_{0.20}}\xspace}

\newcommand{\Caltech}{\affiliation{Theoretical Astrophysics 350-17,
    California Institute of Technology, Pasadena, CA 91125, USA}}
\newcommand{\Cornell}{\affiliation{Center for Radiophysics and Space
    Research, Cornell University, Ithaca, New York 14853, USA}}
\newcommand{\GWPAC}{\affiliation{Gravitational Wave Physics and
    Astronomy Center, California State University Fullerton,
    Fullerton, California 92834, USA}} %

\begin{document}

\title{
Improved methods for simulating nearly extremal
binary black holes
}

\author{Mark~A.~Scheel}\Caltech
\author{Matthew~Giesler}\Caltech\GWPAC
\author{Daniel~A.~Hemberger}\Caltech
\author{Geoffrey~Lovelace}\GWPAC\Caltech
\author{Kevin~Kuper}\GWPAC
\author{Michael~Boyle}\Cornell
\author{B\'{e}la Szil\'{a}gyi}\Caltech
\author{Lawrence~E.~Kidder}\Cornell
\date{\today}

\begin{abstract}
Astrophysical black holes could be nearly extremal (that is,
rotating nearly as fast as possible);
therefore, nearly extremal black holes could be
among the binaries
that current and future
gravitational-wave observatories will detect. Predicting the gravitational
waves emitted by merging black holes requires numerical-relativity
simulations, but these simulations are especially challenging 
when one or both holes have mass $m$ and spin $S$ 
exceeding the Bowen-York limit of 
$S/m^2=0.93$.
We present improved
methods that enable us to simulate merging, nearly extremal black holes more
robustly and more efficiently. We use these methods to
simulate an unequal-mass, precessing binary black hole coalescence,
where the larger black hole has
$S/m^2=0.99$.
We also use these methods to simulate
a non-precessing binary black hole coalescence, where both black holes have
$S/m^2=0.994$, nearly reaching
the Novikov-Thorne upper bound for
holes spun up by thin accretion disks. 
We demonstrate numerical convergence 
and estimate the numerical errors of 
the waveforms;
we compare numerical waveforms from
our
simulations with post-Newtonian 
and effective-one-body waveforms;
we compare the evolution of the black-hole
masses and spins with 
analytic predictions; and we explore 
the effect of increasing spin magnitude on 
the orbital dynamics (the so-called ``orbital hangup'' effect).
\end{abstract}

\pacs{}

\maketitle

\section{Introduction}\label{sec:intro}
Second-generation interferometers
such as Advanced
LIGO, Virgo, and
KAGRA~\cite{Harry:2010zz,aVIRGO,aVIRGO:2012,Somiya:2012} will
soon begin searching for gravitational waves.
To increase the number of gravitational-wave detections and
to maximize what we can learn about the waves' sources,
we require
accurate theoretical models of
the sources and the emitted 
gravitational radiation.

The inspiral, merger, and ringdown of binary black holes (BBHs)
are among
the most promising astrophysical sources of gravitational
waves. As the black holes
orbit,
they lose energy to gravitational radiation, inspiraling until
they collide and merge to form a final black hole (the ``remnant'')
that eventually
settles
to a stationary Kerr state.

A BBH is
characterized by 7 intrinsic
parameters: the spin angular momenta $\vec{S}$ of each hole
and the mass ratio $q$.
The spin magnitude of a black hole is often characterized
by the dimensionless quantity $\chi \equiv S/m^2$, where
$S=|\vec{S}|$, $m$ is the black hole mass, and we use
geometrized units where $c=G=1$.  A black hole with the
maximum possible dimensionless spin $\chi=1$ is called extremal.
There is considerable uncertainty in the expected
mass ratios and spins of 
astrophysical BBHs 
that are likely to be detected
by gravitational-wave interferometers; 
however, there is evidence that nearly extremal black holes exist
in nature.
For instance, recent 
measurements of 
stellar-mass black holes (such as 
Cygnus X-1~\cite{Gou:2011nq,Fabian:2012kv,Gou:2014una}, 
GRS 1915+105~\cite{McClintockEtAl:2006}, and GX 339-4~\cite{Miller:2009cw}) 
and supermassive black holes 
(such as~Swift J0501.9-3239~\cite{Walton:2012aw}) 
suggest that there 
could be a
population of black holes with spins of 
$\chi\sim 1$. 
(See, e.g., Refs.~\cite{McClintock:2013vwa,Reynolds:2013qqa} for 
reviews of astrophysical black-hole spin measurements.)

Post-Newtonian (PN) methods accurately model the binary evolution and
the emitted gravitational radiation during the early
inspiral~\cite{Blanchet2006}, but numerical simulations solving the
full Einstein equations are necessary to model the 
binary through 
late inspiral, merger, and ringdown.
Following breakthroughs in 
2005--2006~\cite{Pretorius2005a,Campanelli2006a, Baker2006a}, 
a number of research groups 
have made tremendous progress toward simulating merging black holes 
with different black-hole masses and spins (see, e.g., 
~\cite{Centrella:2010,Pfeiffer:2012pc,Hannam:2013pra,Tiec:2014lba} 
for recent reviews), 
and several groups are building 
catalogs of BBH simulations~\cite{Ajith:2012az,Hinder:2013oqa,Pekowsky:2013ska,
Mroue:2013PRL,Healy:2014yta,Clark:2014fva}.

So far, the region
of the parameter space with black hole spins 
near the theoretical maximum $\chi = 1$ 
remains almost completely unexplored.
Numerical simulations of nearly extremal, merging black holes 
are especially 
challenging. One reason for this is
that initial data for a BBH must satisfy the 
Einstein constraint equations, but the simplest
method for 
constructing constraint-satisfying initial data, 
the Bowen-York method~\cite{bowen79,Bowen-York:1980,Brandt1997}, 
cannot yield initial data with 
nearly extremal black holes. This
is because the Bowen-York construction assumes 
that the initial spatial geometry is conformally flat 
(i.e., that the initial spatial metric is proportional to the 
metric of flat space). Conformally flat spacetimes cannot represent 
black holes that i) are in equilibrium, and ii) possess 
linear~\cite{york80} or angular~\cite{GaratPrice:2000,Kroon:2004} 
momentum; therefore, conformally-flat spinning black holes
are out of equilibrium and will 
quickly relax to an equilibrium state.
Specifically, Bowen-York puncture initial data can produce 
BBHs with 
initial spins as large as $\chi=0.984$, but when evolved, 
the spins rapidly 
relax to about $\chi=0.93$ or less
(the Bowen-York limit)~\cite{cook90,DainEtAl:2002,HannamEtAl:2009}. 

Even given
initial data containing black holes 
with spins exceeding the Bowen-York limit,
evolving those data through inspiral, merger, and ringdown 
is especially challenging for two reasons, 
as discussed in 
Refs.~\cite{Lovelace:2010ne,Lovelace:2011nu,Lovelace:2013vma}.
First, the portion of the 
spacetime near the horizons requires very high resolution 
(and thus high computational cost), since 
metric gradients are much larger
than for lower spins. 
Second, for evolution methods that excise the singularities inside
each black hole and evolve only the exterior region, 
constructing and maintaining a suitable computational domain that 
keeps each excision boundary just inside the corresponding
apparent horizon becomes more and more
challenging 
as the spin approaches 
extremality. 

In this paper, we use the phrase ``nearly extremal'' to refer to 
$\chi > 0.93$, i.e., to a black hole with a spin above the Bowen-York 
limit. Note that a black hole with 
$\chi=0.93$ is significantly less extremal than a black hole with 
$\chi=0.998$, the Novikov-Thorne upper bound 
for black holes spun up by accretion~\cite{NovikovThorne:1973,Thorne:1974}.
This is because the effects of spin scale 
nonlinearly with increasing $\chi$.
For instance, if the rotational energy
of a Kerr black hole with a fixed mass
is denoted $E_{\rm rot}(\chi)$, then
$E_{\rm rot}(0.93)/E_{\rm rot}(1)$ is only $59\%$, while
$E_{\rm rot}(0.998)/E_{\rm rot}(1)$ is $92.5\%$
(c.f., Fig.~1 of 
Ref.~\cite{Lovelace:2010ne}).
Furthermore, the 
total energy that a 
BBH emits in gravitational waves
also scales nonlinearly with 
$\chi$. For example, for
equal masses and equal spins aligned with the orbital angular 
momentum, a BBH
with $\chi=1$ radiates $10\%$ more energy than 
a BBH with $\chi=0.93$, whereas
a BBH with $\chi=0.07$ radiates 
only $4\%$ 
more energy than a BBH with $\chi=0$
(Eq.~(9) of Ref.~\cite{Hemberger:2013hsa}).
Nonlinear scaling with $\chi$ is also
seen for binaries consisting of a black hole and a neutron
star: the amount of neutron-star matter remaining outside the black hole
just after tidal disruption increases very rapidly with
black-hole spin 
(Fig.~10 of Ref.~\cite{Lovelace:2013vma}).

Several groups have constructed and evolved Bowen-York
puncture initial data with 
spins near (but below) the Bowen-York 
limit~\cite{Rezzolla:2007xa,HannamEtAl:2010,MarronettiEtal:2008,
DainEtAl:2008}.
Recently, Ruchlin
et al.~\cite{Ruchlin:2014zva} 
constructed and evolved 
puncture initial data for a head-on collision of two black holes with 
equal mass and spins of magnitude $\chi=0.99$.
Only four previously published
simulations~\cite{Lovelace:2010ne,Lovelace:2011nu,Mroue:2013PRL} 
out of hundreds published to date
contain the quasi-circular coalescence of BBHs
with $\chi>0.93$.
These four simulations were evolved using the Spectral Einstein 
Code (SpEC)~\cite{SpECwebsite} from
``superposed-Kerr-Schild'' excision initial data~\cite{Lovelace2008} 
and have equal masses and equal spins 
aligned or anti-aligned with the orbital angular 
momentum.

In this paper, we present technical improvements that have enabled us
to simulate BBHs with black-hole spins up to $\chi=0.994$ 
({\it i.e.,} $E_{\rm rot}(0.994)/E_{\rm rot}(1) = 87.1\%$)
and to complete the first
nearly extremal BBH simulation that includes precession.
We compare the results of these simulations with analytical
models describing the remnant properties (e.g. final spin and total
radiated energy) as a function of the initial black hole spins; these
models were constructed using lower-spin simulations and then extrapolated
to higher spins.  
We measure the
slow increase in mass (``tidal heating'') and decrease in spin (``tidal
torquing'') of the individual black holes before merger, and 
we compare
these measurements with perturbative calculations of the same quantities.
We also compare gravitational waveforms from these simulations with
post-Newtonian and effective-one-body~\cite{Taracchini:2013rva}
models.

The methods described here 
allow
us to robustly explore the portion of the BBH 
parameter space
where one or both black holes are nearly extremal.
Simulations using these methods will enable 
us to calibrate and validate analytic waveform models, construct
improved models of the dependence of remnant properties on
the initial masses and spins of the black holes, 
and explore the dynamics of the strongly warped spacetime during
the merger. In a companion paper, we use 
these methods to explore the extremality of apparent horizons in numerical 
simulations~\cite{Lovelace2014}.

The remainder of this paper is organized as follows. We summarize our 
techniques in Sec.~\ref{sec:techniques}, focusing on new improvements to 
our algorithm that enable us to simulate higher black-hole spins more 
robustly. We 
present three new simulations in Sec.~\ref{sec:simulations},
and we present results in Sec.~\ref{sec:results},
including a comparison of the emitted 
gravitational waveforms with analytical 
predictions and also 
a comparison of the evolution of the 
black-hole masses and
spins with analytic predictions.
We briefly conclude in Sec.~\ref{sec:conclusion}.

\section{Techniques}\label{sec:techniques}

We carry out numerical simulations with the Spectral Einstein Code
(SpEC)~\cite{SpECwebsite}.  We construct~\cite{Pfeiffer2003}
quasi-equilibrium~\cite{Caudill-etal:2006,Lovelace2008} 
constraint-satisfying~\cite{York1999} initial data
based on a weighted superposition of two boosted, spinning Kerr-Schild
black holes~\cite{Lovelace2008}.  We use an iterative method
to produce initial data with low
eccentricity~\cite{Pfeiffer-Brown-etal:2007,Buonanno:2010yk,Mroue:2012kv}.

We use a generalized
harmonic 
formulation~\cite{Friedrich1985, Garfinkle2002, Pretorius2005c,Lindblom2006}
of Einstein's equations and damped harmonic 
gauge~\cite{Lindblom2009c,Choptuik:2009ww,Szilagyi:2009qz} to
evolve the initial data.
The adaptively-refined~\cite{Lovelace:2010ne,Szilagyi:2014fna} grid
extends from pure-outflow
excision boundaries just inside the apparent 
horizons~\cite{Scheel2009,Szilagyi:2009qz,Hemberger:2012jz,Ossokine:2013zga}
to an artificial outer boundary, where we enforce constraint-preserving
boundary conditions~\cite{Lindblom2006, Rinne2006, Rinne2007}.
The grid has only one excision 
boundary after the holes merge~\cite{Scheel2009,Hemberger:2012jz}. 
We use a pseudospectral fast-flow algorithm~\cite{Gundlach1998} 
to find apparent horizons, and we compute spins on these apparent horizons
using the approximate Killing vector formalism of Cook, Whiting, and
Owen~\cite{Cook2007, OwenThesis}.

In the remainder of this section, 
we describe new techniques that allow simulations of binaries
with large black-hole spins.
Large spins are
difficult for two reasons.  First, the metric gradients 
near the black-hole horizons become larger with larger spin, 
making increased numerical resolution 
necessary in this region.
Second, black-hole excision is more difficult: in SpEC,
we remove the physical singularity inside each black hole
by placing an artificial {\em excision boundary} just inside each
apparent horizon and evolve only the region 
exterior to all
excision boundaries.  We find that the maximum required
coordinate distance between an excision boundary and the corresponding
horizon becomes smaller with larger spin, so that our algorithm for
dynamically adjusting the excision boundaries to track the size, shape,
and motion of the horizons must be more accurate.
We consider both of these difficulties 
below.

Not all of the improvements discussed here were necessary for all of the
simulations described in Sec.~\ref{sec:simulations}. For example,
the simulation discussed in Sec.~\ref{sec:Spin99} succeeded without
some of the grid and control system improvements; however, these
improvements became necessary when
simulating even larger black-hole spins (Sec.~\ref{sec:Spin994})
or allowing generic spin directions and unequal
masses (Sec.~\ref{sec:spin-0.99-precessing}).

\subsection{Grid improvements}

Meeting the need for high resolution near the horizons is 
accomplished
via spectral adaptive mesh refinement~\cite{Szilagyi:2014fna}. This
includes both $p$-type refinement (changing the number of collocation
points in a given spectral subdomain) and $h$-type refinement (adding,
removing, or changing
the distribution of subdomains).  The simulations described here
used the algorithm detailed 
in~\cite{Szilagyi:2014fna}, with adjustments to default
parameters so as to allow for higher resolution.  In
particular:  We increased the number of radial collocation points
in a spherical subdomain that forces $h$-refinement from 20 points
to 40, we disabled angular $h$-refinement in the spherical subdomains
that touch the excision boundary so as to retain a single 
spherical subdomain at this boundary, and we increased the allowed number of
spherical-harmonic coefficients in spherical shells from $L=40$ to
$L=80$.  Note that most of these changes (such as allowing up to
$L=80$) were necessary only for a small portion of the simulation when
the horizons are highly distorted, such
as during the initial
``junk radiation'' transients (when spurious  
gravitational radiation is emitted as the BBH relaxes to equilibrium)
and near the moment of merger.

We also reduced the initial distance between the apparent horizons and
the excision boundaries.  To understand this change, note that when
solving elliptic equations for initial data, the excision boundaries
are made to coincide exactly with the apparent horizons via boundary
conditions imposed on those surfaces.  But for the evolution, the
excision boundaries must be slightly inside the horizons, so that the
horizons are fully contained in the computational domain and therefore
can be
determined by the apparent horizon finder.  
To accomplish this, 
after the initial data have been determined, these data
are extrapolated slightly inside the horizons to a new excision boundary,
before the evolution begins.  For large spins, this extrapolation occurs
in the region where metric gradients are growing rapidly as $r$ decreases,
so placing the excision boundary at a larger $r$ reduces those gradients.
To carry out some of the simulations
shown here, we moved the initial excision
boundary radius from $94\%$ to $98\%$ of the
initial horizon radius.

\subsection{Control system improvements}

Several of the improvements necessary for handling high spins involve
control systems used to adjust mappings between coordinate systems.
These control systems and the mappings are described in detail
in~\cite{Hemberger:2012jz}. Here we briefly summarize important points,
and we discuss key differences from~\cite{Hemberger:2012jz}. 

\subsubsection{Summary of size and shape control systems}

In SpEC, we remove the physical singularity inside each black hole
by placing an artificial excision boundary
just inside each
apparent horizon, 
evolving only the region exterior to all
excision boundaries.
We use multiple coordinate systems to handle excision of black
holes that are moving and changing
shape~\cite{Scheel2006,Boyle2007,Scheel2009,Szilagyi:2009qz,Buchman:2012dw,%
  Chu2009,Hemberger:2012jz}.  
We call ``inertial coordinates''
$\inertial{x}^i$ those 
asymptotically inertial 
coordinates in which the black holes orbit each
other, have a distorted and dynamical shape, and approach each other
as energy is lost to gravitational radiation.  We apply spectral
methods in a different coordinate system, ``grid coordinates''
$\grid{x}^i$, in which the excision boundaries are spherical and
time-independent.  We connect these two coordinate systems with an
analytic mapping function $\map{}\!: \grid{x}^i \to \inertial{x}^i$ that
depends on a set of time-dependent parameters $\lambda(t)$.  These
parameters $\lambda(t)$ are adjusted automatically by feedback control
systems so that, as the apparent horizons of the black holes move and
change shape (in the inertial frame), the excision boundaries are
mapped to inertial-coordinate surfaces that follow this motion and
remain just inside the apparent horizons.

The control of all parameters $\lambda(t)$ is 
accomplished in the
same way, using a general control system we have developed, as described
in~\cite{Hemberger:2012jz}.  The part of the algorithm that distinguishes
one $\lambda(t)$ from another is the
specification of the control error $Q(t)$, which is different for each
control parameter. For example, the $\lambda\sub{Scaling}(t)$
that represents the distance between the excision boundaries has a different
$Q(t)$ than the matrix $\lambda\sub{Rotation}(t)$
that represents the rotation of the
inertial coordinates with respect to the grid coordinates.
If there exists a desired value of $\lambda(t)$, call it
$\lambda\sub{target}$, which depends on observables $A$ (such
as the positions or shapes of the apparent horizons)
but does not depend on $\lambda$ itself, then we define
\begin{equation}
Q(t) = \lambda\sub{target}(A) - \lambda(t).
\label{eq:Qgeneric}
\end{equation}
For more general situations in which $\lambda\sub{target}$ depends on
$\lambda$ itself, we generalize the above definition: 
we require that $\lambda$ takes on its desired value when $Q=0$, and
we require that 
\begin{equation}
\frac{\partial Q}{\partial \lambda} = -1 + {\cal O}(Q).
\label{eq:Qgenericnonlinear}
\end{equation}
Given $Q(t)$, our algorithm adjusts the corresponding $\lambda(t)$ 
so that $Q(t)$ is driven towards zero; this driving occurs on a
timescale $\tau_d$ that is determined dynamically and that is 
different for each control system.

The full map from grid to inertial coordinates is
$\inertial{x}^i = \mathcal{M} \grid{x}^i$, where
\begin{equation}
  \label{eq:MapSequence}
\begin{array}{ll}
\mathcal{M} =&
\map{Translation}
\circ\map{Rotation}
\circ\map{Scaling}  \\
& \circ\map{Skew}
\circ\map{CutX}
\circ\map{Shape}.
\end{array}
\end{equation}
Each of these maps is described in detail in
Sec. 4 of~\cite{Hemberger:2012jz}.

\textbf{Shape control.}
Here we are concerned only with the last map, \map{Shape}, which is defined as:
\begin{equation}
  x^i \mapsto x^i\left(1-\sum_H\frac{f_H(r_H,\theta_H,\phi_H)}{r_H}
  \sum_{\ell m} Y_{\ell m}(\theta_H,\phi_H)\lambda^H_{\ell m}(t)\right).
\label{eq:GridToShape}
\end{equation}
The index $H$ goes over each of the two excised regions $A$ and $B$,
and the map is applied to the grid-frame coordinates. The polar
coordinates $(r_H,\theta_H,\phi_H)$ centered about excised region $H$
are defined in the usual way, the quantities $Y_{\ell
  m}(\theta_H,\phi_H)$ are spherical harmonics, and $\lambda^H_{\ell
  m}(t)$ are expansion coefficients that parameterize the map near
excision region $H$; these $\lambda^H_{\ell
  m}(t)$ are the coefficients that we adjust using a
control system.  The function $f_H(r_H,\theta_H,\phi_H)$ is
chosen to be unity near excision region $H$ and zero near the other
excision region, so that the distortion maps for the two black holes
are decoupled; see Eq.~72 and Fig.~4 of~\cite{Hemberger:2012jz} 
for a precise definition of
$f_H(r_H,\theta_H,\phi_H)$. 
In the following, the control systems
for each excised region $H$, while independent,
 are identical in operation, so we will omit the
$H$ labels for simplicity.

We control $\lambda_{\ell m}(t)$ so that each excision boundary is
driven to the same shape as the corresponding apparent horizon; this
results in conditions on $\lambda_{\ell m}(t)$ for $\ell>0$, but
leaves $\lambda_{00}(t)$ unconstrained~\cite{Hemberger:2012jz}.  

\textbf{Size control.} 
The size of the excision boundary, as encoded in the remaining
coefficient $\lambda_{00}(t)$, must satisfy two conditions.  

{\it Horizon tracking.} The
first is that the excision boundary remains inside the apparent
horizon.  To satisfy this condition, we first write the shape of
each apparent horizon as an expansion in spherical harmonics, 
parameterized in terms of polar coordinates about
the center of the corresponding excision boundary, 
\begin{equation}
\distorted{r}^{\rm AH}(\distorted{\theta},\distorted{\phi}) = \sum_{\ell m} 
\distorted{S}_{\ell m} Y_{\ell m}(\distorted{\theta}, \distorted{\phi}),
\label{eq:StrahlkorperAH}
\end{equation}
where the intermediate frame
$\distorted{x}^i$ is connected to the grid frame by the map
\begin{equation}
\map{Distortion} = \map{CutX}\circ\map{Shape}
\label{eq:DistortionMapSymbol}.
\end{equation}
By construction, $\map{Distortion}$ leaves invariant the 
centers of the excision
boundaries, and the angles with respect to these centers. 
Then we choose
\begin{equation}
Q = \dot{\distorted{S}}_{00}(\Delta r-1)-\dot\lambda_{00}
\label{eq:State2Q}
\end{equation}
where 
\begin{equation}
\Delta r = 1 - \frac{\langle\distorted{r}_{\rm EB}\rangle}
                    {\langle\distorted{r}_{\rm AH}\rangle}
\label{eq:DeltaR}
\end{equation}
is the relative difference between the average radius of the apparent
horizon (in the intermediate frame) and the average radius of the excision
boundary. The angle brackets in Eq.~(\ref{eq:DeltaR}) represent
averaging over angles.  Choosing $Q(t)$ according to 
Eq.~(\ref{eq:State2Q}) drives
$d/dt(\Delta r)$ towards zero, so that the excision boundary remains
a fixed (relative) distance inside the apparent horizon.

{\it Characteristic speed tracking.} 
The second condition that must be satisfied by $\lambda_{00}(t)$ involves
characteristic speeds of the evolved Einstein equations: well-posedness
of our system of equations requires that all of the characteristic speeds
be non-negative, i.e.~characteristic fields must flow into the black
hole.  The minimum characteristic speed at each excision boundary is given by
\begin{equation}
  v = -\alpha - \inertial{\beta}^i \inertial{n}_i - \inertial{n}_i
  \frac{\partial \inertial{x}^i}{\partial t},\label{eq:CharSpeedDef}
\end{equation}
where $\alpha$ is the lapse, $\inertial{\beta}^i$ is the shift, and
$\inertial{n}_i$ is the normal to the excision boundary pointing {\it out
  of the computational domain}, i.e., toward the center of the hole.
It is possible to write~\cite{Hemberger:2012jz}
\begin{equation}
   v = v_0 + \distorted{n}_i \frac{x^i}{r} Y_{00} \dot\lambda_{00},
\end{equation}
where $v_0$ collects all terms that are independent of $\dot\lambda_{00}$.
Therefore, a control system that controls $\dot\lambda_{00}$ 
and drives $v$ to some target value $v_T$
can be constructed by defining the control error
\begin{equation}
Q = ({\rm min}(v)-v_T)/\langle-\Xi\rangle,
\label{eq:CharSpeedControlQ}
\end{equation}
where
\begin{equation}
\Xi = \distorted{n}_i \frac{x^i}{r} Y_{00},
\end{equation}
and the minimum is over the excision boundary.  Note that $\Xi<0$ because
$\distorted{n}_i$ and $x^i/r$ point in opposite directions.

{\it Switching between horizon and characteristic speed tracking.} 
Note that Eqs.~(\ref{eq:State2Q})
and~(\ref{eq:CharSpeedControlQ}) specify two {\em different} control
systems that control the same degree of freedom, $\lambda_{00}$: the
first control system, which we call ``horizon tracking'', 
adjusts $\dot\lambda_{00}$ to control $\Delta r$, and the
other, which we call ``characteristic speed tracking'', 
adjusts $\dot\lambda_{00}$ to control $v$.  Both $\Delta r$ and $v$
must remain nonnegative for a successful evolution, but we cannot use
both Eqs.~(\ref{eq:State2Q}) and~(\ref{eq:CharSpeedControlQ})
simultaneously.  Furthermore, changes in $\dot\lambda_{00}$ affect
$\Delta r$ and $v$ in the opposite direction: if $\dot\lambda_{00}$ increases,
$\Delta r$ increases, but the characteristic speed $v$ decreases.

In practice (Sec.~\ref{sec:switching}), 
we now alternate between the two control systems,
Eqs.~(\ref{eq:State2Q}) and~(\ref{eq:CharSpeedControlQ}).  We monitor
both $v$ and $\Delta r$ as functions of time and predict whether
either of these quantities is likely to become negative in the
immediate future; if so, we estimate the timescale $\tau_v$ or
$\tau_{\Delta r}$ on which this will occur.  If $\tau_{\Delta r}$ is
small enough, i.e. $\Delta r$ is in imminent danger of becoming
negative, we use Eq.~(\ref{eq:State2Q}) to control $\Delta r$.  If
$\tau_v$ is small enough that $v$ is in danger of becoming negative,
we use Eq.~(\ref{eq:CharSpeedControlQ}) to control $v$.  The details of
how we make these decisions have been improved since the description
in Sec.~5.3 of~\cite{Hemberger:2012jz}, so we describe the improved algorithm
below. 

\subsubsection{Improvements in gain scheduling}
\label{sec:impr-gain-sched}

We now describe improvements in the control systems
that were necessary for our new high-spin simulations to succeed.

\textbf{Comoving characteristic speed as a control system diagnostic.} 
We define a new quantity $v_c$ which 
we call the \emph{comoving characteristic speed}:
\begin{align}
 v_c =& -\alpha - \inertial{\beta}^i \inertial{n}_i
        - \inertial{n}_i
        \frac{\partial \inertial{x}^i}{\partial \distorted{t}}
        \nonumber \\
 &+ \distorted{n}_i \frac{x^i}{r}
  \left[Y_{00}\dot{\distorted{S}}_{00}(\Delta r-1)+
  \sum_{\ell>0} Y_{\ell m}(\theta,\phi)\dot\lambda_{\ell m}(t)\right].
  \label{eq:ComovingCharSpeedFormula}
\end{align}
The comoving characteristic speed $v_c$ is what the characteristic speed $v$ 
{\em would be} if $Q(t)$ in 
Eq.~(\ref{eq:State2Q}) were exactly zero, i.e. if $\Delta r$ were constant in
time.  In other words, if we
turn on horizon tracking, the control system drives $v$ toward $v_c$.
This tells us (for instance) that if we find $v_c<0$, 
we should not use horizon tracking, since horizon tracking would drive
$v$ to a negative value. The instantaneous value of $v_c$ is
independent of $\dot{\lambda}_{00}$ and roughly independent of $\lambda_{00}$; 
the only dependence on $\lambda_{00}$
comes from the smooth spatial variation of the metric functions.  Hence,
$v_c$ is a useful quantity for separating the effects of
the control system for $\dot{\lambda}_{00}$ from the effects of other
control systems.

One way we use $v_c$ is in determining whether our control system for
$\dot\lambda_{00}$ will fail.  During a simulation, $v_c$ is usually
positive, but it routinely becomes negative for short periods of time,
particularly when the shapes of the horizons are changing rapidly, for
example near $t=0$ when the black holes are ringing down from initial
``junk radiation'' transients.  However, if $v_c$ becomes negative and
remains so indefinitely, our control system for $\dot\lambda_{00}$
must eventually fail.  This is because for $v>0$ and $v_c<0$, 
$\Delta
r$ must be decreasing, so if we keep $v>0$ the excision boundary will
eventually intersect the apparent horizon.

\textbf{Control error damping timescale improvements.} 
For many of the high-spin SpEC simulations that 
failed before we made the improvements 
described in this paper, we observed that $v_c<0$
for an extended period of time.  This was caused by inaccurate control
systems for the $\lambda(t)$ parameters {\em other than}
$\lambda_{00}$; in particular, the shape parameters $\lambda_{\ell m}$
for $\ell>0$.  In other words, the shape and position of the excision
boundary differed from the shape and position of the horizon by a
sufficient amount that it was not possible to make both $v>0$ and
$\Delta r>0$ everywhere by adjusting only the radial motion of the
excision boundary, $\lambda_{00}$.  

This particular problem was fixed by changing the algorithm for 
setting the
tolerance on the control error $Q(t)$, for all $Q(t)$ except $Q_{00}$.
Associated with each of our control systems is a timescale parameter
$\tau_d$ which is adjusted dynamically.  The control error $Q(t)$ is
damped like $e^{-t/\tau_d}$, under the assumption that $\tau_d$ is
smaller than all other timescales in the problem.  Therefore
decreasing $\tau_d$ results in smaller values of $Q(t)$.
The previous
method of adjusting $\tau_d$ is described 
in Sec.~3.3 of~\cite{Hemberger:2012jz}:
at regular time intervals $t_i$, the timescale is changed according to
\begin{equation}
\tau_d^{i+1} = \beta \tau_d^i,
\label{eq:TimescaleTunerOld1}
\end{equation}
where
\begin{equation}
\label{eq:TimescaleTunerOld2}
\beta = \left\{
  \begin{array}{ll}
    0.99, & {\rm ~if~} \dot{Q}/Q > -1/2\tau_d {\rm ~and~} |Q| {\rm ~or~} |\dot{Q}\tau_d| > Q_t^{\rm Max} \\
    1.01, & {\rm ~if~} |Q| < Q_t^{\rm Min} {\rm ~and~} |\dot{Q}\tau_d| < Q_t^{\rm Min} \\
    1, & {\rm ~otherwise.}
  \end{array}\right.
\end{equation}
Here $Q_t^{\rm Min}$ and $Q_t^{\rm Max}$ are thresholds for the control
error $Q$, set to constant values
\begin{align}
Q_t^{\rm Max} &= \frac{2\xten{-3}}{m_A/m_B + m_B/m_A}\label{eq:QtMax} \\
Q_t^{\rm Min} &= \frac{1}{4}Q_t^{\rm Max}.
\end{align}

\begin{figure}
\includegraphics[width=0.95\columnwidth]{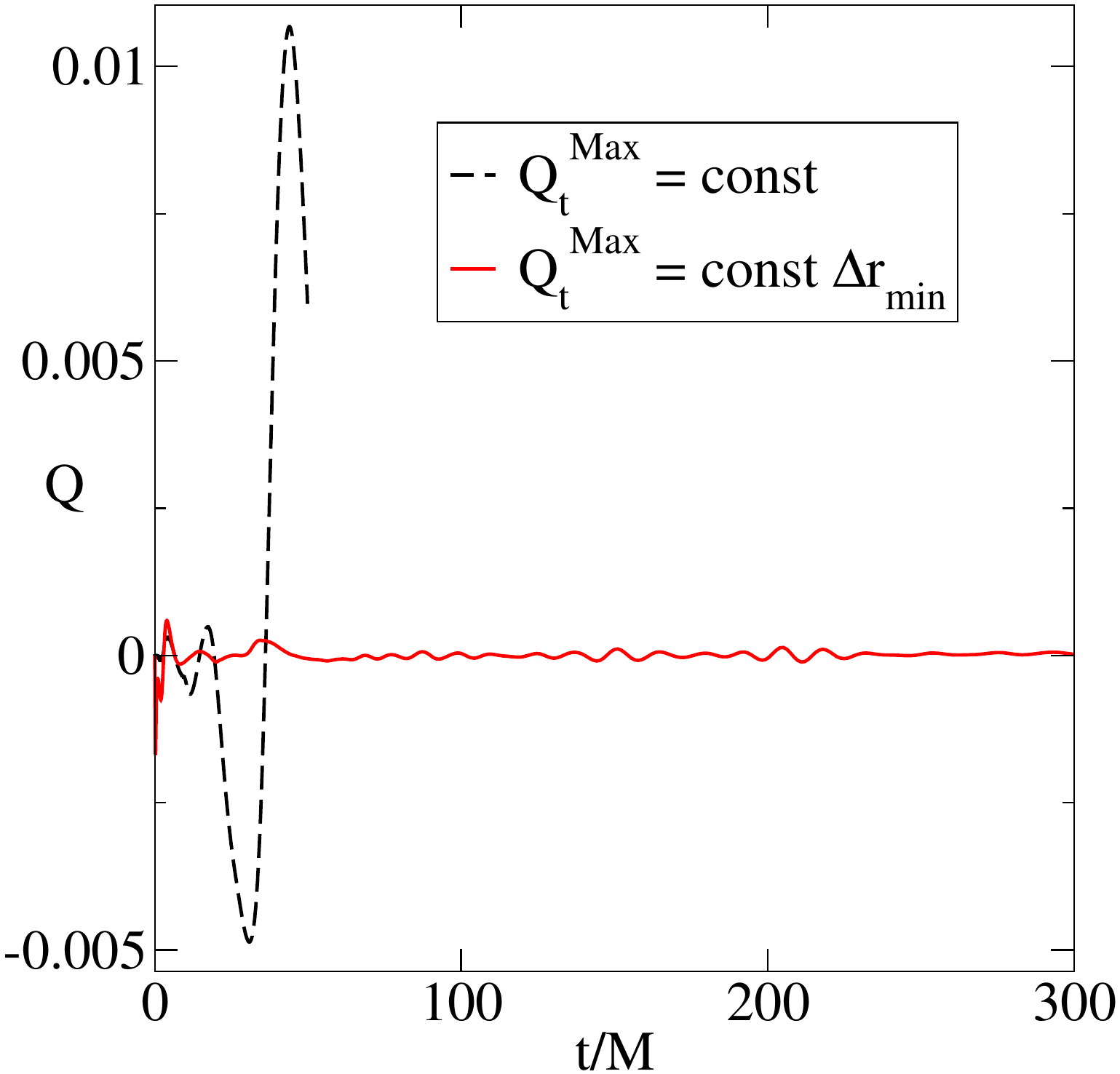}
\caption{The control error $Q(t)$ associated with one particular control 
  system
  for two different \spin{994} simulations which
  differ only in the treatment of $Q_t^{\rm Max}$
  for that control system.  The black dashed curve shows the case in which
  $Q_t^{\rm Max}$ is a constant, given by Eq.~(\ref{eq:QtMax}), and the
  red solid curve shows the case in which $Q_t^{\rm Max}$ is chosen to
  be $0.05\Delta r\sub{min}$, with $\Delta r\sub{min}$ 
  given by Eq.~(\ref{eq:MinDeltaR}).  The former simulation crashes at
  $t\sim 50M$.  
  \label{fig:SmoothRAhQ}
}
\end{figure}

In the new algorithm, we make three
changes.  The first is that
$Q_t^{\rm Max}$ is no longer a constant: instead, it is chosen
to be $Q_t^{\rm Max} = a \Delta r\sub{min}$, where
$a$ is a constant (typically chosen to be 0.05$(m_A+m_B)$ for those
$Q$ values with dimensions of length, and 0.05 for those $Q$ values
that are dimensionless)
and $\Delta r\sub{min}$ is
the minimum relative distance between the
excision boundary and the apparent horizon:
\begin{equation}
\Delta r\sub{min} = \min_{\distorted{\theta},\distorted{\phi}}\left(1 - 
\frac{\distorted{r}_{\rm EB}(\distorted{\theta},\distorted{\phi})}
     {\distorted{r}_{\rm AH}(\distorted{\theta},\distorted{\phi})}\right).
\label{eq:MinDeltaR}
\end{equation}
The second change is that we define an estimate of the time 
that the horizon will cross the excision surface
\begin{equation}
\tau_{\Delta r {\rm cross}} = 
-\Delta r\sub{min} \left(\frac{d}{dt}\Delta r\sub{min}\right)^{-1},
\label{eq:DeltaRCross}
\end{equation}
and if $\tau_{\Delta r {\rm cross}}>0$ and $\tau^i_d >\tau_{\Delta r {\rm cross}}$,
then we set $\tau^{i+1}_d = \tau_{\Delta r {\rm cross}}$ instead of using
Eq.~(\ref{eq:TimescaleTunerOld1}).

Both of the above
changes force each $Q(t)$ to be closer to zero
when the excision boundary approaches the horizon.  
A third, minor, change we make in the
algorithm affects only the behavior of $Q(t)$ at early times: 
the initial values of each $\tau_d$ were decreased so that each
$Q(t)$ is smaller at earlier times; these initial values are specified
separately from the tolerances $Q_t^{\rm Max}$ that determine when
$\tau_d$ is modified.
The effect of the first change, setting $Q_t^{\rm Max}$ 
proportional to $\Delta r\sub{min}$, is illustrated in 
Fig.~\ref{fig:SmoothRAhQ} for one particular control 
system.\footnote{
  The $Q(t)$ illustrated here is the one for the
  control system that
  computes a smooth approximation $\distorted{r}_{\rm AH}^{appx}(t)$ 
  to the average
  horizon radius; this approximate value is used to compute 
  $\dot{\distorted{S}}_{00}$ and $\Delta r$
  in Eq.~(\ref{eq:State2Q}), in order to reduce
  the number of calls to the computationally expensive horizon finder
  (see section 7 and Eq.~(108) of~\cite{Hemberger:2012jz} for details).}
In Fig.~\ref{fig:SmoothRAhQ} and the remainder of the paper, 
$M\equiv m_A+m_B$ is the sum of the Christodoulou masses of the two
black holes at the time $t_{\rm relax}$ when the
initial ``junk radiation'' transients have decayed away.

\subsubsection{Size control: switching between Eqs.~(\ref{eq:State2Q}) and~(\ref{eq:CharSpeedControlQ}).}\label{sec:switching}

At every time step, the control system for $\lambda_{00}$
is governed by a $Q$ given by either 
Eq.~(\ref{eq:State2Q}) or Eq.~(\ref{eq:CharSpeedControlQ}), with
an associated damping timescale $\tau_d$ and (if using characteristic
speed control) a target speed $v_T$.  At regular intervals (typically
every time step), the algorithm has an opportunity to change
from using Eq.~(\ref{eq:State2Q}) to using Eq.~(\ref{eq:CharSpeedControlQ})
or vice versa,
and to choose a new value of $\tau_d$ and (if using characteristic
speed control) $v_T$.  Here we describe how we make these choices.
A previous version of this algorithm
was described in~\cite{Hemberger:2012jz}, but many improvements have
been made since then.

Because the goal of the $\lambda_{00}$ control system is to keep both
$v$ and $\Delta r\sub{min}$ positive, we regularly monitor $v$ and
$\Delta r\sub{min}$ as functions of time.  We predict whether either
of these quantities is likely to become negative in the immediate
future, and if so, we estimate the timescale $\tau_v$ or $\tau_{\Delta
  r\sub{min}}$ on which this will occur, using the method described in
Appendix C of~\cite{Hemberger:2012jz}. Because the sign of $v_c$ is important
to the success of horizon tracking, we also monitor $v_c$ as a function
of time, and if it is positive and decreasing, we predict
the timescale $\tau_{v_c}$ on which it will become negative.
If $v$, $v_c$,
or $\Delta r\sub{min}$ are increasing instead of decreasing, we define
the corresponding timescale $\tau_v$, $\tau_{v_c}$, or 
$\tau_{\Delta r\sub{min}}$ to be infinite.

We begin by determining whether $v$ is in imminent danger of becoming 
negative, so that some immediate 
action must be taken to prevent this
from occurring.  We
regard $v$ to be in danger if $\tau_v<\tau_d$ and $\tau_v<\tau_{\Delta
  r\sub{min}}$.
Furthermore, if characteristic speed control is in
effect, we additionally require 
$\tau_v<\sigma_3 \tau_d$ and $v<\sigma_4 v_T$ 
to
deem $v$ in danger; 
here $\sigma_3\lesssim 1$ and $\sigma_4\sim 1$ are constants,
typically chosen to be 0.99 and 1.1, 
respectively\footnote{Labels for control system constants like
$\sigma_i$ and $\eta$ are consistent with the notation
in Ref.~\cite{Hemberger:2012jz}.}. The 
first requirement, $\tau_v<\sigma_3 \tau_d$, prevents the algorithm
from switching back and forth between characteristic speed control and
horizon tracking on each time step.
The second requirement, $v<\sigma_4 v_T$, prevents the
control system from rapidly {\em decreasing} the characteristic speed
to achieve a target $v_T$ that is less than $v$. 

If $v$ is deemed to be in danger, the action taken to
prevent $v$ from becoming negative depends on the current state of
the control system.  If characteristic speed
control is in effect, then it remains in effect, and $\tau_d$ is set
equal to $\tau_v$ in order to drive $v$ towards $v_T$ more quickly.
If horizon tracking is in effect, and if $v_c<0$ or $v_c$ is
decreasing, then characteristic speed control goes into effect, with
$v_T=\sigma_5 v$, and $\tau_d$ is left unchanged. The 
constant $\sigma_5$, typically 1.01,
prevents the control system from switching back and forth on each
timestep.  Finally, if horizon tracking is in effect, and if $v_c>0$
and $v_c$ is nondecreasing, then horizon tracking remains
in effect and we reduce $\tau_d$ by a factor of $\sigma_6<1$ (typically
0.99). This change is all that is required
because horizon tracking will
drive $v$ toward $v_c$, which is in no danger of becoming negative.

If $v$ is deemed not to be in danger, then we check whether $\Delta
r\sub{min}$ is in danger of soon becoming negative. We regard $\Delta
r\sub{min}$ to be in danger if $\tau_{\Delta r\sub{min}}<\tau_v$ and
if $\tau_{\Delta r\sub{min}}<\sigma_1\tau_d$, where $\sigma_1$ is
a constant typically chosen to be 20. Furthermore, if horizon
tracking is in effect, we additionally require $\tau_{\Delta
  r\sub{min}}<\sigma_7\tau_d$ to deem $\tau_{\Delta r\sub{min}}$ in
danger, where $\sigma_7<1$ is usually chosen to be 0.99; 
this condition prevents the control system from switching
on every time step.

\begin{figure}
\includegraphics[width=0.95\columnwidth]{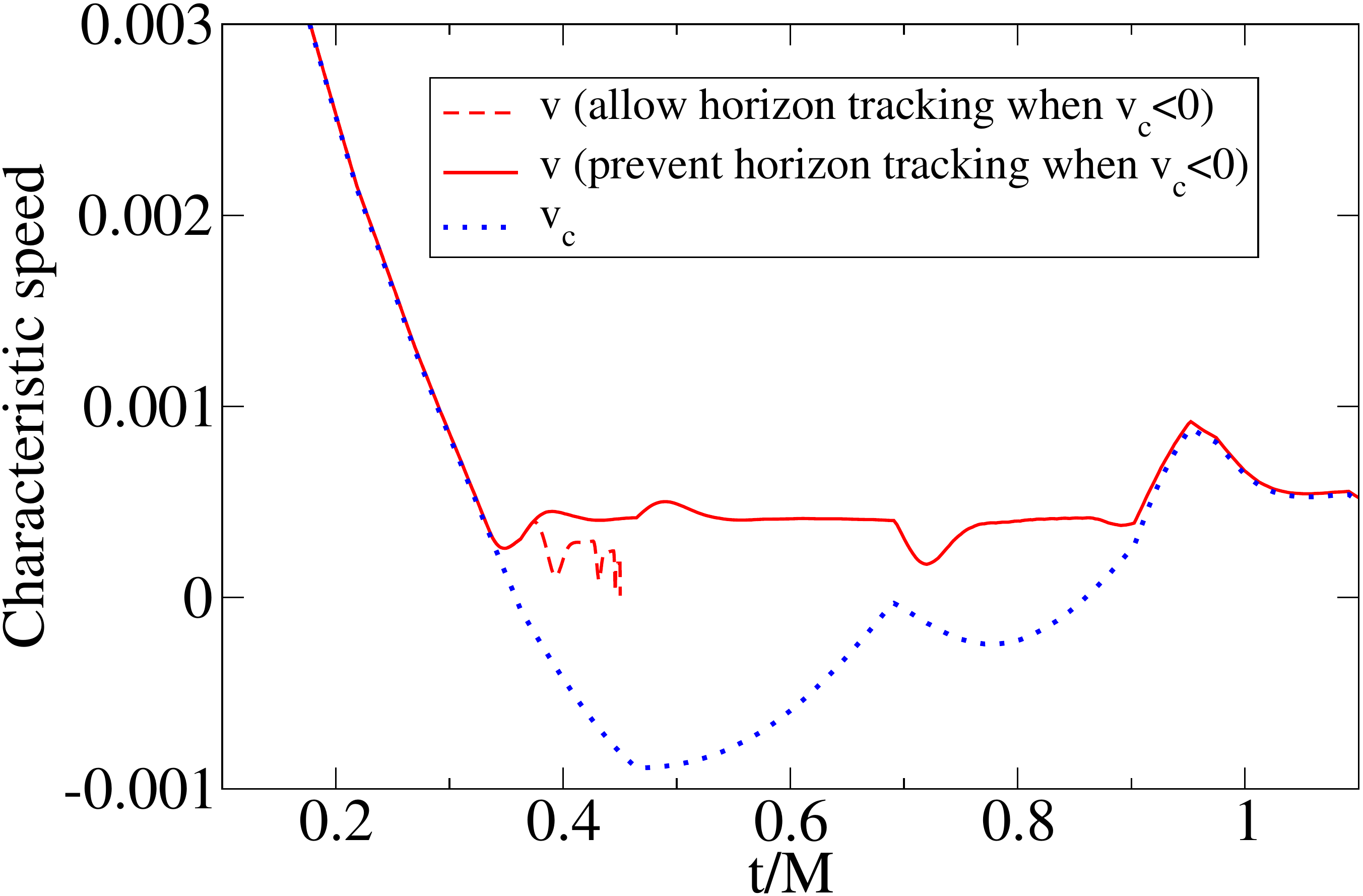}
\caption{Characteristic speed $v$ and comoving characteristic speed $v_c$
  for two different \spin{994} simulations that
  differ only in the algorithm for treating the situation
  in which $\Delta r\sub{min}$ is deemed in danger while characteristic
  speed control is in effect and $v_c<0$.
  The dashed red curve shows $v$ for a simulation in which 
  horizon tracking becomes active in this situation; the code
  crashes early, at only $t\sim 0.4M$.
  The solid red curve shows $v$ for a simulation in which
  for this situation characteristic
  speed control remains in effect, but the target characteristic
  speed is reduced as described in the text. The quantity $v_c$ is
  the same for both simulations.
  \label{fig:995State1Or2}
}
\end{figure}

If $\Delta r\sub{min}$ is in danger, the action again depends on the
state of the control system and other variables.  If horizon tracking
is in effect, then it remains in effect, and $\tau_d$ is set equal to
$\tau_{\Delta r\sub{min}}$ in order to drive $\Delta r\sub{min}$ to a
constant more quickly.  If characteristic speed control is in effect
and if $v_c>0$, then horizon tracking goes into effect, and $\tau_d$
is set equal to $\tau_{\Delta r\sub{min}}$.  We require $v_c>0$ to
activate horizon tracking because horizon tracking drives $v$ towards
$v_c$, and we wish to maintain $v>0$; if horizon tracking
becomes active even if $v_c<0$, the simulation often fails,
as shown in Fig.~\ref{fig:995State1Or2}.  To solve
the problem illustrated by Fig.~\ref{fig:995State1Or2},
when the code finds that $\Delta r\sub{min}$ is in danger while 
characteristic speed control
is in effect and if $v_c<0$, then 
the code allows characteristic speed control to remain in
effect, but it sets the
new $\tau_d$ to $\min(\tau_d,\tau_{\Delta r\sub{min}})$, and
it reduces $v_T$ to $\eta v$, 
where $\eta<1$ is a constant typically chosen to be 0.125.
Reducing the target $v_T$ will reduce $v$ but will increase 
$\Delta r\sub{min}$.  If $v_c<0$ for
an extended period of time, several such reductions of $v_T$ will
occur as needed.  As mentioned above, if $v_c<0$ and remains so, this
algorithm must eventually fail; the way to prevent such
a failure is to adjust the control systems {\em other than} the
one for $\lambda_{00}$ to attempt to make $v_c$ positive, 
as discussed in Sec.~\ref{sec:impr-gain-sched}.

If neither $v$ nor $\Delta r\sub{min}$ are in imminent danger of
becoming negative, then the system attempts to find an equilibrium
using horizon tracking.  If characteristic speed control is in effect,
and if $v_c>0$, $\dot{v_c} \ge 0$, 
and either $v>v_T$ or $v_c>v$,
then horizon tracking goes into effect, using the current $\tau_d$.
However, if both $v$ and $v_c$ are decreasing, horizon tracking does
not go into effect unless $v$ is decreasing faster than $v_c$ and 
$\tau_{v_c}>\sigma_2 \tau_d$, where $\sigma_2$ is a constant
we usually set to 5.
The purpose of these various conditions on $v$, $v_c$, and their
derivatives and predicted zero-crossing times
is to prevent horizon tracking from going into effect
when it is likely that a switch back to characteristic
speed control will soon be necessary. For example, if $\dot{v_c}<0$ and $v_c$
is decreasing faster than $v$, then we anticipate that $v_c$ will soon become 
negative, in which case horizon tracking is inappropriate because it would
drive $v$ towards zero.  

The behavior of the control system depends on various
constants $\sigma_i$ $(1<i<7)$ and $\eta$ described above; these constants
govern decisions made by the algorithm.  These constants
have restricted values (e.g. $\eta$ should not be greater than unity),
but they were
chosen without any fine tuning.  Changing their values slightly will
change details such as the exact value of $\tau_d$ at a particular
timestep, but we expect that small changes in parameters
will not change whether a simulation succeeds or fails,
and will change physical results only at the level of 
truncation error (because the
control system changes the grid coordinates). 

Occasionally when horizon tracking is in effect, we find that the
value of $\Delta r\sub{min}$ is excessively large or small.  If it is
excessively small, then $\tau_d$ becomes small, and we are forced to
reduce the timestep in the evolution equations to keep the control
system stable, resulting in a large computational expense.  If it is
too large, then the excision boundary lies deep inside the horizon,
and excessive computational resources are needed to resolve the large
gradients.  Therefore, we allow a drift term to sometimes be added to
Eq.~(\ref{eq:State2Q}), as discussed in~\cite{Hemberger:2012jz}.

\section{Simulations}\label{sec:simulations}

We present three new simulations, summarized in Table~\ref{table:runs}.
We will refer to quantities defined in Table~\ref{table:runs}
throughout the remainder of this paper.
The techniques described in Sec.~\ref{sec:techniques} were essential
to the successful completion of these simulations.

\begin{table*}
\begin{tabular}{lc|rrccccccccc|rc|cc}
Name & Catalog ID & $t_{\rm relax}$ & $q^r$ & $m_A^r$ & $m_B^r$ & $M\omega_{\rm orb}^r$ & $\chi_A^r$ & $\theta_A^r/\pi$ & $\phi_A^r/\pi$ & $\chi_B^r$ & $\theta_B^r/\pi$ & $\phi_B^r/\pi$  & $10^{4}e$ & $N$ & $M_f$ & $\chi_f$ \\
\hline \hline
$S^{++}_{0.99}$ & SXS:BBH:0177 & 320.0 & 1.0 & 0.5 & 0.5 & 0.0154 & 0.989 & 0.00 & -- & 0.989 & 0.00 & -- & 12.6 & 25.4 & 0.888 & 0.949 \\
$S^{++}_{0.994}$ & SXS:BBH:0178 & 640.0 & 1.0 & 0.5 & 0.5 & 0.0157 & 0.994 & 0.00 & -- & 0.994 & 0.00 & -- & 8.6 & 25.4 & 0.887 & 0.950 \\
$S^{0.99}_{0.20}$ & SXS:BBH:0179 & 380.0 & 1.5 & 0.6 & 0.4 & 0.0148 & 0.991 & 0.00 & 0.73 & 0.200 & 0.24 & 0.23 & 322.4 & 23.8 & 0.922 & 0.897 \\

\end{tabular}
\caption{Summary of physical simulation parameters. Data are
    publicly
    available
    online~\cite{SXSCatalog} indexed by their Catalog ID.
    Quantities with an $r$ superscript are reported at time $t=t_{\rm relax}$, 
    the time after the initial ``junk radiation'' transients have
    settled down:
    $q$ is the mass ratio, $m_H$ is
    the Christodoulou mass
    of an
    individual black hole (where $H$ represents black hole $A$
    or $B$), $M\omega_{\rm orb}$ is the orbital frequency,
    $\chi_H$ is the dimensionless spin,
    $\theta_H$ is the angle between $\vec{\omega}_{\rm orb}$ and $\vec{\chi}_H$,
    and $\phi_H$ is the angle between the separation vector
    and the component of $\vec{\chi}_H$
    in the orbital plane.
    The remaining quantities are eccentricity $e$, number of
    orbits $N$ from $t=0$ to merger, final
    Christodoulou
    mass $M_f$, and final
    spin magnitude $\chi_f$.
}
\label{table:runs}
\end{table*}

\subsection{Equal-mass, aligned spins $\chi=0.99$}
\label{sec:Spin99}

\begin{figure}
\includegraphics[width=0.95\columnwidth]{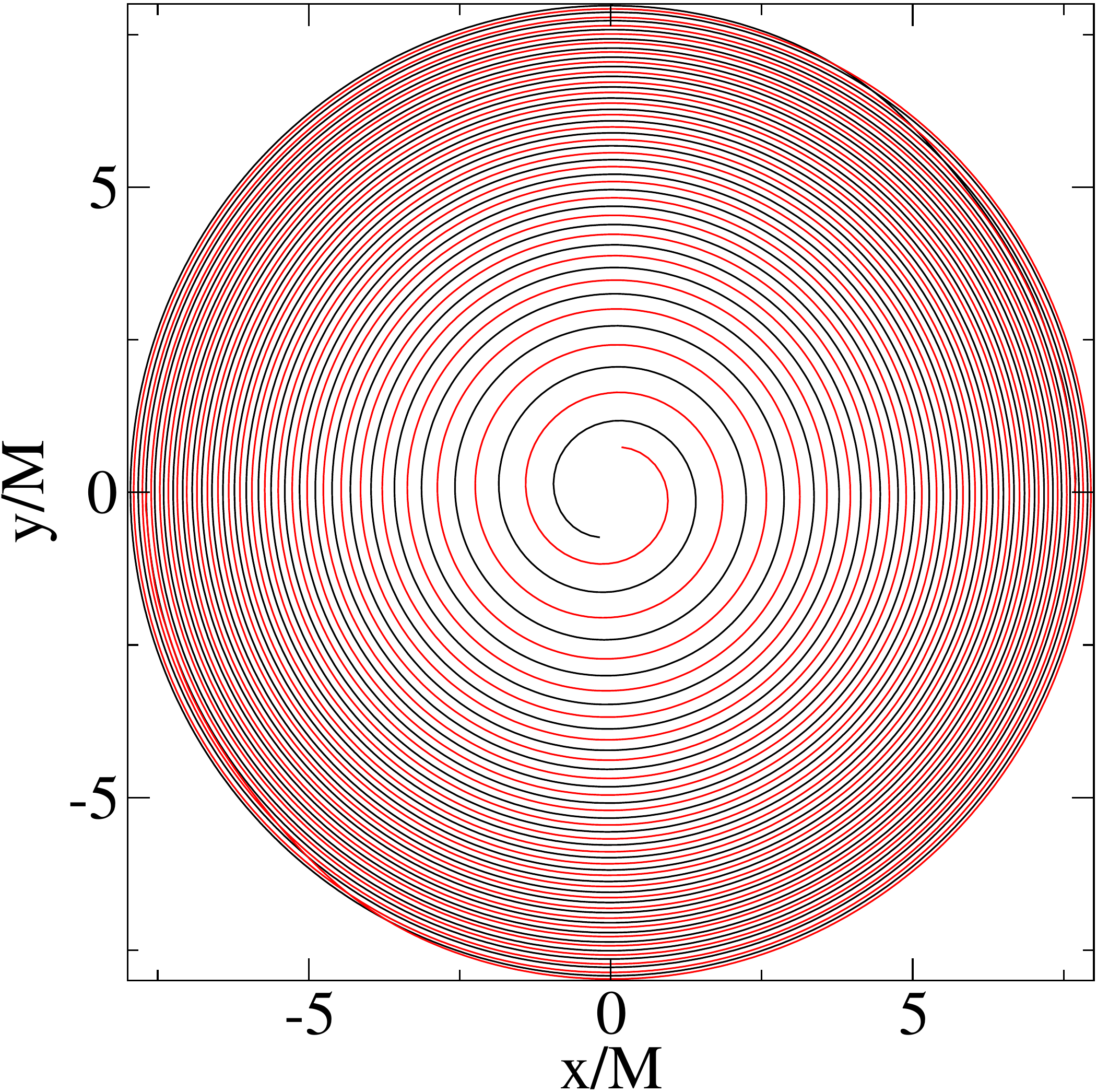}
\caption{The trajectories of the centers of the individual apparent horizons
  for the highest resolution of \spin{99}.
  \label{fig:XYTrajectory99}
}
\end{figure}

The first simulation we present, and refer to as \spin{99}, 
is an equal-mass case in which each
black hole has a spin of $\chi=0.99$ aligned with the orbital angular
momentum.  At $t=t_{\rm relax}$ the simulation
has $M\omega_{\rm orb}=0.0154$, where 
$M$ is the sum of the relaxed
Christodoulou masses. The binary then
evolves through 25 orbits, merger
and ringdown.  This simulation took 83 days on
48 cores for the highest resolution.

\begin{figure}
\includegraphics[width=0.95\columnwidth]{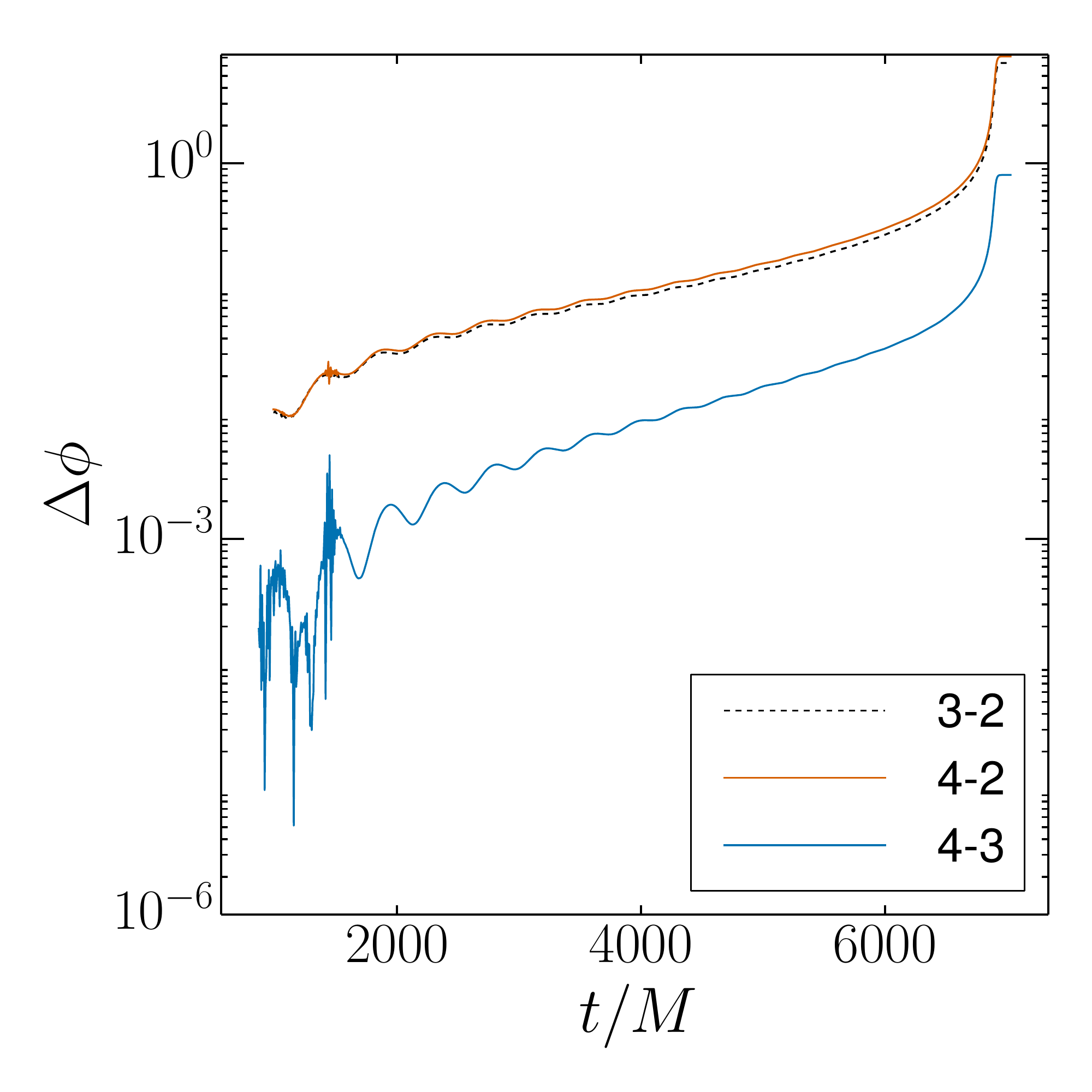}
\caption{Convergence test for \spin{99}.
  Shown are gravitational-wave phase differences between $\Psi_4$
  computed using different values of the numerical resolution parameter
  $N$.  Several differences are shown, and labeled by the values of $N$
  that are compared, e.g. ``3-2'' means $N=3$ versus $N=2$.
  Waveforms are extracted at a finite radius $r=465M$, and no alignment
  of waveforms was performed.
  \label{fig:Conv99}
}
\end{figure}

To assess numerical convergence, we perform several
simulations that are identical except for the numerical
resolution, which we label by an integer $N$.  Larger $N$ corresponds
to finer resolution, but the absolute scale of $N$ is different for
different physically distinct simulations.  The value of $N$ enters
the simulation through the tolerance in adaptive mesh refinement
(AMR): the AMR truncation error
tolerance is chosen to be proportional to $e^{-N}$.
For each value of $N$, we compute the complex phase $\phi$ of
the $\ell=2,m=2$ component of $\Psi_4$.  We then take the difference 
$\Delta\phi$ between 
$\phi$ computed using otherwise-identical simulations 
using different values of $N$. 

Figure~\ref{fig:Conv99} shows these differences for \spin{99}.
No alignment of the
waveforms in time or phase has been performed.  Note the rapid
convergence: $\Delta\phi$ between $N=3$ and $N=4$
(labeled ``4-3") is significantly
smaller than $\Delta\phi$ between the two lower resolutions.  Also
note that the difference
``3-2" is nearly the same as
``4-2",
indicating that this difference effectively
measures the numerical truncation error in the $N=2$ simulation.
Similarly, the difference ``4-3"
represents the numerical truncation error in the $N=3$ simulation.
Furthermore, one would expect that the truncation error in the $N=4$
simulation is smaller than the ``4-3'' curve by another order of
magnitude (although it would be necessary to run an $N=5$ simulation
to actually measure this).

\begin{figure}
\includegraphics[width=0.95\columnwidth]{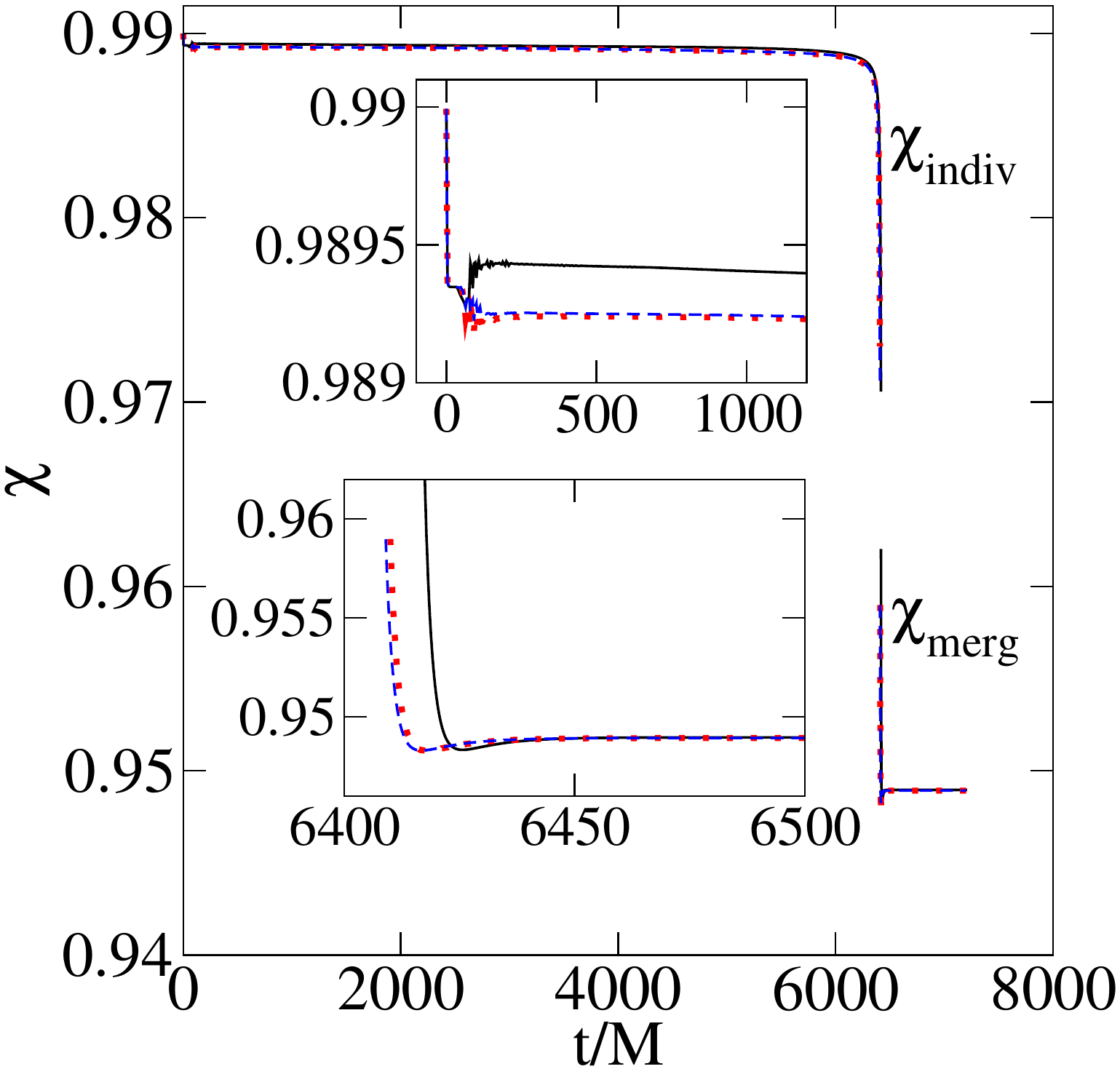}
\caption{Spin magnitude as a function of time for \spin{99}.
  At early times, the spin of one of the apparent horizons is shown
  at resolutions $N=2$ (black solid), $N=3$ (red dotted) and 
  $N=4$ (blue dashed).  A closeup of early times is shown in the
  upper inset.  At late times, the spin of the merged apparent horizon
  is shown as a function of time for the same resolutions, and a closeup
  of late times is shown in the lower inset.
  \label{fig:ChiVsTime99}
}
\end{figure}

In the \spin{99} initial data, the spin of each black
hole is $0.99$.
When the system is evolved, the spins
decrease very slightly for the first $\roughly 10M$ as initial transients
propagate away from the horizons, as shown in the upper inset of
Fig.~\ref{fig:ChiVsTime99}. Then the spins level off and become
roughly constant, but with a small negative slope.  All values of
resolution $N$ agree quite well, and the higher two resolutions are
indistinguishable in Fig.~\ref{fig:ChiVsTime99}.  The spins decrease
more rapidly just before merger ($t\sim 6400 M$). The common horizon
first appears with a spin greater than the final value, and then
relaxes as the remnant black hole settles down, as shown in the lower
inset of Fig.~\ref{fig:ChiVsTime99}.  The final spin is
$\chi_f=0.948927(3)$,
where the
uncertainty is the
difference between the two highest resolution simulations.  

The radiated energy fraction $E_{\rm rad}$
is the relative
change in energy of the binary
from $t=-\infty$ to $t=\infty$ and can be computed from
\begin{equation}
  E_{\rm rad} \equiv 1 - \frac{E_{\infty}}{E_{-\infty}} = 1 - \frac{M_f}{M}.
\end{equation}
The final Christodoulou mass $M_f$ is the
energy of the system at $t=\infty$,
because the remnant is in equilibrium at the end of the simulation;
the total Christodoulou mass $M$ at $t=t_{\rm relax}$
is the energy of the system at $t=-\infty$,
because the individual black-hole masses
change by less than one part in $10^6$ between $t=-\infty$
and $t=t_{\rm relax}$ (see, e.g. Eq.~14 in Ref.~\cite{Alvi:2001mx}).
We find that
$E_{\rm rad} = 11.26593(3)\%$, where the uncertainty is again the
difference between
the two highest resolutions.

The formulas
from Ref.~\cite{Hemberger:2013hsa} predict
$\chi_f = 0.94933(8)$ and a radiated energy fraction $E_{\rm rad} = 11.24(2)\%$,
in good agreement with the simulations.
While the fractional differences between the measured and predicted values
are small, their uncertainty intervals are disjoint, i.e. our measurements
lie outside the uncertainty interval of the formulas.
This is because the error estimates in Ref.~\cite{Hemberger:2013hsa}
did not account for the observed correlated trends in the fit residuals
(as seen in the lower panels of Figs.~6 and~8 of Ref.~\cite{Hemberger:2013hsa}).
As a result,
extrapolating these formulas to initial spins above $\chi=0.97$
is expected to overestimate
the final spin (see Fig.~6 in Ref.~\cite{Hemberger:2013hsa}) and
underestimate the final radiated energy
(see Fig.~8 in Ref.~\cite{Hemberger:2013hsa}), and this is what we
find
with \spin{99}.

\subsection{Equal-mass, aligned spins $\chi=0.994$}
\label{sec:Spin994}

\begin{figure}
\includegraphics[width=0.95\columnwidth]{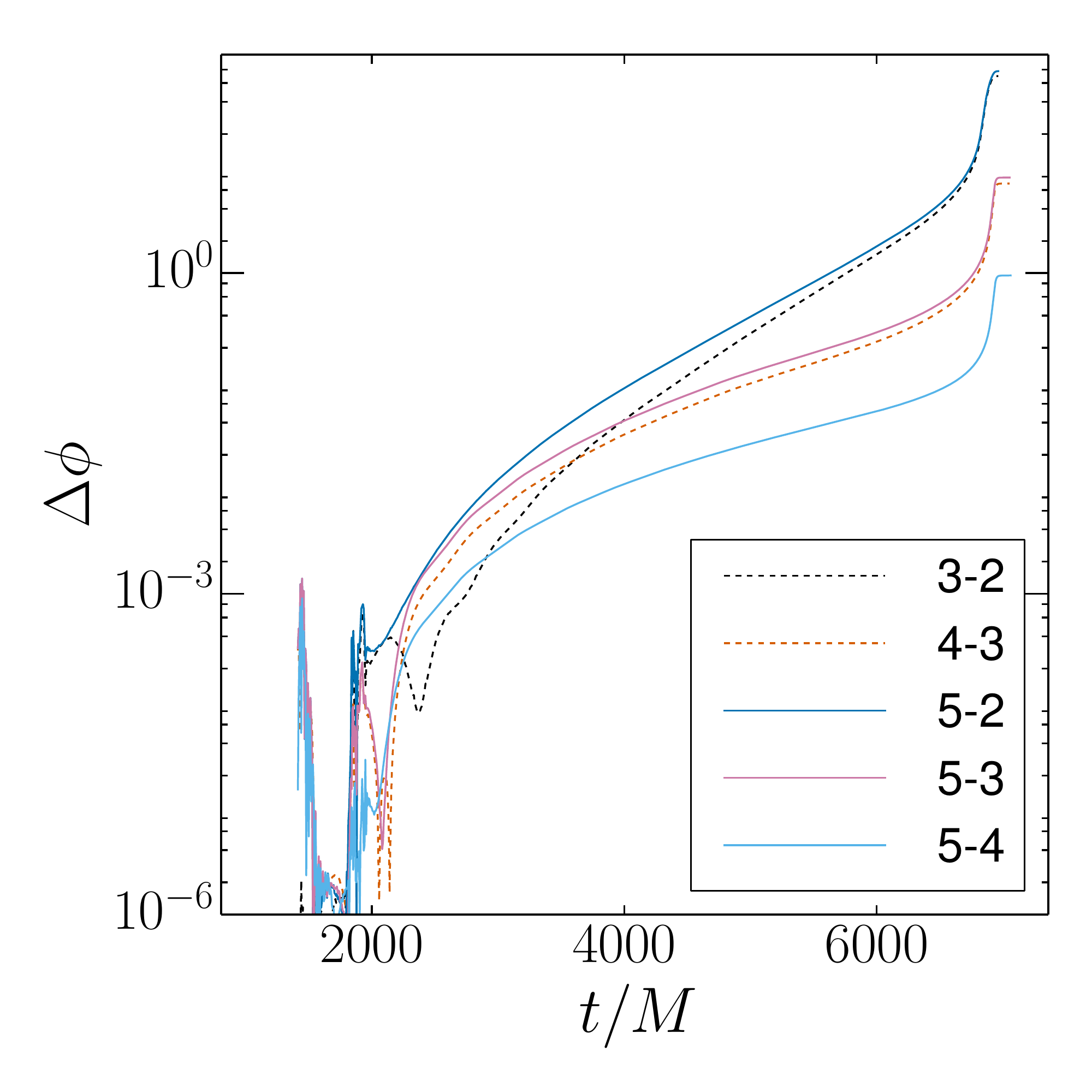}
\caption{Convergence test for \spin{994}.
  Labels are the same as for Fig.~\ref{fig:Conv99}.
  For $N\neq 5$, the simulations were started at $t_{\rm branch}=1414M$, 
  using the 
  $N=5$ solution as initial data.
  \label{fig:Conv995v2}
}
\end{figure}

We repeated the equal-mass aligned-spin simulation above, but with a
larger spin.  
We refer
to this case as \spin{994}. 
The initial data were chosen with $\chi=0.995$ for each
black hole, but the spins drop to $\chi=0.9942$ after about $t=10M$
of evolution time, a much smaller timescale than the relaxation
time $t_{\rm relax}$ (this rapid initial decrease in spin
can also be seen for \spin{99} in the upper inset of 
Fig.~\ref{fig:ChiVsTime99}).
The simulation \spin{994} represents
the largest spin ever simulated for a black-hole binary.  
It has
$M\omega_{\rm orb}=0.0157$
at $t=t_{\rm relax}$, and then
proceeds through 25
orbits, merger, and ringdown.
The highest resolution completed in
approximately
71 days on 48 cores.
Note that this simulation, \spin{994}, was computationally cheaper
than the lower-spin simulation, \spin{99}, and achieved a smaller
overall phase error (see Figs.~\ref{fig:Conv99} and~\ref{fig:Conv995v2}).
This is due to code optimization that was done between the time
that the \spin{99} and \spin{994} simulations were carried out; for the
same version of SpEC, there is actually a steep increase in
computational cost as a function of spin.

Obtaining convergence was more difficult for this simulation than for
\spin{99}.  The reason is that it is difficult to fully
resolve the initial transients, sometimes called ``junk radiation'',
that result from imperfect initial data.  If these transients are
unresolved, then the small changes in masses, spins, and trajectories
caused by these transients are effectively random, and therefore
otherwise-identical simulations run with different values of
resolution $N$ will differ by random small amounts that will not
converge with increasing $N$.  So to investigate convergence, we
remove the initial transients in the following way.  We first carry
out a simulation with one value of $N$, call it $N_{\rm base}$.  In
the case of \spin{994}, $N_{\rm base}$ represents the highest
resolution.  Then we choose some fiducial time $t=t_{\rm branch}>t_{\rm relax}$ at
which we decide that the transients have decayed away.  We then carry
out simulations with $N \neq N_{\rm base}$ starting at $t=t_{\rm branch}$, 
using the $N=N_{\rm base}$ solution as initial data.  This
procedure removes the effects of the transients
from our convergence tests.

However, this procedure alone was insufficient to achieve convergence.
When convergence is rapid enough in a particular subdomain so that
adding a single grid point results in a large decrease in truncation error,
it is possible for two different AMR truncation error tolerances,
e.g. $e^N$ and $e^{N-1}$, to result in the same number of grid points
for that subdomain.
This makes the truncation error in that subdomain identical
for two different values of $N$, 
which spoils convergence tests for simulations
with those values of $N$.
To remedy this problem in such cases, we increase the spacing
in truncation error tolerance as a function of level $N$:
the truncation error tolerance is proportional to
$10^N$ instead of $e^N$.
This, combined with the procedure
to remove the effect of transients, results in good
convergence, as shown in Fig.~\ref{fig:Conv995v2}.

The spin of the remnant black hole is
$\chi_f = 0.949931(5)$
and the radiated energy fraction is
$E_{\rm rad} = 11.351(5)\%$.  The formulas in
Ref.~\cite{Hemberger:2013hsa} predict
$\chi_f = 0.95021(8)$ and
$E_{\rm rad} = 11.30(2)\%$,
in good agreement with the simulations. However, the
uncertainty intervals of the measured and predicted values
are disjoint
for the same reason as
explained in Sec.~\ref{sec:Spin99}.

\subsection{Unequal-mass, precessing}
\label{sec:spin-0.99-precessing}

The final simulation we present is an unequal-mass case with $q=1.5$,
in which the larger black hole has a spin of $\chi = 0.99$ aligned with
the orbital angular momentum, while the smaller black hole has a
spin magnitude of $\chi = 0.2$ in an arbitrary direction
misaligned
with the orbital angular momentum.
We will refer to this case as 
\generic, using a notation similar to that introduced earlier.
The simulation has
$M\omega_{\rm orb}=0.0148$
at $t=t_{\rm relax}$, and then
proceeds through 23 orbits, merger, and ringdown.
This simulation took approximately
26 days on 48 cores for the highest resolution
using
the same optimized version of SpEC as the
\spin{994} case described in
Sec.~\ref{sec:Spin994}.

\begin{figure}
\includegraphics[width=0.95\columnwidth]{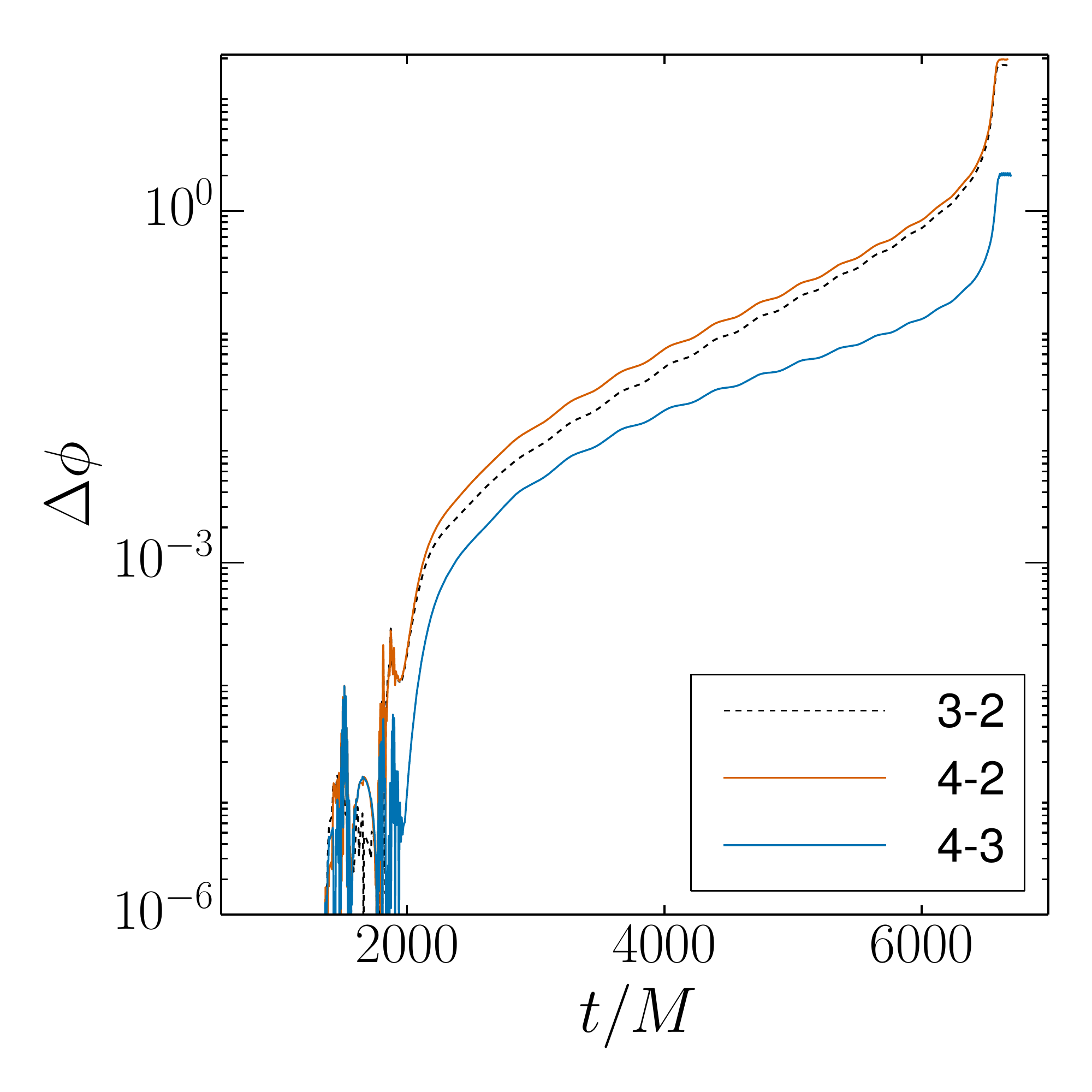}
\caption{Convergence test for \generic.
  Labels are the same as for Figure~\ref{fig:Conv99}.
  For $N\neq 4$, the simulations were started at $t_{\rm branch}=1362M$, 
  using the $N=4$ solution as initial data.
  \label{fig:ConvPostUltimate}
}
\end{figure}

We found that for \generic\ we needed to remove the effect of
unresolved initial transients
and increase the spacing in AMR truncation error tolerance
to obtain acceptable convergence results.
To do this we followed
the same procedure as for \spin{994}, described in
Sec.~\ref{sec:Spin994}. Figure~\ref{fig:ConvPostUltimate} shows
good convergence of the gravitational-wave
phase difference when using this procedure.

\begin{figure}
\includegraphics[width=0.95\columnwidth]{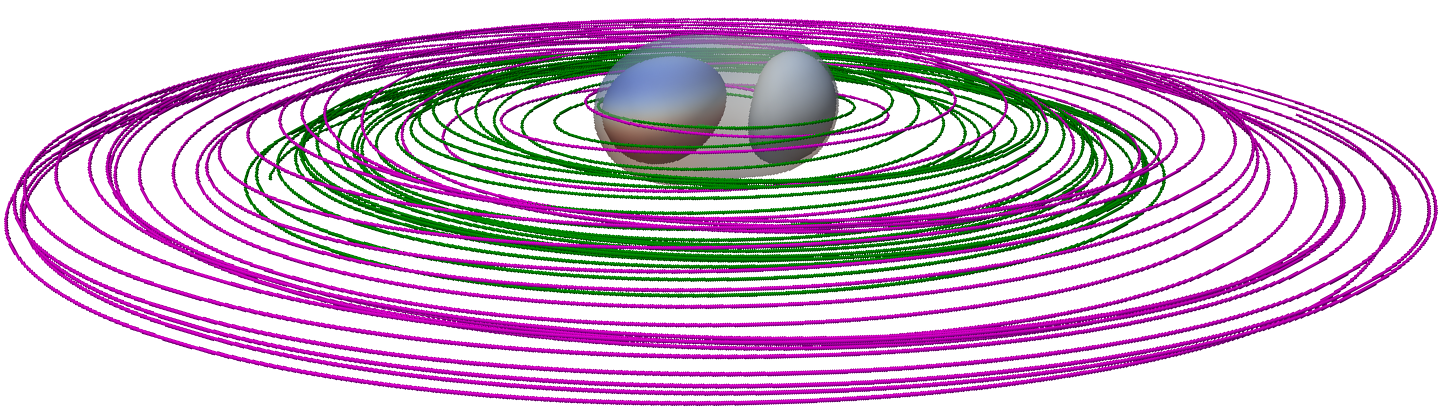}
\caption{Coordinate trajectories (green and purple lines)
  of the black holes and coordinate
  shapes of the individual and common
  apparent horizons (surfaces) at the moment of merger, for
  \generic.
  The horizons are colored according to
  their
  vorticity~\cite{OwenEtAl:2011}. 
  \label{fig:PostUltimateTraj1}
}
\end{figure}

Figure~\ref{fig:PostUltimateTraj1} shows the trajectories of the centers of the
apparent horizons for this simulation, as well as the individual
apparent horizons and the common apparent horizon at the moment when
the common horizon first appears.  Trajectories and horizon shapes are
shown in the asymptotically inertial
coordinate system used in the simulation.  Because the
spin of the smaller hole $\vec{\chi}_B$
is not aligned with the orbital angular
momentum, the system precesses, so the trajectories do not lie in a
plane.  

\begin{figure}
\includegraphics[width=\columnwidth,trim={0 1cm 0 1cm},clip=true]{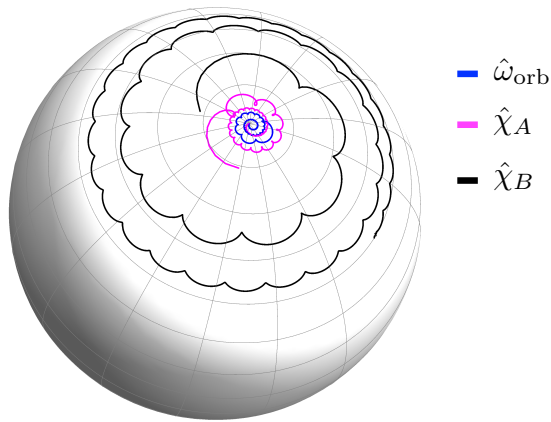}
\caption{Precession of the spins and orbital frequency
  for the highest-resolution simulation $N=4$ of \generic.
  The unit vector spins,  $\hat{\chi}_A$ and $\hat{\chi}_B$,
  and orbital frequency $\hat{\omega}_{\rm orb}$ trace the precession
  on the unit sphere.
  The precession
    curves for $N=2$ and $N=3$ converge to the $N=4$ curves
    shown here, and the curves for $N=3$ and $N=4$ are nearly 
    indistinguishable.
  \label{fig:precession}
}
\end{figure}

Figure~\ref{fig:precession} shows the precession of the spin and
orbital frequency vectors in \generic.
The spin $\vec{\chi}_A$ and orbital frequency $\vec{\omega}_{\rm orb}$
initially point along the $z$-axis.
Because the misaligned spin $\vec{\chi}_B$ is on the smaller black hole and
is much smaller in magnitude than $\vec{\chi}_A$, it has a minimal effect
on the orbital dynamics, so $\vec{\chi}_A$ and
$\vec{\omega}_{\rm orb}$ remain near the $z$-axis throughout the simulation.
Therefore, we consider the precession to be mild.
As angular momentum is carried away by gravitational radiation,
the opening angles of the precession cones change.
The angles of
$\vec{\omega}_{\rm orb}$ and $\vec{\chi}_A$ with respect to the $z$-axis
increase from $0^{\circ}$ at $t=0$ to
$6^{\circ}$ and $12^{\circ}$, respectively, at the time of merger.
In contrast, the angle of $\vec{\chi}_B$ with respect to
the $z$-axis decreases from $45^{\circ}$ to $12^{\circ}$.
The spins $\vec{\chi}_A$ and $\vec{\chi}_B$ complete 2.1 and 2.5 precession
cycles, respectively, and 
$\vec{\omega}_{\rm orb}$ completes 2.4
precession cycles.

The spin of the remnant black hole is $\chi_f = 0.89692(5)$, and
the radiated energy fraction is $E_{\rm rad} = 7.8560(8)\%$.
The formulas from Healy et al. (2014)~\cite{Healy:2014yta} predict
$\chi_f = 0.89686$ and $E_{\rm rad} = 7.8365\%$.
Even though these predictions lie outside the numerical uncertainty of the
measured values, the agreement is quite good.\footnote{
To evaluate the quantities $S_{||}$ and $\Delta_{||}$
in Ref.~\cite{Healy:2014yta}, we used the
$z$-component of $\vec{S}$ and $\vec{\Delta}$
at $t_{\rm relax}$, which should be strictly valid only
for non-precessing binaries. 
Also, the formula for $\chi_f$
in Ref.~\cite{Healy:2014yta} requires evaluating certain quantities at the 
innermost stable circular orbit (ISCO) of a Kerr black hole with a spin of 
$\chi_f$, so that $\chi_f$ is not given in closed form; for simplicity
we evaluate the ISCO quantities using the measured $\chi_f$ from the
simulation.
}

\section{Results}\label{sec:results}

\subsection{Spin evolution during inspiral}

During the inspiral,
the tidal field of each black hole affects its companion,
and this interaction slowly changes the black-hole
masses and spins
as a function of time.  For aligned spins,
Alvi~\cite{Alvi:2001mx} has derived perturbative expressions for the
time rate of change of the mass and spin of a black hole in a binary.
Chatziioannou, Poisson, and Yunes (hereafter
CPY)~\cite{Chatziioannou:2013},
have recently extended these expressions to
higher order in perturbation theory.
Although CPY's expressions in Ref.~\cite{Chatziioannou:2013} are
computed to 1.5PN beyond leading order
(i.e. terms in $dS/dt$ proportional to $v^{15}$ and terms
in $dM/dt$ proportional to $v^{18}$,
where $v^2=M/r$ is the PN expansion parameter), 
their 1.5PN terms are incorrect and will
be corrected soon~\cite{Chatziioannou:2014}; so here we will truncate
CPY's expressions to 1PN order.

In our simulations we track the apparent horizons as a function of
time, and at frequent time intervals we measure both the surface area
and the spin of the horizons.  The spin computation is carried out
using the approximate Killing vector formalism of Cook, Whiting, and
Owen~\cite{Cook2007, OwenThesis}.  The mass of the black hole is then
computed using Christodoulou's formula.
We compare our numerical results to the
analytic results of Alvi and CPY.

\begin{figure}
\includegraphics[width=0.95\columnwidth]{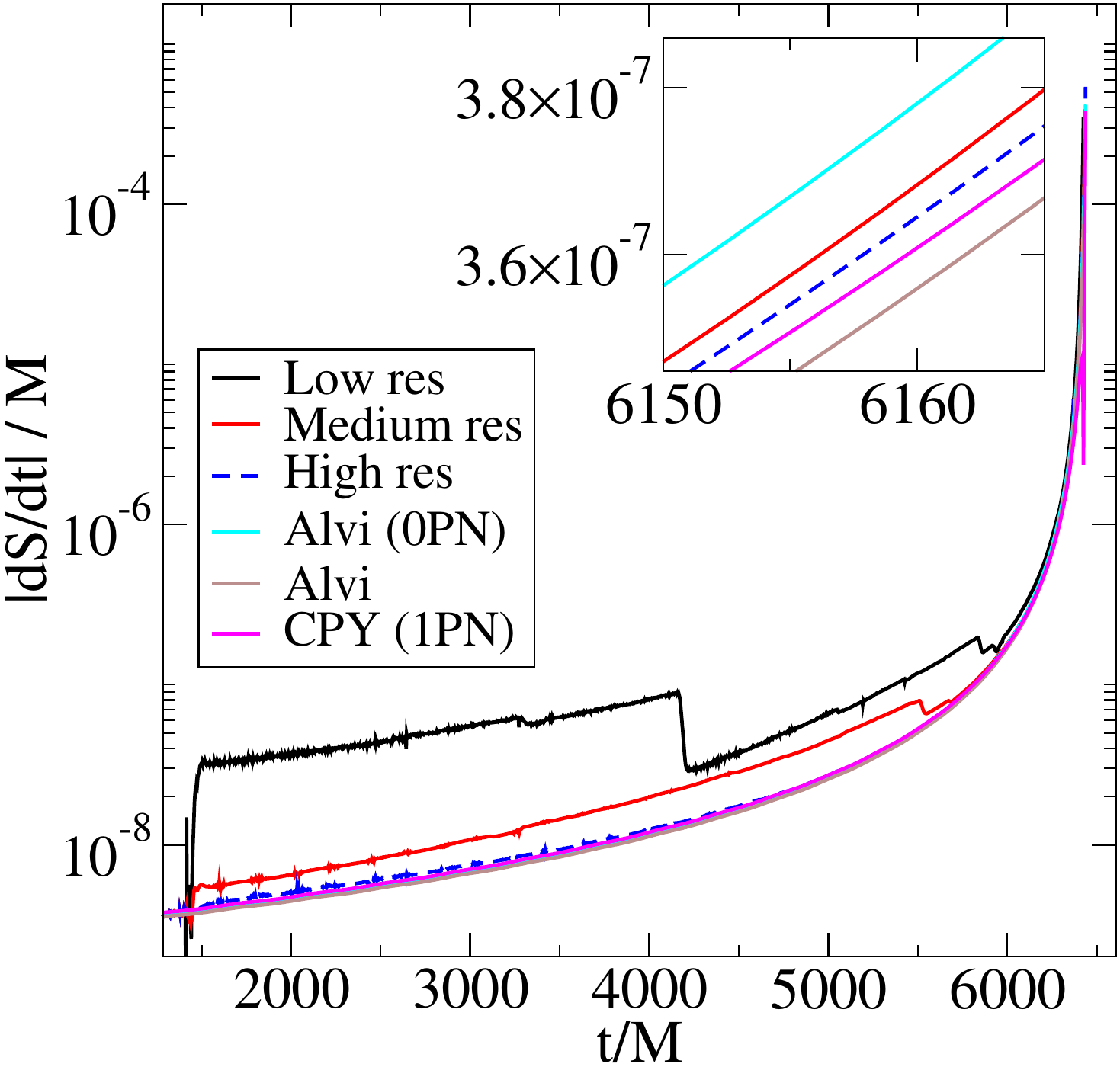}
\caption{Magnitude of $dS/dt$ of one of the black holes from \spin{994}.
  Shown are three numerical resolutions,
  Alvi's expression as written (Eq.~(11) of~\cite{Alvi:2001mx}),
  Alvi's expression truncated to leading order, and the CPY 
  expression\cite{Chatziioannou:2013,Chatziioannou:2014} to 1PN order.  
  The inset zooms closer to the high-resolution
  numerical curve.
  \label{fig:dSdtComparison}
}
\end{figure}

\begin{figure}
\includegraphics[width=0.95\columnwidth]{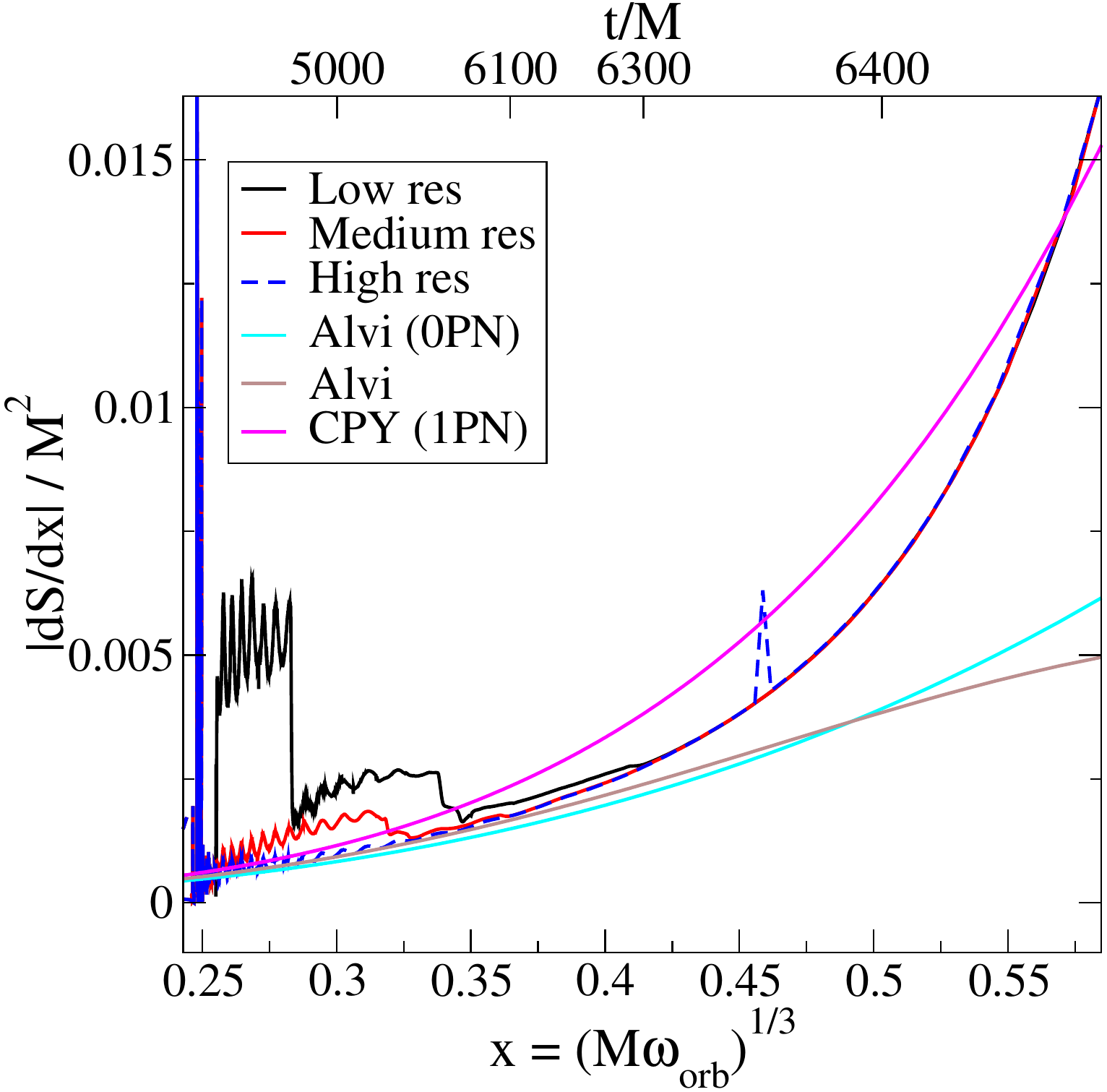}
\caption{Magnitude of $dS/dx$, where $x\equiv(M\omega_{\rm orb})^{1/3}$, 
  of one of the black holes from \spin{994}.
  The top horizontal axis shows $t/M$ of
    the highest-resolution numerical simulation, for comparison
    with values of $x$.
  \label{fig:dSdxComparison}
}
\end{figure}

To compare a black-hole mass or spin from a numerical simulation to
that of a perturbative expression, the two quantities must be compared
at the same event along the black hole trajectory.  Although {\it
  waveform} quantities at future null infinity
computed by numerical simulations
are routinely compared with waveforms computed by PN
expansions, it is not straightforward to compare near-zone quantities
like black-hole masses and spins
because of gauge ambiguities.  Here we make two comparisons.
The first compares quantities at the same numerical and perturbative
$t$ coordinate. The second assumes
that the orbital angular
velocity $\omega_{\rm orb}=d\phi/dt$ of the black hole in the 
numerical simulation can
be equated with that of the perturbative expression.  
Note that in
both the numerical and perturbative cases, the $t$ coordinate becomes
the Minkowski $t$ at infinity, and the $\phi$ coordinate is periodic.
Because of the approximate helical Killing vector 
$d/dt + \omega_{\rm orb} d/d\phi$, $\omega_{\rm orb}$ is approximately
an angular velocity at infinity.
Therefore, one might hope that equating the perturbative and numerical 
$\omega_{\rm orb}$ yields
better agreement than, e.g., equating the radial coordinate
$r$ of the simulation with that of perturbation theory.

Figure~\ref{fig:dSdtComparison} compares the magnitude of $dS/dt$ of
one of the black holes for \spin{994} with the
expressions of both CPY and Alvi.  We include numerical results for
three resolutions in Fig.~\ref{fig:dSdtComparison} because the
magnitude of $dS/dt$ is extremely small and difficult to
resolve. Indeed, the lowest resolution fails to resolve $dS/dt$ until
around $t=6000M$, when $dS/dt$ grows to about $10^{-7} M$, and the medium
resolution fails to resolve $dS/dt$ only slightly earlier. Note that
Alvi's expression 
includes some 1.5PN terms,
but ignores
1PN effects such as magnetic-type tidal perturbations and the
difference between the global PN time coordinate
and the local time coordinate of a frame moving along with one
of the black holes.  Therefore, we plot both Alvi's expression in its
entirety, and Alvi's expression truncated to lowest (0PN) order.  The
CPY expression includes 0PN and 1PN
terms.  The CPY and Alvi expressions agree to 0PN order.

Figure~\ref{fig:dSdtComparison} shows overall excellent
agreement between the PN
and numerical simulation results.
All the perturbative curves agree within our
our numerical error up to $t\sim 6100M$, but
for $t > 6150M$ none of the perturbative
approximations agree with the numerical result
within numerical error.
This disagreement
at late times is not surprising
since all the perturbative expressions
should lose accuracy shortly before merger.  
We can eliminate the time coordinate, a possible
source of gauge dependence, by instead
plotting $dS/dx$ versus $x$, where
$x \equiv (M\omega_{\rm orb})^{1/3}$.
This is shown in Fig.~\ref{fig:dSdxComparison}.  
To obtain $dS/dx$ from $dS/dt$ and to obtain $x$ from $t$, it is necessary
to have some function $x(t)$.  For the numerical curves, this function
is obtained from the numerical time coordinate and the numerical orbital
frequency.  For the perturbative curves, this function is
the PN expression for $x(t)$ 
derived from Eq. (4.14) of Ref.~\cite{kidder95}.
Thus, all the numerical curves in Figs.~\ref{fig:dSdtComparison}
and~\ref{fig:dSdxComparison} are independent of any perturbative
assumptions,
and all the perturbative curves in Figs.~\ref{fig:dSdtComparison}
and~\ref{fig:dSdxComparison} are independent of the numerical data,
except that the 
perturbative and numerical $t$ coordinates 
are both represented by the same
horizontal axis of Fig.~\ref{fig:dSdtComparison}, and the
perturbative and numerical $\omega_{\rm orb}$ are
both represented by the same horizontal axis of Fig.~\ref{fig:dSdxComparison}.

In
Fig.~\ref{fig:dSdxComparison}, 
the perturbative and numerical expressions agree early in the inspiral,
but not at late times; this is expected because perturbative expressions
become inaccurate for large $x$.  Alvi's full expression appears
to agree with the numerical simulations slightly better than the others
for small $x$, but that expression diverges from the numerical result
at larger $x$ earlier than the others.
Note that Fig.~\ref{fig:dSdxComparison} emphasizes late times because
the frequency increases very rapidly with time.

\begin{figure}
\includegraphics[width=0.95\columnwidth]{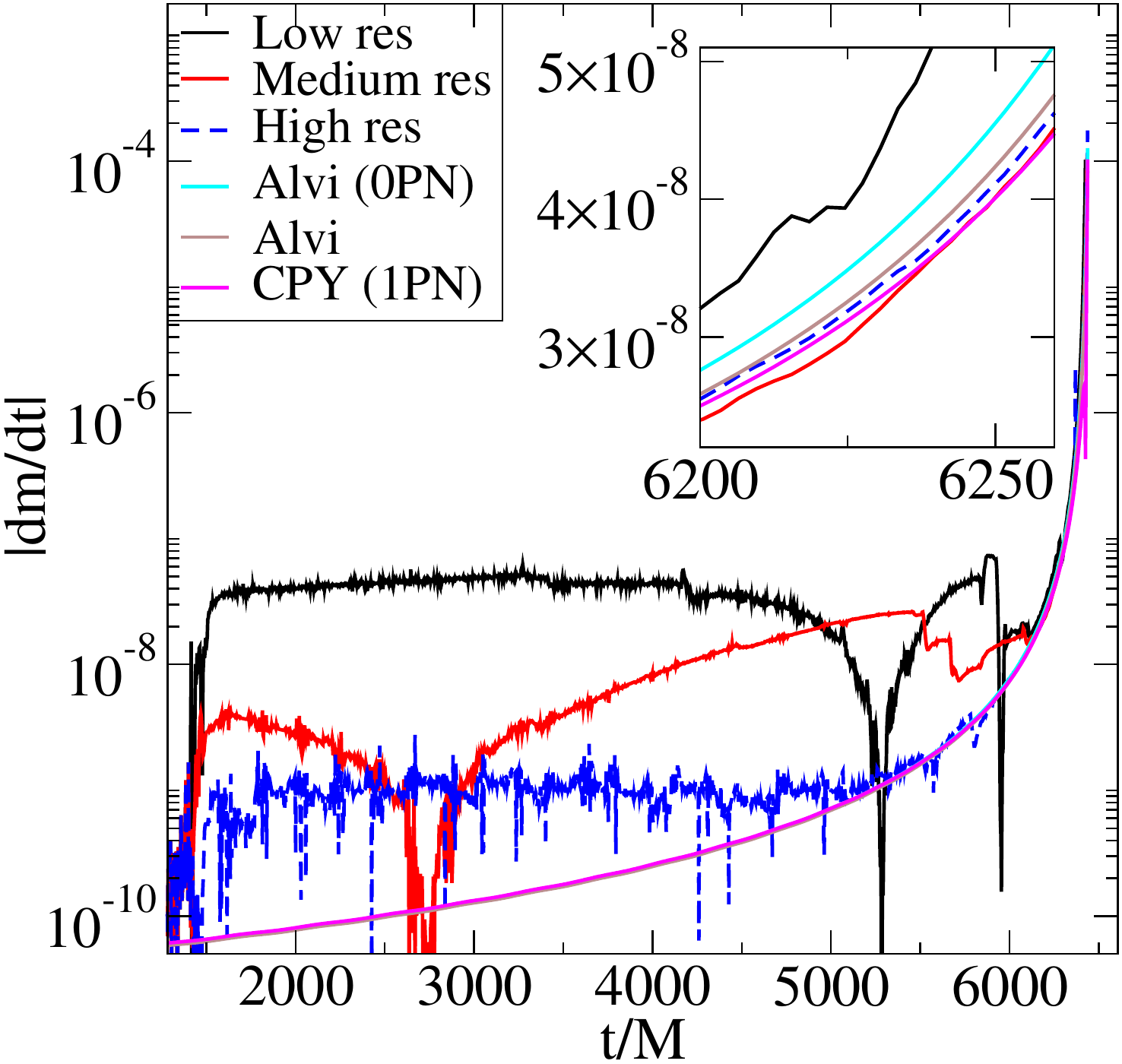}
\caption{Magnitude of $dm/dt$ of one of the black holes from \spin{994}.
  Shown are three numerical resolutions,
  Alvi's expression as written (Eq.~(11) of~\cite{Alvi:2001mx}),
  Alvi's expression truncated to leading order, and the CPY 
  expression\cite{Chatziioannou:2013,Chatziioannou:2014} truncated
  to 1PN order.  
  The inset zooms closer to the high-resolution
  numerical curve.
  \label{fig:dMdtComparison}
}
\end{figure}

\begin{figure}
\includegraphics[width=0.95\columnwidth]{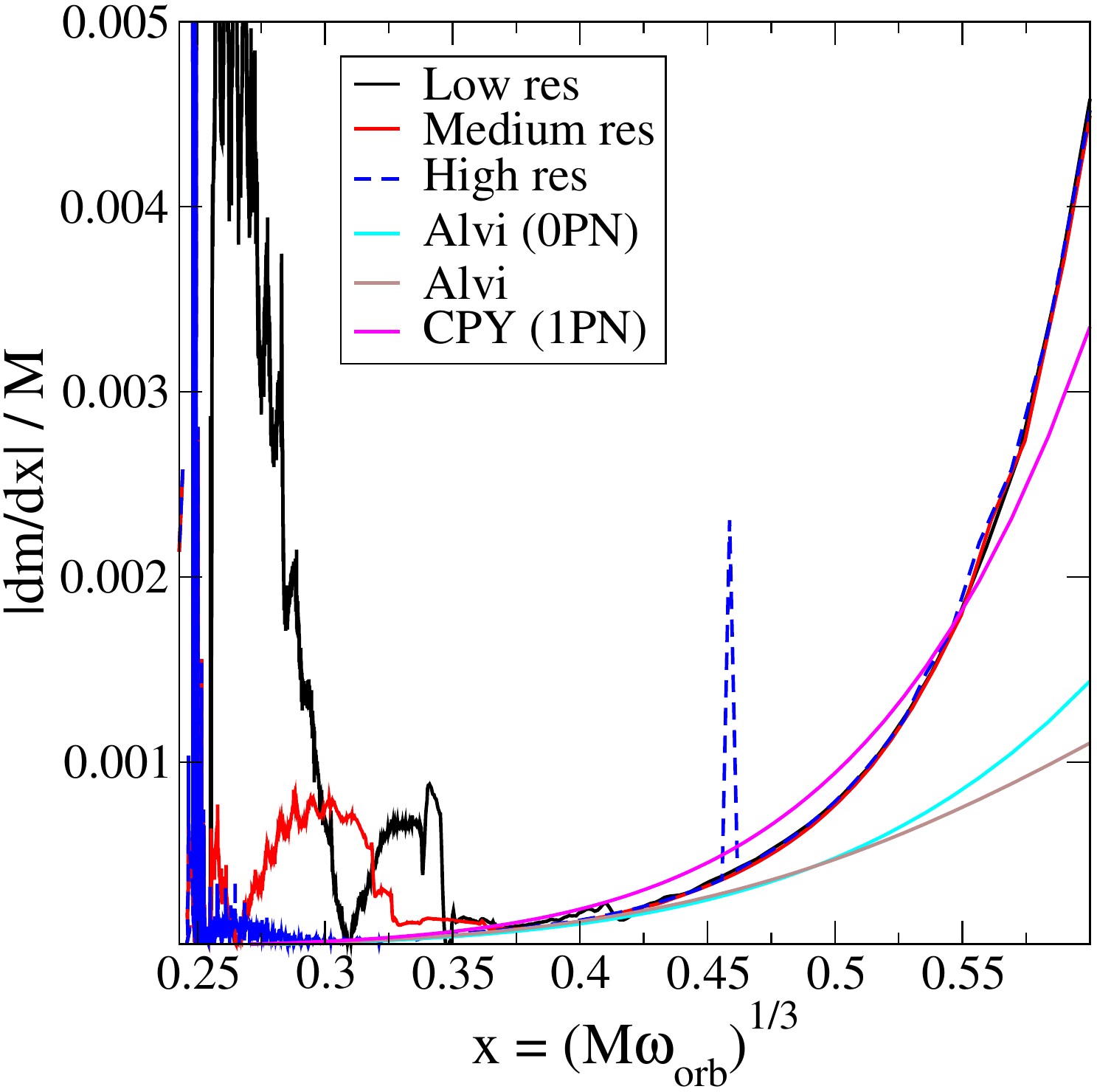}
\caption{Magnitude of $dm/dx$, where $x\equiv(M\omega_{\rm orb})^{1/3}$, 
  of one of the black holes from \spin{994}.
  \label{fig:dMdxComparison}
}
\end{figure}

Figures~\ref{fig:dMdtComparison} and~\ref{fig:dMdxComparison} are
similar to Figs.~\ref{fig:dSdtComparison}
and~\ref{fig:dSdxComparison} except that they show the change in
Christodoulou mass instead of the change in dimensionful spin.  
As was the case
for the spin comparisons, both
the Alvi and CPY formulas agree well with each other and
with the numerical result early in the inspiral, but do not
agree at late times. Note that
since the derivative of the mass
is smaller (by a factor of $v^3$ in PN) than the derivative of the
spin, $dm/dt$ is more difficult to resolve numerically than $dS/dt$,
as seen by the larger numerical errors in
Figs.~\ref{fig:dMdtComparison} and~\ref{fig:dMdxComparison}
compared with the numerical errors in Figs.~\ref{fig:dSdtComparison}
and~\ref{fig:dSdxComparison}.

\subsection{Orbital hangup}

During a BBH inspiral, the orbital frequency $\omega_{\rm orb}$
secularly evolves along with the black-hole
masses and spins. 
For equal-mass binaries with equal spins aligned (or antialigned)
with the orbital angular momentum, the number of orbits until merger
increases as a function of $S\cdot L$. Damour~\cite{Damour01c}
observed this effect, today commonly called ``orbital hangup'', 
in an effective-one-body model of the holes' motion;
the effect is a consequence of 
post-Newtonian spin-orbit coupling~\cite{Kidder:1995zr}.
Campanelli, Lousto, and Zlochower~\cite{Campanelli2006c} first 
demonstrated orbital hangup in numerical simulations of merging BBHs.

Instead of examining the number of orbits from the trajectories, we infer the
number of orbits from the dominant $\ell=m=2$ mode of the emitted 
gravitational waves\footnote{Specifically, we extrapolate the gravitational 
waves measured on a series of concentric shells to $r \rightarrow \infty$, 
as discussed in detail in Sec.~\ref{sec:GW}.}. We do this because it is easier
to define a gauge-invariant time of merger 
from the waveforms than from the trajectories; specifically, 
we define the time of merger as the
time when the waveform amplitude is at a maximum.

\begin{figure}
\includegraphics[width=0.95\columnwidth]{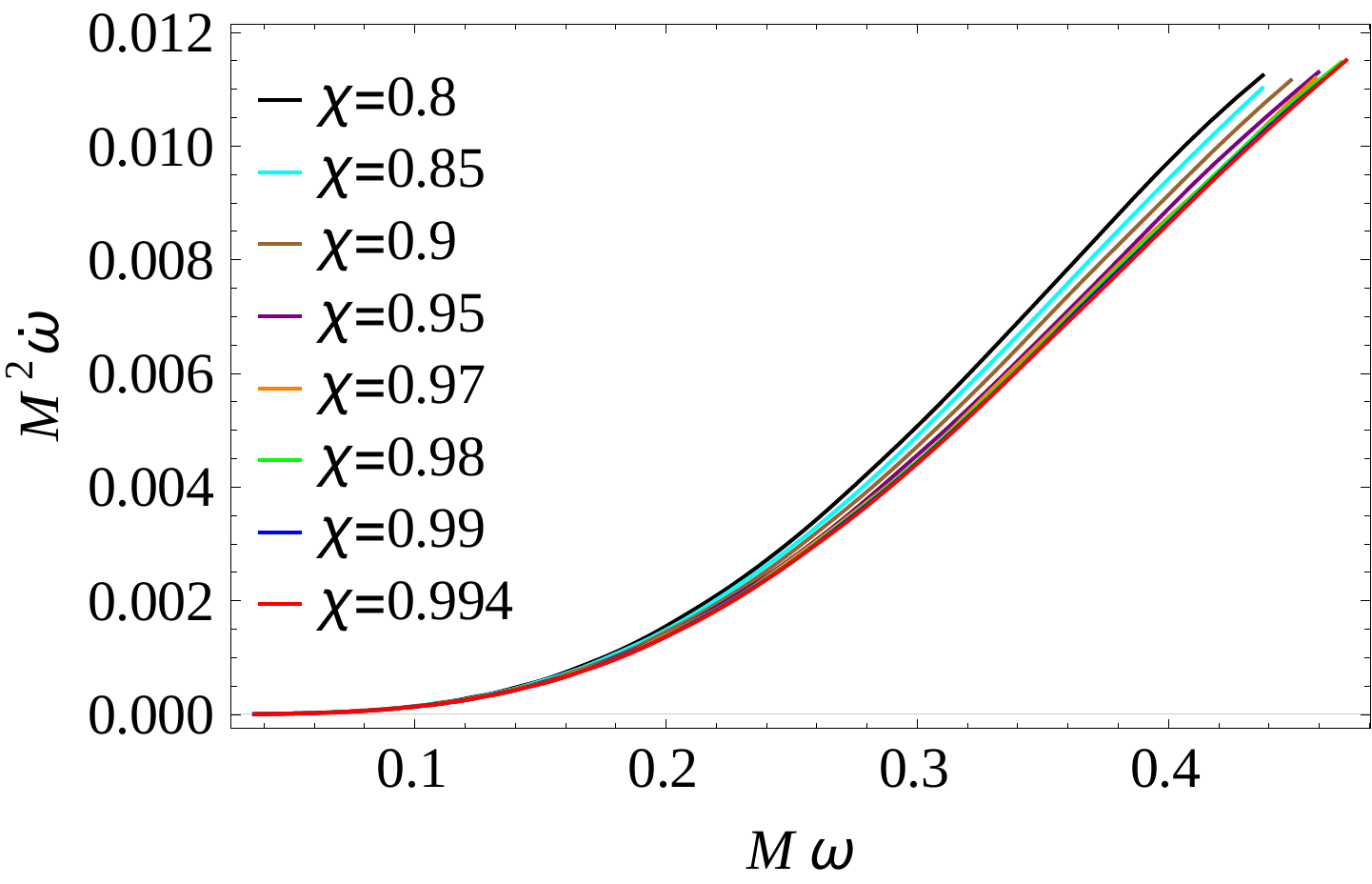}
\caption{The evolution of the derivative of
the gravitational-wave frequency $\dot{\omega}_{22}=d\omega_{22}/dt$, 
for simulations  
\spin{99} and \spin{994} and 
(for comparison) simulations \spin{8}~\cite{Hemberger:2013hsa}, 
\spin{85}~\cite{Hemberger:2013hsa}, 
\spin{9}~\cite{Hemberger:2013hsa}, \spin{95}~\cite{Hemberger:2013hsa},
\spin{97}~\cite{Lovelace:2011nu}, and \spin{98}~\cite{Mroue:2013PRL}.
\label{fig:DFrequencies}}
\end{figure}

Let $h_{22}(t)$ be the ${}_{-2}Y_{22}$ spin-weighted spherical harmonic
mode of the gravitational wave strain $h(t)$, and let $\omega_{22}$ be
the frequency of $h_{22}(t)$.
Figure~\ref{fig:DFrequencies} shows the time
evolution of $d\omega_{22}/dt$ for simulations \spin{99} and \spin{994}.
For comparison, we also show 
results for other simulations with equal masses and equal spins aligned 
with the orbital angular 
momentum~\cite{Hemberger:2013hsa,Lovelace:2011nu,Mroue:2013PRL}. 
Note that $d\omega_{22}/dt$
is positive and steadily
increasing: the frequency 
does not slow down or momentarily remain constant, as a 
literal
interpretation of the term ``orbital hangup"
might suggest. 

\begin{figure}
\includegraphics[width=0.95\columnwidth]{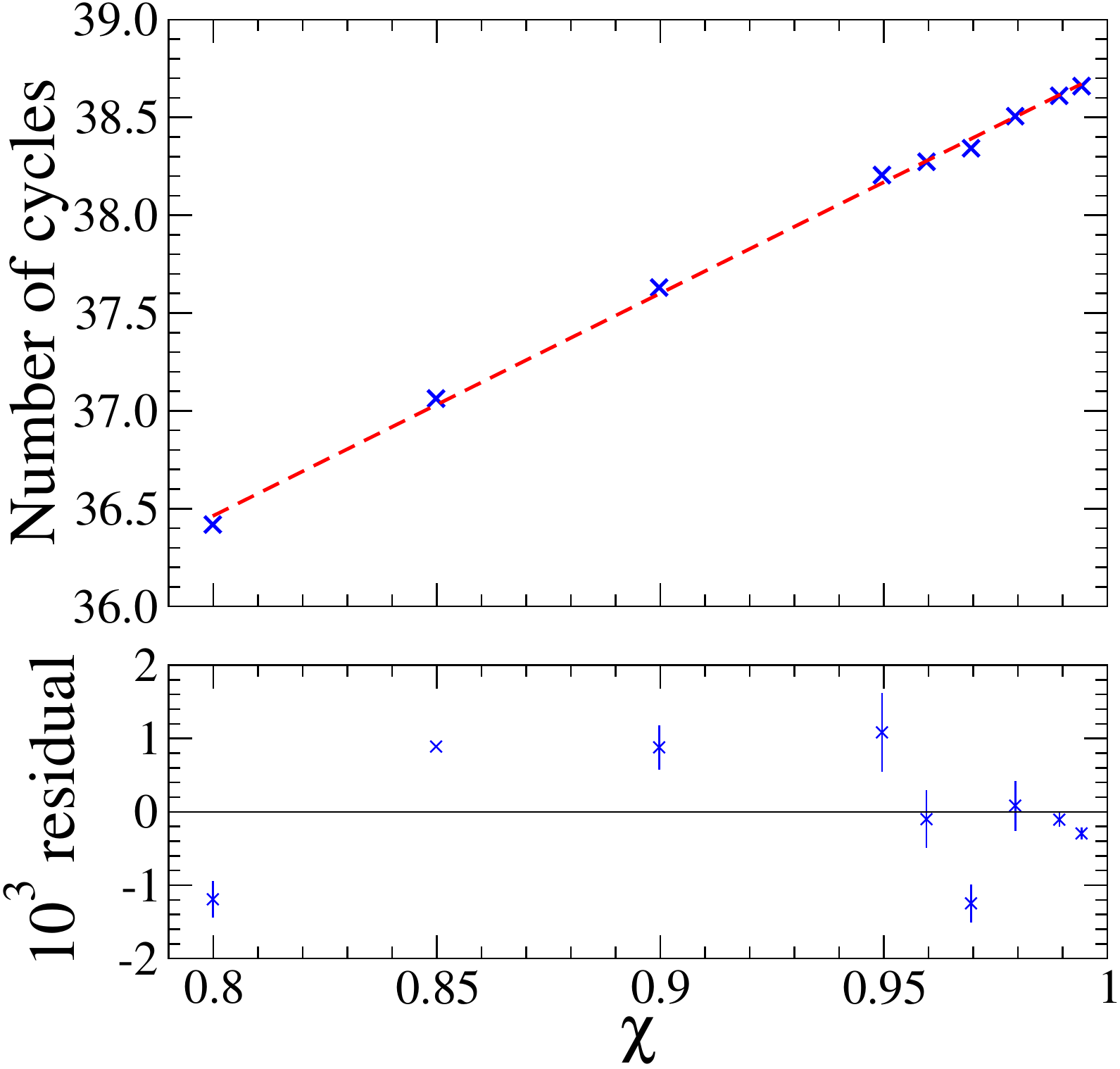}
\caption{The number of 
  gravitational-wave cycles as a function of 
  the initial spin $\chi$, measured after the initial relaxation, 
  for simulations  
\spin{99} and \spin{994} and 
(for comparison) simulations \spin{8}~\cite{Hemberger:2013hsa}, 
\spin{85}~\cite{Hemberger:2013hsa}, 
\spin{9}~\cite{Hemberger:2013hsa}, \spin{95}~\cite{Hemberger:2013hsa},
\spin{96}~\cite{Lovelace2014},   
\spin{97}~\cite{Lovelace:2011nu}, and \spin{98}~\cite{Mroue:2013PRL}.
  {\it Upper panel:} 
  The number of  
    gravitational-wave cycles of $h_{22}$ accumulated 
  between a gravitational-wave 
  frequency $M\omega_{22}=0.036$ and merger (i.e., the time when the  
  amplitude of $h_{22}$ peaks). The 
  dashed line is a linear fit to the data.
  {\it Lower panel:} Fractional difference (``residual'') 
  between our results and the linear fit, with uncertainties 
  for simulations except \spin{85} (which we ran at only one resolution) 
  estimated as 
  differences between medium and high numerical resolutions.
  \label{fig:OrbitsPlot}
}
\end{figure}

Figure~\ref{fig:OrbitsPlot} shows the gravitational-wave cycles
accumulated between an initial 
gravitational-wave frequency of $M\omega_{22}=0.036$ (i.e., an initial orbital 
frequency of $M\omega_{\rm orb} = 0.018$) and merger 
(when the amplitude of $h_{22}$ peaks). Simulations \spin{99} and \spin{994} 
reveal that the orbital hangup 
depends approximately linearly on
the initial spin $\chi$, even at spins that are nearly extremal; however, 
most of our simulations only agree with the linear fit to $\mathcal{O}(0.1\%)$,
which is often larger than our estimated numerical uncertainties. 
This linearity implies that even near extremality, the orbital 
hangup effect is dominated by spin-orbit coupling; 
resolving nonlinear features in Fig.~\ref{fig:OrbitsPlot} 
would require more simulations with higher accuracy.

\subsection{Comparison with analytic approximants}\label{sec:GW}

We compare the gravitational waveforms
from our simulations to several analytic
waveform approximants.
The numerical waveforms were computed by performing Regge-Wheeler-Zerilli
extraction~\cite{Sarbach2001,Rinne2008b}
at a sequence of radii between 100$M$ and 465$M$,
and then extrapolating to $\mathscr{I}^+$ using
the open-source \texttt{GWFrames} software package~\cite{Boyle:2013a,
  Boyle:2014, OssokineEtAl:2014}.  The TaylorT1, TaylorT4, and TaylorT5
approximants were constructed using the \texttt{PostNewtonian} module in
\texttt{GWFrames}.\footnote{To our knowledge, the \texttt{PostNewtonian}
  module includes all terms currently found in the literature.  Non-spin
  terms are given up to 4.0 PN order for the binding
  energy~\cite{Blanchet2006, BiniDamour:2013}; 3.5 PN with incomplete 4.0 PN
  information for the flux~\cite{Blanchet2006}; and 3.5 PN for the waveform
  modes~\cite{BFIS, FayeEtAl:2012, FayeEtAl:2014}.  The spin-orbit terms are
  given to 4.0 PN in the binding energy~\cite{BoheEtAl:2012}; 3.5 PN with
  incomplete 4.0 PN terms in flux~\cite{MarsatEtAl:2013}; and 2.0 PN in the
  waveform modes~\cite{Boyle:2014}.  Terms quadratic in spin are given to
  2.0 PN order in the binding energy and flux~\cite{Will96, Arun:2009}, and
  waveform modes~\cite{Will96, BuonannoFayeHinderer:2013, Boyle:2014}.}
The EOB approximants were constructed using
\texttt{SEOBNRv2}~\cite{Taracchini:2013rva}
from the LIGO Algorithm Library, with the function 
\texttt{SimIMRSpinAlignedEOBWaveform} modified to return 
$h_{22}(t)$.
Physical parameters for the approximants were taken from the highest
resolution from each simulation at the relaxation time.
Because \texttt{SEOBNRv2}
is strictly valid only for non-precessing systems, and therefore
accepts only scalar values of the spins as input, 
it is not obvious what to input for the case of \generic.  We
pass the $z$-component of the spins
into \texttt{SimIMRSpinAlignedEOBWaveform}.
If instead we pass the spin magnitudes, we see larger disagreements between
the EOB and numerical waveforms for \generic, likely
due to a change in the strength of spin-orbit coupling.
We will see below that non-precessing
EOB agrees remarkably well with \generic despite
the mild precession of this simulation.

\begin{figure}
\includegraphics[width=\columnwidth]{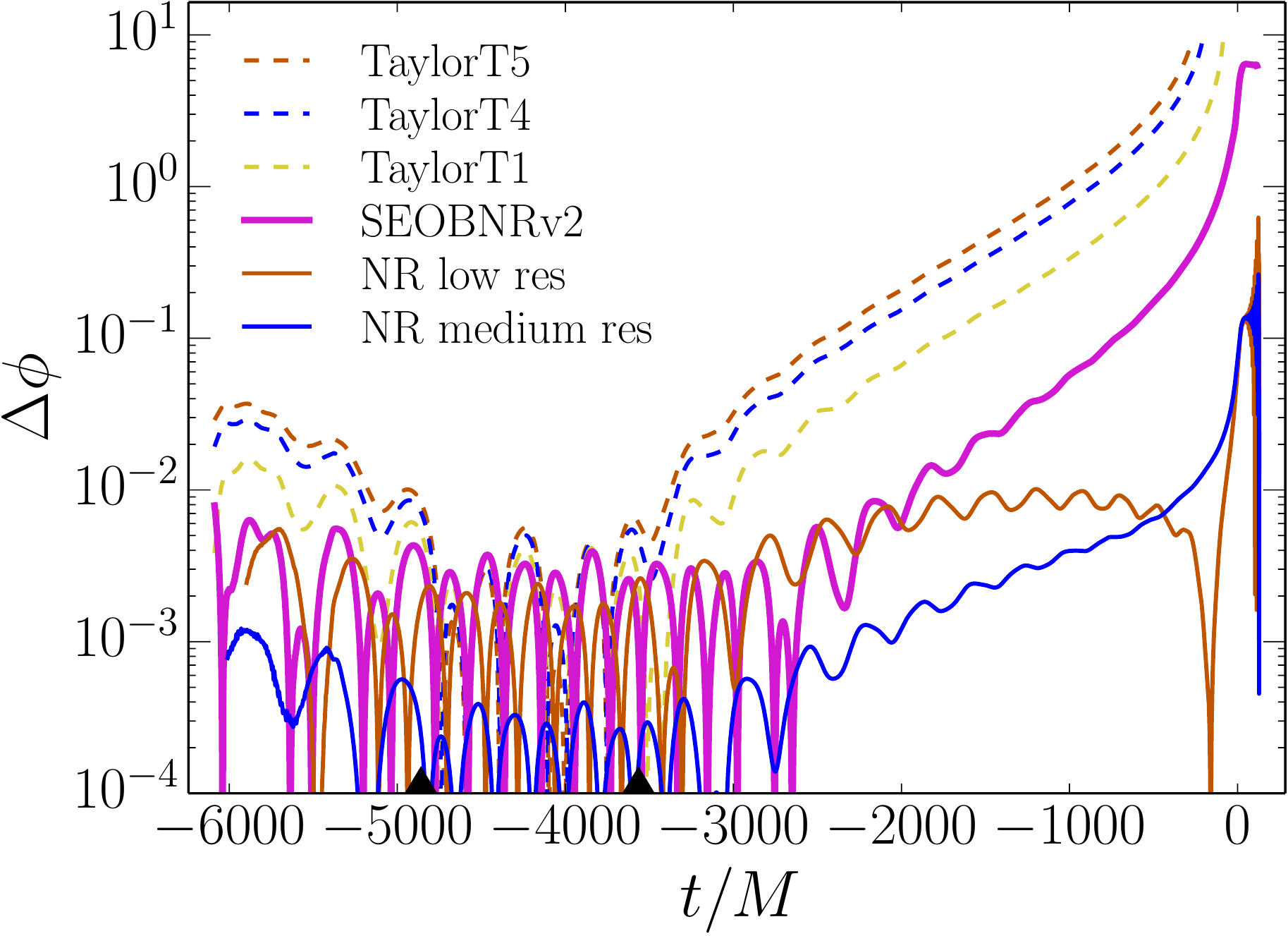}
\caption{Phase differences $\Delta\phi$ of
  $h_{22}$ as a function
  of retarded time before merger for \spin{99}.
  Shown are differences between the highest numerical resolution
  and several analytic approximants. Differences between
  the highest numerical resolution and other
  numerical resolutions are shown for comparison.
  The waveforms are aligned in the time interval delimited
  by the black triangles.
  \label{fig:990DeltaPhi}
}
\end{figure}

\begin{figure}
\includegraphics[width=\columnwidth]{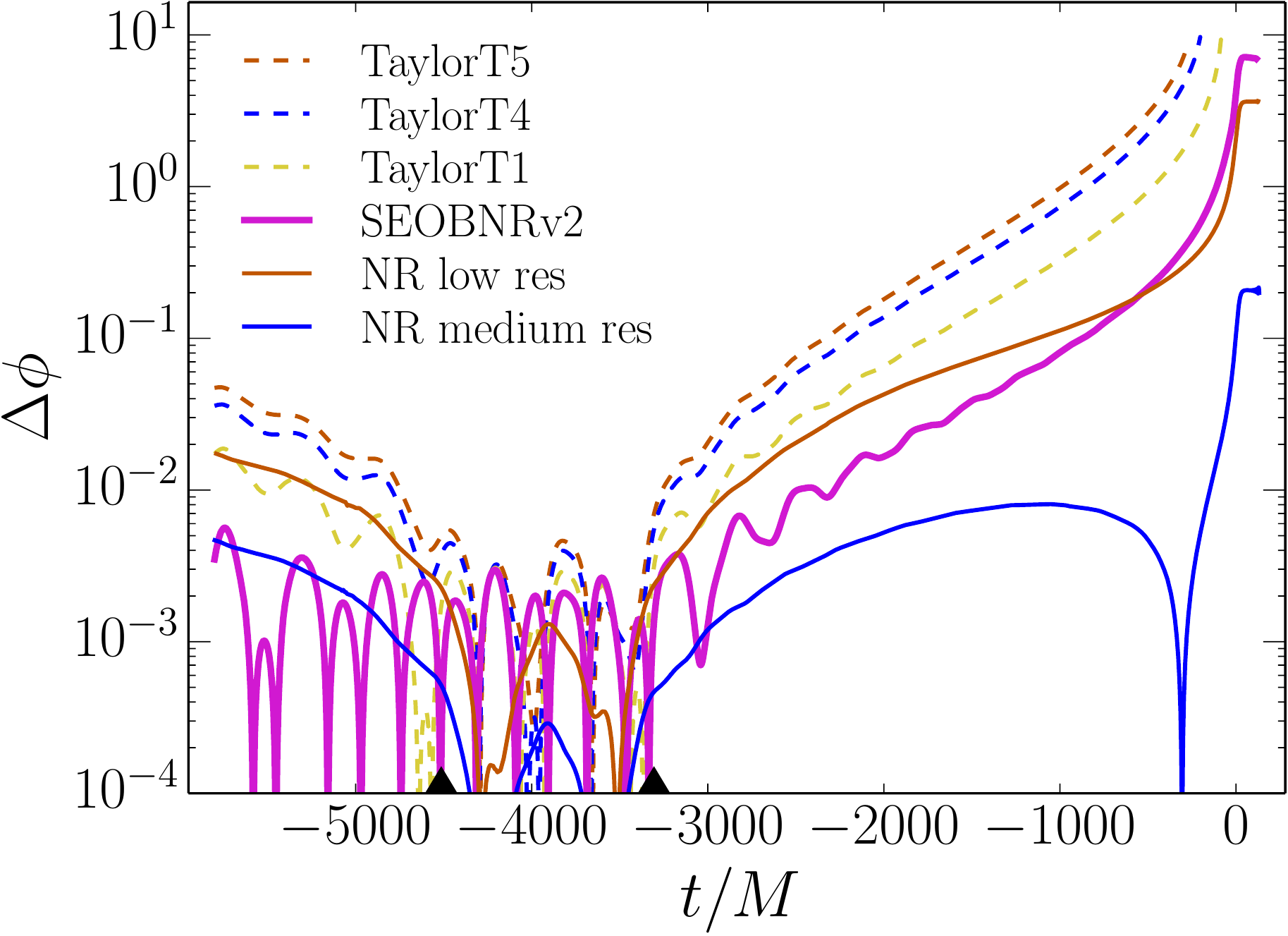}
\caption{Phase differences $\Delta\phi$ of $h_{22}$
  between numerical and approximant data
  for \spin{994}.
  Labels are the same as
  for Fig.~\ref{fig:990DeltaPhi}.
  \label{fig:995DeltaPhi}
}
\end{figure}

\begin{figure}
\includegraphics[width=\columnwidth]{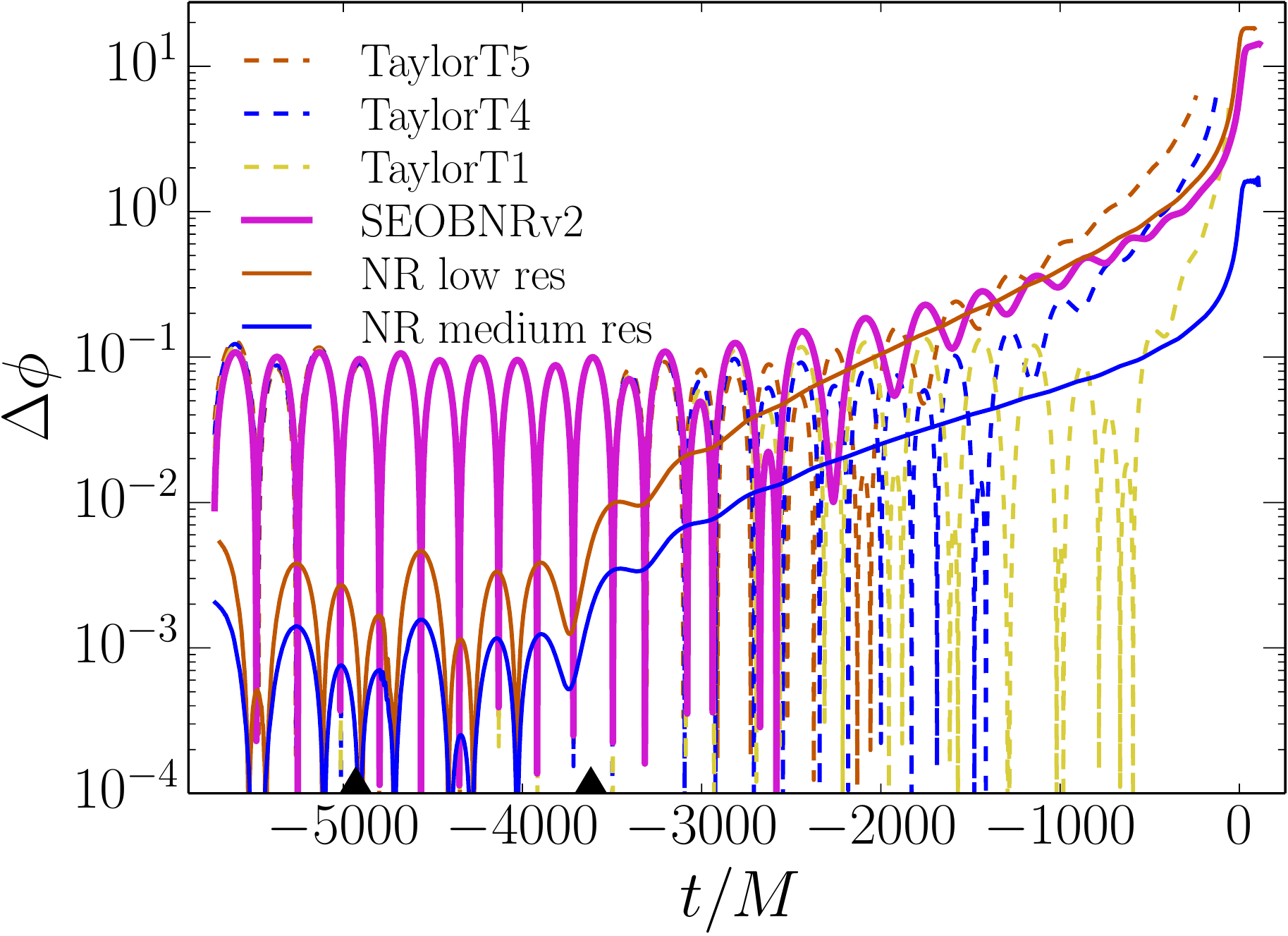}
\caption{Phase differences $\Delta\phi$ of $h_{22}$
  between numerical and approximant data
  for \generic.
  Labels are the same as
  for Fig.~\ref{fig:990DeltaPhi}.
  Note that the Taylor models include precession but \texttt{SEOBNRv2}
    does not. However, the precession of \generic is mild so the
    numerical waveform
    still agrees reasonably well with \texttt{SEOBNRv2}.
  \label{fig:991DeltaPhi}
}
\end{figure}

In Figs.~\ref{fig:990DeltaPhi}, \ref{fig:995DeltaPhi}, and
\ref{fig:991DeltaPhi}, we show for \spin{99}, \spin{994}, and 
\generic {}
(respectively) the phase difference $\Delta\phi$
of $h_{22}$
between the highest numerical resolution and the PN and EOB
approximants. We also include $\Delta\phi$ between the highest
numerical resolution and other numerical
resolutions for comparison.
To compute $\Delta\phi$, we first align each waveform with the highest
resolution numerical-relativity (NR)
waveform using the procedure prescribed in
Ref.~\cite{Boyle:2008}:
we find the time offset $\delta t$ and phase offset $\delta\phi$
that minimize $\Phi(\delta t, \delta\phi)$,
a measure of the phase difference in 
$h_{22}$, given by
\begin{equation}
\Phi(\delta t, \delta\phi) \equiv \int_{t_1}^{t_2} \left[\phi_a(t) - \phi_b(t+\delta t) + \delta\phi\right]^2 dt,
\end{equation}
where $\delta\phi$ can be computed analytically from $\delta t$
\begin{equation}
\delta\phi(\delta t) = \frac{1}{t_2-t_1}\int_{t_1}^{t_2}\left[\phi_a(t)-\phi_b(t+\delta t)\right]dt.
\end{equation}
The alignment interval $t\in[t_1, t_2]$ is the same for all comparisons
with a particular simulation. 
The lower bound $t_1$ is chosen such that
the junk radiation has left the computational domain for all numerical
resolutions, specifically $t_1 = \max[t_0 + 3(t_{\rm relax}-t_0)]$,
where $t_0$ is the time at the beginning of the waveform.
The upper bound $t_2$ is chosen such that the gravitational-wave
frequency changes by at
least 10\% during the interval 
$[t_1,t_2]$, as suggested in Ref.~\cite{MacDonald:2011ne}.

We have also computed $\Delta\phi$ with a few
other alignment methods,
including the three-dimensional minimization of complex $h_{22}$ differences
in Ajith et al.~2008 (Eq.~4.9 in Ref.~\cite{Ajith-Babak-Chen-etal:2007b})
and the four-dimensional minimization over time and frame-rotation degrees
of freedom in Boyle~2013 (Eq.~22 in Ref.~\cite{Boyle:2013a}).
We have found that our
results are qualitatively independent of alignment method.

The TaylorT family of
PN approximants shows the largest discrepancy with our
highest numerical resolution. Phase errors between PN
and NR waveforms grow to several radians
before the merger in every case. The smallest phase errors outside
the alignment interval occur for \generic, which is likely a consequence
of the smaller black hole having a moderate spin.
We find the best agreement with TaylorT1, in contrast to PN comparisons
for other nearly extremal systems~\cite{Lovelace:2011nu}, which found the
best agreement with TaylorT4 for spins aligned with the orbital
angular momentum; note that the PN waveforms
considered in Ref.~\cite{Lovelace:2011nu} include
fewer higher-order PN terms than we do here.
This is further evidence that agreement with a particular PN
approximant in the TaylorT family depends sensitively on the
PN order. Agreement with a particular PN approximant
also depends on the parameters of the simulation 
(e.g., Ref.~\cite{Lovelace:2011nu}).

The EOB approximant performs significantly
better than the PN approximants for \spin{99} and \spin{994},
which is impressive 
considering that the parameters of these waveforms
are outside the range
in which \texttt{SEOBNRv2} was calibrated to NR. Only about 5 radians of
phase error is accumulated in \spin{99} and \spin{994}.
\footnote{Note that \texttt{SEOBNRv2} was calibrated by minimizing
unfaithfulness rather than phase error; it is possible to
have relatively large phase errors even when the unfaithfulness is
small~\cite{Taracchini:2013rva}.}
Phase error increases to a little over 10 radians in
\generic, but this case is precessing, and \texttt{SEOBNRv2}
is only valid for non-precessing systems.
However, the precession is mild (cf. Figs.~\ref{fig:PostUltimateTraj1}
and~\ref{fig:precession}),
which could account for the relatively good agreement.

The analytic approximants show much larger
$\Delta\phi$
at early times for \generic (see Fig.~\ref{fig:991DeltaPhi})
than for \spin{99} and \spin{994}. We conjecture that this is due to
the relatively large eccentricity of \generic (see Table~\ref{table:runs}),
whereas the PN and EOB models used here are non-eccentric.
Note that we use precessing PN models for comparing to \generic.

The phase errors 
between numerical waveforms computed at different resolutions are 
convergent. Because of the rapid convergence, 
the difference between the two highest numerical
resolutions represents the numerical error in the {\em second highest} 
resolution; to determine the numerical error of the highest resolution
waveform, we would need to perform a simulation at an even higher
resolution.  As a conservative estimate of the numerical error of the
highest resolution waveform, we use the difference between the
two highest-resolution waveforms as an upper bound. The upper bound of the
numerical phase error of the highest resolution simulation, computed
in this way, is thus about 0.2 radians for \spin{99} and 
\spin{994} and about 1 radian
for \generic. 

In Figs.~\ref{fig:990DeltaPhi}, \ref{fig:995DeltaPhi}, and
\ref{fig:991DeltaPhi}, the
larger numerical phase errors in the
{\em lower} resolutions of \spin{994} and \generic {}
are expected, 
because
these simulations use a larger spacing in AMR truncation error
tolerance as described in Sec.~\ref{sec:Spin994}. The larger spacing
increases relative phase errors between successive
numerical resolutions.
Nevertheless, our comparisons show that numerical errors are
much smaller than the errors in the PN and EOB waveforms for
systems with nearly extremal black holes, indicating that
these numerical waveforms will be useful for calibrating
and extending the regime of validity for approximate waveforms.

\section{Conclusion}\label{sec:conclusion}
We 
have presented improved methods for simulating 
the binary evolution of nearly extremal black holes,
i.e., black holes with spins above the Bowen-York limit of $\chi=0.93$.
These techniques enable robust
simulations in
the portion of 
BBH parameter space where the black holes have very large spins.
Because 
nearly extremal black holes might exist in
astrophysical binaries, these simulations will be
important for helping to maximize what we can learn 
from gravitational-wave experiments.

We have applied our new methods to carry out
the first unequal-mass, mildly-precessing BBH
simulation containing a nearly extremal black hole,
and to extend aligned-spin BBH simulations to spin magnitudes
that begin to approach the Novikov-Thorne limit of $\chi=0.998$.
From these new simulations, we have learned that perturbative predictions
for tidal heating and tidal torquing agree well with the numerics
at low frequency,
even for nearly extremal spins.
However, we find that our numerical errors are still large enough
that we cannot reliably distinguish between
0PN and 1PN predictions. Doing so would require further investigation 
with more accurate simulations.

While many physical quantities depend on $\chi$ in an extremely nonlinear
fashion, we find that
the number of orbits starting from a chosen 
orbital frequency (i.e., the orbital hangup) scales approximately 
linearly with $\chi$.
Finally, after demonstrating numerical 
convergence, 
we have found that our numerical waveforms agree with
\texttt{SEOBNRv2} much better than with TaylorT PN approximants,
even though the parameters for these simulations are outside
the range in which \texttt{SEOBNRv2} was calibrated.  However,
even the \texttt{SEOBNRv2} waveforms disagree with our numerical
waveforms by more than our numerical truncation error. This indicates
that these simulations are sufficiently accurate to
validate and 
further improve analytical waveform approximants for future gravitational-wave 
observations.
How significant these improvements will be for Advanced LIGO is the
subject of future work.

\begin{acknowledgments}
We are grateful to 
Eric Poisson, Nicolas Yunes,
and Katerina Chatziioannou for
detailed discussions about perturbative expressions for tidal torquing
and about the problems inherent in comparing numerical and post-Newtonian
expressions for near-field quantities. 
We thank Alessandra Buonanno and Sebastiano Bernuzzi for helpful
discussions.
Simulations used in this work
were computed with SpEC~\cite{SpECwebsite}.
This work was supported in part by 
the Sherman Fairchild Foundation; 
NSF grants
PHY-1440083 and AST-1333520 at Caltech, 
NSF
grants PHY-1306125 and AST-1333129 at Cornell, 
and 
NSF grant PHY-1307489 at California State University Fullerton; 
a 2013--2014 California State University Fullerton 
Junior Faculty Research Grant.
Computations
were performed on the Zwicky cluster at Caltech, which is supported by
the Sherman Fairchild Foundation and by NSF award PHY-0960291; on the
NSF XSEDE network under grant TG-PHY990007N; 
on the Orca cluster supported by  
NSF award NSF-1429873 and by California State University Fullerton; 
and on the GPC
supercomputer at the SciNet HPC Consortium~\cite{scinet}. SciNet is
funded by: the Canada Foundation for Innovation under the auspices of
Compute Canada; the Government of Ontario; Ontario Research
Fund--Research Excellence; and the University of Toronto. 
\end{acknowledgments}

\bibliography{References/References}

\begin{thebibliography}{103}%
\makeatletter
\providecommand \@ifxundefined [1]{%
 \@ifx{#1\undefined}
}%
\providecommand \@ifnum [1]{%
 \ifnum #1\expandafter \@firstoftwo
 \else \expandafter \@secondoftwo
 \fi
}%
\providecommand \@ifx [1]{%
 \ifx #1\expandafter \@firstoftwo
 \else \expandafter \@secondoftwo
 \fi
}%
\providecommand \natexlab [1]{#1}%
\providecommand \enquote  [1]{``#1''}%
\providecommand \bibnamefont  [1]{#1}%
\providecommand \bibfnamefont [1]{#1}%
\providecommand \citenamefont [1]{#1}%
\providecommand \href@noop [0]{\@secondoftwo}%
\providecommand \href [0]{\begingroup \@sanitize@url \@href}%
\providecommand \@href[1]{\@@startlink{#1}\@@href}%
\providecommand \@@href[1]{\endgroup#1\@@endlink}%
\providecommand \@sanitize@url [0]{\catcode `\\12\catcode `\$12\catcode
  `\&12\catcode `\#12\catcode `\^12\catcode `\_12\catcode `\%12\relax}%
\providecommand \@@startlink[1]{}%
\providecommand \@@endlink[0]{}%
\providecommand \url  [0]{\begingroup\@sanitize@url \@url }%
\providecommand \@url [1]{\endgroup\@href {#1}{\urlprefix }}%
\providecommand \urlprefix  [0]{URL }%
\providecommand \Eprint [0]{\href }%
\providecommand \doibase [0]{http://dx.doi.org/}%
\providecommand \selectlanguage [0]{\@gobble}%
\providecommand \bibinfo  [0]{\@secondoftwo}%
\providecommand \bibfield  [0]{\@secondoftwo}%
\providecommand \translation [1]{[#1]}%
\providecommand \BibitemOpen [0]{}%
\providecommand \bibitemStop [0]{}%
\providecommand \bibitemNoStop [0]{.\EOS\space}%
\providecommand \EOS [0]{\spacefactor3000\relax}%
\providecommand \BibitemShut  [1]{\csname bibitem#1\endcsname}%
\let\auto@bib@innerbib\@empty
\bibitem [{\citenamefont {Harry}(2010)}]{Harry:2010zz}%
  \BibitemOpen
  \bibfield  {author} {\bibinfo {author} {\bibfnamefont {G.~M.}\ \bibnamefont
  {Harry}} (\bibinfo {collaboration} {LIGO Scientific Collaboration}),\ }\href
  {\doibase 10.1088/0264-9381/27/8/084006} {\bibfield  {journal} {\bibinfo
  {journal} {Class. Quant. Grav.}\ }\textbf {\bibinfo {volume} {27}},\ \bibinfo
  {pages} {084006} (\bibinfo {year} {2010})}\BibitemShut {NoStop}%
\bibitem [{\citenamefont {{The Virgo Collaboration}}(2009)}]{aVIRGO}%
  \BibitemOpen
  \bibfield  {author} {\bibinfo {author} {\bibnamefont {{The Virgo
  Collaboration}}},\ }\href {https://tds.ego-gw.it/ql/?c=6589} {\enquote
  {\bibinfo {title} {{Advanced Virgo Baseline Design}},}\ } (\bibinfo {year}
  {2009}),\ \bibinfo {note} {{[VIR-0027A-09]}}\BibitemShut {NoStop}%
\bibitem [{\citenamefont {{The Virgo Collaboration}}(2012)}]{aVIRGO:2012}%
  \BibitemOpen
  \bibfield  {author} {\bibinfo {author} {\bibnamefont {{The Virgo
  Collaboration}}},\ }\href {https://tds.ego-gw.it/ql/?c=6940} {\enquote
  {\bibinfo {title} {{Advanced Virgo Technical Design Report}},}\ } (\bibinfo
  {year} {2012}),\ \bibinfo {note} {{[VIR-0128A-12]}}\BibitemShut {NoStop}%
\bibitem [{\citenamefont {Somiya}\ and\ \citenamefont {the
  {KAGRA}~Collaboration}(2012)}]{Somiya:2012}%
  \BibitemOpen
  \bibfield  {author} {\bibinfo {author} {\bibfnamefont {K.}~\bibnamefont
  {Somiya}}\ and\ \bibinfo {author} {\bibnamefont {the
  {KAGRA}~Collaboration}},\ }\href {\doibase 10.1088/0264-9381/29/12/124007}
  {\bibfield  {journal} {\bibinfo  {journal} {Class.\ Quantum Grav.}\ }\textbf
  {\bibinfo {volume} {29}},\ \bibinfo {pages} {124007} (\bibinfo {year}
  {2012})}\BibitemShut {NoStop}%
\bibitem [{\citenamefont {Gou}\ \emph {et~al.}(2011)\citenamefont {Gou},
  \citenamefont {McClintock}, \citenamefont {Reid}, \citenamefont {Orosz},
  \citenamefont {Steiner}, \citenamefont {Narayan}, \citenamefont {Xiang},
  \citenamefont {Remillard}, \citenamefont {Arnaud},\ and\ \citenamefont
  {Davis}}]{Gou:2011nq}%
  \BibitemOpen
  \bibfield  {author} {\bibinfo {author} {\bibfnamefont {L.}~\bibnamefont
  {Gou}}, \bibinfo {author} {\bibfnamefont {J.~E.}\ \bibnamefont {McClintock}},
  \bibinfo {author} {\bibfnamefont {M.~J.}\ \bibnamefont {Reid}}, \bibinfo
  {author} {\bibfnamefont {J.~A.}\ \bibnamefont {Orosz}}, \bibinfo {author}
  {\bibfnamefont {J.~F.}\ \bibnamefont {Steiner}}, \bibinfo {author}
  {\bibfnamefont {R.}~\bibnamefont {Narayan}}, \bibinfo {author} {\bibfnamefont
  {J.}~\bibnamefont {Xiang}}, \bibinfo {author} {\bibfnamefont {R.~A.}\
  \bibnamefont {Remillard}}, \bibinfo {author} {\bibfnamefont {K.~A.}\
  \bibnamefont {Arnaud}}, \ and\ \bibinfo {author} {\bibfnamefont {S.~W.}\
  \bibnamefont {Davis}},\ }\href {\doibase 10.1088/0004-637X/742/2/85}
  {\bibfield  {journal} {\bibinfo  {journal} {Astrophys.~J.}\ }\textbf
  {\bibinfo {volume} {742}},\ \bibinfo {pages} {85} (\bibinfo {year} {2011})},\
  \Eprint {http://arxiv.org/abs/1106.3690} {arXiv:1106.3690 [astro-ph.HE]}
  \BibitemShut {NoStop}%
\bibitem [{\citenamefont {Fabian}\ \emph {et~al.}(2012)\citenamefont {Fabian},
  \citenamefont {Wilkins}, \citenamefont {Miller}, \citenamefont {Reis},
  \citenamefont {Reynolds} \emph {et~al.}}]{Fabian:2012kv}%
  \BibitemOpen
  \bibfield  {author} {\bibinfo {author} {\bibfnamefont {A.}~\bibnamefont
  {Fabian}}, \bibinfo {author} {\bibfnamefont {D.}~\bibnamefont {Wilkins}},
  \bibinfo {author} {\bibfnamefont {J.}~\bibnamefont {Miller}}, \bibinfo
  {author} {\bibfnamefont {R.}~\bibnamefont {Reis}}, \bibinfo {author}
  {\bibfnamefont {C.}~\bibnamefont {Reynolds}},  \emph {et~al.},\ }\href@noop
  {} {\bibfield  {journal} {\bibinfo  {journal} {MNRAS}\ }\textbf {\bibinfo
  {volume} {424}},\ \bibinfo {pages} {217} (\bibinfo {year} {2012})},\ \Eprint
  {http://arxiv.org/abs/1204.5854} {arXiv:1204.5854 [astro-ph.HE]} \BibitemShut
  {NoStop}%
\bibitem [{\citenamefont {Gou}\ \emph {et~al.}(2014)\citenamefont {Gou},
  \citenamefont {McClintock}, \citenamefont {Remillard}, \citenamefont
  {Steiner}, \citenamefont {Reid} \emph {et~al.}}]{Gou:2014una}%
  \BibitemOpen
  \bibfield  {author} {\bibinfo {author} {\bibfnamefont {L.}~\bibnamefont
  {Gou}}, \bibinfo {author} {\bibfnamefont {J.~E.}\ \bibnamefont {McClintock}},
  \bibinfo {author} {\bibfnamefont {R.~A.}\ \bibnamefont {Remillard}}, \bibinfo
  {author} {\bibfnamefont {J.~F.}\ \bibnamefont {Steiner}}, \bibinfo {author}
  {\bibfnamefont {M.~J.}\ \bibnamefont {Reid}},  \emph {et~al.},\ }\href
  {\doibase 10.1088/0004-637X/790/1/29} {\bibfield  {journal} {\bibinfo
  {journal} {Astrophys.J.}\ }\textbf {\bibinfo {volume} {790}},\ \bibinfo
  {pages} {29} (\bibinfo {year} {2014})}\BibitemShut {NoStop}%
\bibitem [{\citenamefont {McClintock}\ \emph {et~al.}(2006)\citenamefont
  {McClintock}, \citenamefont {Shafee}, \citenamefont {Narayan}, \citenamefont
  {Remillard}, \citenamefont {Davis},\ and\ \citenamefont
  {Li}}]{McClintockEtAl:2006}%
  \BibitemOpen
  \bibfield  {author} {\bibinfo {author} {\bibfnamefont {J.~E.}\ \bibnamefont
  {McClintock}}, \bibinfo {author} {\bibfnamefont {R.}~\bibnamefont {Shafee}},
  \bibinfo {author} {\bibfnamefont {R.}~\bibnamefont {Narayan}}, \bibinfo
  {author} {\bibfnamefont {R.~A.}\ \bibnamefont {Remillard}}, \bibinfo {author}
  {\bibfnamefont {S.~W.}\ \bibnamefont {Davis}}, \ and\ \bibinfo {author}
  {\bibfnamefont {L.-X.}\ \bibnamefont {Li}},\ }\href@noop {} {\bibfield
  {journal} {\bibinfo  {journal} {Astrophys.\ J.}\ }\textbf {\bibinfo {volume}
  {652}},\ \bibinfo {pages} {518} (\bibinfo {year} {2006})}\BibitemShut
  {NoStop}%
\bibitem [{\citenamefont {Miller}\ \emph {et~al.}(2009)\citenamefont {Miller},
  \citenamefont {Reynolds}, \citenamefont {Fabian}, \citenamefont {Miniutti},\
  and\ \citenamefont {Gallo}}]{Miller:2009cw}%
  \BibitemOpen
  \bibfield  {author} {\bibinfo {author} {\bibfnamefont {J.}~\bibnamefont
  {Miller}}, \bibinfo {author} {\bibfnamefont {C.}~\bibnamefont {Reynolds}},
  \bibinfo {author} {\bibfnamefont {A.}~\bibnamefont {Fabian}}, \bibinfo
  {author} {\bibfnamefont {G.}~\bibnamefont {Miniutti}}, \ and\ \bibinfo
  {author} {\bibfnamefont {L.}~\bibnamefont {Gallo}},\ }\href {\doibase
  10.1088/0004-637X/697/1/900} {\bibfield  {journal} {\bibinfo  {journal}
  {Astrophys.J.}\ }\textbf {\bibinfo {volume} {697}},\ \bibinfo {pages} {900}
  (\bibinfo {year} {2009})},\ \Eprint {http://arxiv.org/abs/0902.2840}
  {arXiv:0902.2840 [astro-ph.HE]} \BibitemShut {NoStop}%
\bibitem [{\citenamefont {Walton}\ \emph {et~al.}(2013)\citenamefont {Walton},
  \citenamefont {Nardini}, \citenamefont {Fabian}, \citenamefont {Gallo},\ and\
  \citenamefont {Reis}}]{Walton:2012aw}%
  \BibitemOpen
  \bibfield  {author} {\bibinfo {author} {\bibfnamefont {D.}~\bibnamefont
  {Walton}}, \bibinfo {author} {\bibfnamefont {E.}~\bibnamefont {Nardini}},
  \bibinfo {author} {\bibfnamefont {A.}~\bibnamefont {Fabian}}, \bibinfo
  {author} {\bibfnamefont {L.}~\bibnamefont {Gallo}}, \ and\ \bibinfo {author}
  {\bibfnamefont {R.}~\bibnamefont {Reis}},\ }\href@noop {} {\bibfield
  {journal} {\bibinfo  {journal} {MNRAS}\ }\textbf {\bibinfo {volume} {428}},\
  \bibinfo {pages} {2901} (\bibinfo {year} {2013})},\ \Eprint
  {http://arxiv.org/abs/1210.4593} {arXiv:1210.4593 [astro-ph.HE]} \BibitemShut
  {NoStop}%
\bibitem [{\citenamefont {McClintock}\ \emph {et~al.}(2014)\citenamefont
  {McClintock}, \citenamefont {Narayan},\ and\ \citenamefont
  {Steiner}}]{McClintock:2013vwa}%
  \BibitemOpen
  \bibfield  {author} {\bibinfo {author} {\bibfnamefont {J.~E.}\ \bibnamefont
  {McClintock}}, \bibinfo {author} {\bibfnamefont {R.}~\bibnamefont {Narayan}},
  \ and\ \bibinfo {author} {\bibfnamefont {J.~F.}\ \bibnamefont {Steiner}},\
  }\href {\doibase 10.1007/s11214-013-0003-9} {\bibfield  {journal} {\bibinfo
  {journal} {Space Sci.Rev.}\ }\textbf {\bibinfo {volume} {183}},\ \bibinfo
  {pages} {295} (\bibinfo {year} {2014})},\ \Eprint
  {http://arxiv.org/abs/1303.1583} {arXiv:1303.1583 [astro-ph.HE]} \BibitemShut
  {NoStop}%
\bibitem [{\citenamefont {Reynolds}(2013)}]{Reynolds:2013qqa}%
  \BibitemOpen
  \bibfield  {author} {\bibinfo {author} {\bibfnamefont {C.~S.}\ \bibnamefont
  {Reynolds}},\ }\href@noop {} {\  (\bibinfo {year} {2013})},\ \Eprint
  {http://arxiv.org/abs/1302.3260} {arXiv:1302.3260 [astro-ph.HE]} \BibitemShut
  {NoStop}%
\bibitem [{\citenamefont {Blanchet}(2006)}]{Blanchet2006}%
  \BibitemOpen
  \bibfield  {author} {\bibinfo {author} {\bibfnamefont {L.}~\bibnamefont
  {Blanchet}},\ }\href@noop {} {\bibfield  {journal} {\bibinfo  {journal}
  {Living Rev.Rel.}\ }\textbf {\bibinfo {volume} {9}},\ \bibinfo {pages} {4}
  (\bibinfo {year} {2006})}\BibitemShut {NoStop}%
\bibitem [{\citenamefont {Pretorius}(2005{\natexlab{a}})}]{Pretorius2005a}%
  \BibitemOpen
  \bibfield  {author} {\bibinfo {author} {\bibfnamefont {F.}~\bibnamefont
  {Pretorius}},\ }\href {\doibase 10.1103/PhysRevLett.95.121101} {\bibfield
  {journal} {\bibinfo  {journal} {Phys.\ Rev.\ Lett.}\ }\textbf {\bibinfo
  {volume} {95}},\ \bibinfo {pages} {121101} (\bibinfo {year}
  {2005}{\natexlab{a}})},\ \Eprint {http://arxiv.org/abs/gr-qc/0507014}
  {arXiv:gr-qc/0507014 [gr-qc]} \BibitemShut {NoStop}%
\bibitem [{\citenamefont {Campanelli}\ \emph
  {et~al.}(2006{\natexlab{a}})\citenamefont {Campanelli}, \citenamefont
  {Lousto}, \citenamefont {Marronetti},\ and\ \citenamefont
  {Zlochower}}]{Campanelli2006a}%
  \BibitemOpen
  \bibfield  {author} {\bibinfo {author} {\bibfnamefont {M.}~\bibnamefont
  {Campanelli}}, \bibinfo {author} {\bibfnamefont {C.}~\bibnamefont {Lousto}},
  \bibinfo {author} {\bibfnamefont {P.}~\bibnamefont {Marronetti}}, \ and\
  \bibinfo {author} {\bibfnamefont {Y.}~\bibnamefont {Zlochower}},\ }\href
  {\doibase 10.1103/PhysRevLett.96.111101} {\bibfield  {journal} {\bibinfo
  {journal} {Phys.\ Rev.\ Lett.}\ }\textbf {\bibinfo {volume} {96}},\ \bibinfo
  {pages} {111101} (\bibinfo {year} {2006}{\natexlab{a}})},\ \Eprint
  {http://arxiv.org/abs/gr-qc/0511048} {arXiv:gr-qc/0511048 [gr-qc]}
  \BibitemShut {NoStop}%
\bibitem [{\citenamefont {Baker}\ \emph {et~al.}(2006)\citenamefont {Baker},
  \citenamefont {Centrella}, \citenamefont {Choi}, \citenamefont {Koppitz},\
  and\ \citenamefont {van Meter}}]{Baker2006a}%
  \BibitemOpen
  \bibfield  {author} {\bibinfo {author} {\bibfnamefont {J.~G.}\ \bibnamefont
  {Baker}}, \bibinfo {author} {\bibfnamefont {J.}~\bibnamefont {Centrella}},
  \bibinfo {author} {\bibfnamefont {D.-I.}\ \bibnamefont {Choi}}, \bibinfo
  {author} {\bibfnamefont {M.}~\bibnamefont {Koppitz}}, \ and\ \bibinfo
  {author} {\bibfnamefont {J.}~\bibnamefont {van Meter}},\ }\href {\doibase
  10.1103/PhysRevLett.96.111102} {\bibfield  {journal} {\bibinfo  {journal}
  {Phys.\ Rev.\ Lett.}\ }\textbf {\bibinfo {volume} {96}},\ \bibinfo {pages}
  {111102} (\bibinfo {year} {2006})},\ \Eprint
  {http://arxiv.org/abs/gr-qc/0511103} {arXiv:gr-qc/0511103 [gr-qc]}
  \BibitemShut {NoStop}%
\bibitem [{\citenamefont {Centrella}\ \emph {et~al.}(2010)\citenamefont
  {Centrella}, \citenamefont {Baker}, \citenamefont {Kelly},\ and\
  \citenamefont {van Meter}}]{Centrella:2010}%
  \BibitemOpen
  \bibfield  {author} {\bibinfo {author} {\bibfnamefont {J.}~\bibnamefont
  {Centrella}}, \bibinfo {author} {\bibfnamefont {J.~G.}\ \bibnamefont
  {Baker}}, \bibinfo {author} {\bibfnamefont {B.~J.}\ \bibnamefont {Kelly}}, \
  and\ \bibinfo {author} {\bibfnamefont {J.~R.}\ \bibnamefont {van Meter}},\
  }\href {\doibase 10.1103/RevModPhys.82.3069} {\bibfield  {journal} {\bibinfo
  {journal} {Rev.\ Mod.\ Phys.}\ }\textbf {\bibinfo {volume} {82}},\ \bibinfo
  {pages} {3069} (\bibinfo {year} {2010})}\BibitemShut {NoStop}%
\bibitem [{\citenamefont {Pfeiffer}(2012)}]{Pfeiffer:2012pc}%
  \BibitemOpen
  \bibfield  {author} {\bibinfo {author} {\bibfnamefont {H.~P.}\ \bibnamefont
  {Pfeiffer}},\ }\href {\doibase 10.1088/0264-9381/29/12/124004} {\bibfield
  {journal} {\bibinfo  {journal} {Class. Quant. Grav.}\ }\textbf {\bibinfo
  {volume} {29}},\ \bibinfo {pages} {124004} (\bibinfo {year} {2012})},\
  \Eprint {http://arxiv.org/abs/1203.5166} {arXiv:1203.5166 [gr-qc]}
  \BibitemShut {NoStop}%
\bibitem [{\citenamefont {Hannam}(2013)}]{Hannam:2013pra}%
  \BibitemOpen
  \bibfield  {author} {\bibinfo {author} {\bibfnamefont {M.}~\bibnamefont
  {Hannam}},\ }\href@noop {} {\  (\bibinfo {year} {2013})},\ \Eprint
  {http://arxiv.org/abs/1312.3641} {arXiv:1312.3641 [gr-qc]} \BibitemShut
  {NoStop}%
\bibitem [{\citenamefont {Le~Tiec}(2014)}]{Tiec:2014lba}%
  \BibitemOpen
  \bibfield  {author} {\bibinfo {author} {\bibfnamefont {A.}~\bibnamefont
  {Le~Tiec}},\ }\href {\doibase 10.1142/S0218271814300225} {\bibfield
  {journal} {\bibinfo  {journal} {Int.~J.~Mod.~Phys.~D}\ }\textbf {\bibinfo
  {volume} {23}},\ \bibinfo {pages} {1430022} (\bibinfo {year} {2014})},\
  \Eprint {http://arxiv.org/abs/1408.5505} {arXiv:1408.5505 [gr-qc]}
  \BibitemShut {NoStop}%
\bibitem [{\citenamefont {Ajith}\ \emph {et~al.}(2012)\citenamefont {Ajith},
  \citenamefont {Boyle}, \citenamefont {Brown}, \citenamefont {Brugmann},
  \citenamefont {Buchman} \emph {et~al.}}]{Ajith:2012az}%
  \BibitemOpen
  \bibfield  {author} {\bibinfo {author} {\bibfnamefont {P.}~\bibnamefont
  {Ajith}}, \bibinfo {author} {\bibfnamefont {M.}~\bibnamefont {Boyle}},
  \bibinfo {author} {\bibfnamefont {D.~A.}\ \bibnamefont {Brown}}, \bibinfo
  {author} {\bibfnamefont {B.}~\bibnamefont {Brugmann}}, \bibinfo {author}
  {\bibfnamefont {L.~T.}\ \bibnamefont {Buchman}},  \emph {et~al.},\
  }\href@noop {} {\bibfield  {journal} {\bibinfo  {journal} {Class.\ Quantum
  Grav.}\ }\textbf {\bibinfo {volume} {29}},\ \bibinfo {pages} {124001}
  (\bibinfo {year} {2012})}\BibitemShut {NoStop}%
\bibitem [{\citenamefont {Hinder}\ \emph {et~al.}(2014)\citenamefont {Hinder}
  \emph {et~al.}}]{Hinder:2013oqa}%
  \BibitemOpen
  \bibfield  {author} {\bibinfo {author} {\bibfnamefont {I.}~\bibnamefont
  {Hinder}} \emph {et~al.} (\bibinfo {collaboration} {The NRAR
  Collaboration}),\ }\href@noop {} {\bibfield  {journal} {\bibinfo  {journal}
  {Classical and Quantum Gravity}\ }\textbf {\bibinfo {volume} {31}},\ \bibinfo
  {pages} {025012} (\bibinfo {year} {2014})},\ \Eprint
  {http://arxiv.org/abs/1307.5307} {arXiv:1307.5307 [gr-qc]} \BibitemShut
  {NoStop}%
\bibitem [{\citenamefont {Pekowsky}\ \emph {et~al.}(2013)\citenamefont
  {Pekowsky}, \citenamefont {O'Shaughnessy}, \citenamefont {Healy},\ and\
  \citenamefont {Shoemaker}}]{Pekowsky:2013ska}%
  \BibitemOpen
  \bibfield  {author} {\bibinfo {author} {\bibfnamefont {L.}~\bibnamefont
  {Pekowsky}}, \bibinfo {author} {\bibfnamefont {R.}~\bibnamefont
  {O'Shaughnessy}}, \bibinfo {author} {\bibfnamefont {J.}~\bibnamefont
  {Healy}}, \ and\ \bibinfo {author} {\bibfnamefont {D.}~\bibnamefont
  {Shoemaker}},\ }\href@noop {} {\bibfield  {journal} {\bibinfo  {journal}
  {Phys.\ Rev.\ D}\ }\textbf {\bibinfo {volume} {88}},\ \bibinfo {pages}
  {024040} (\bibinfo {year} {2013})},\ \Eprint {http://arxiv.org/abs/1304.3176}
  {arXiv:1304.3176 [gr-qc]} \BibitemShut {NoStop}%
\bibitem [{\citenamefont {Mroue}\ \emph {et~al.}(2013)\citenamefont {Mroue},
  \citenamefont {Scheel}, \citenamefont {Szilagyi}, \citenamefont {Pfeiffer},
  \citenamefont {Boyle}, \citenamefont {Hemberger}, \citenamefont {Kidder},
  \citenamefont {Lovelace}, \citenamefont {Ossokine}, \citenamefont {Taylor},
  \citenamefont {Zenginoglu}, \citenamefont {Buchman}, \citenamefont {Chu},
  \citenamefont {Foley}, \citenamefont {Giesler}, \citenamefont {Owen},\ and\
  \citenamefont {Teukolsky}}]{Mroue:2013PRL}%
  \BibitemOpen
  \bibfield  {author} {\bibinfo {author} {\bibfnamefont {A.~H.}\ \bibnamefont
  {Mroue}}, \bibinfo {author} {\bibfnamefont {M.~A.}\ \bibnamefont {Scheel}},
  \bibinfo {author} {\bibfnamefont {B.}~\bibnamefont {Szilagyi}}, \bibinfo
  {author} {\bibfnamefont {H.~P.}\ \bibnamefont {Pfeiffer}}, \bibinfo {author}
  {\bibfnamefont {M.}~\bibnamefont {Boyle}}, \bibinfo {author} {\bibfnamefont
  {D.~A.}\ \bibnamefont {Hemberger}}, \bibinfo {author} {\bibfnamefont {L.~E.}\
  \bibnamefont {Kidder}}, \bibinfo {author} {\bibfnamefont {G.}~\bibnamefont
  {Lovelace}}, \bibinfo {author} {\bibfnamefont {S.}~\bibnamefont {Ossokine}},
  \bibinfo {author} {\bibfnamefont {N.~W.}\ \bibnamefont {Taylor}}, \bibinfo
  {author} {\bibfnamefont {A.}~\bibnamefont {Zenginoglu}}, \bibinfo {author}
  {\bibfnamefont {L.~T.}\ \bibnamefont {Buchman}}, \bibinfo {author}
  {\bibfnamefont {T.}~\bibnamefont {Chu}}, \bibinfo {author} {\bibfnamefont
  {E.}~\bibnamefont {Foley}}, \bibinfo {author} {\bibfnamefont
  {M.}~\bibnamefont {Giesler}}, \bibinfo {author} {\bibfnamefont
  {R.}~\bibnamefont {Owen}}, \ and\ \bibinfo {author} {\bibfnamefont {S.~A.}\
  \bibnamefont {Teukolsky}},\ }\href@noop {} {\bibfield  {journal} {\bibinfo
  {journal} {Phys.\ Rev.\ Lett.}\ }\textbf {\bibinfo {volume} {111}},\ \bibinfo
  {pages} {241104} (\bibinfo {year} {2013})},\ \Eprint
  {http://arxiv.org/abs/1304.6077} {arXiv:1304.6077 [gr-qc]} \BibitemShut
  {NoStop}%
\bibitem [{\citenamefont {Healy}\ \emph {et~al.}(2014)\citenamefont {Healy},
  \citenamefont {Lousto},\ and\ \citenamefont {Zlochower}}]{Healy:2014yta}%
  \BibitemOpen
  \bibfield  {author} {\bibinfo {author} {\bibfnamefont {J.}~\bibnamefont
  {Healy}}, \bibinfo {author} {\bibfnamefont {C.~O.}\ \bibnamefont {Lousto}}, \
  and\ \bibinfo {author} {\bibfnamefont {Y.}~\bibnamefont {Zlochower}},\
  }\href@noop {} {\bibfield  {journal} {\bibinfo  {journal} {Phys.\ Rev.\ D}\
  }\textbf {\bibinfo {volume} {89}},\ \bibinfo {pages} {104052} (\bibinfo
  {year} {2014})},\ \Eprint {http://arxiv.org/abs/1406.7295} {arXiv:1406.7295
  [gr-qc]} \BibitemShut {NoStop}%
\bibitem [{\citenamefont {Clark}\ \emph {et~al.}(2014)\citenamefont {Clark},
  \citenamefont {Cadonati}, \citenamefont {Healy}, \citenamefont {Heng},
  \citenamefont {Logue} \emph {et~al.}}]{Clark:2014fva}%
  \BibitemOpen
  \bibfield  {author} {\bibinfo {author} {\bibfnamefont {J.}~\bibnamefont
  {Clark}}, \bibinfo {author} {\bibfnamefont {L.}~\bibnamefont {Cadonati}},
  \bibinfo {author} {\bibfnamefont {J.}~\bibnamefont {Healy}}, \bibinfo
  {author} {\bibfnamefont {I.~S.}\ \bibnamefont {Heng}}, \bibinfo {author}
  {\bibfnamefont {J.}~\bibnamefont {Logue}},  \emph {et~al.},\ }\href@noop {}
  {\  (\bibinfo {year} {2014})},\ \Eprint {http://arxiv.org/abs/1406.5426}
  {arXiv:1406.5426 [gr-qc]} \BibitemShut {NoStop}%
\bibitem [{\citenamefont {Bowen}(1979)}]{bowen79}%
  \BibitemOpen
  \bibfield  {author} {\bibinfo {author} {\bibfnamefont {J.~M.}\ \bibnamefont
  {Bowen}},\ }\href@noop {} {\bibfield  {journal} {\bibinfo  {journal} {Gen.\
  Relativ.\ Gravit.}\ }\textbf {\bibinfo {volume} {11}},\ \bibinfo {pages}
  {227} (\bibinfo {year} {1979})}\BibitemShut {NoStop}%
\bibitem [{\citenamefont {Bowen}\ and\ \citenamefont {{York,
  Jr.}}(1980)}]{Bowen-York:1980}%
  \BibitemOpen
  \bibfield  {author} {\bibinfo {author} {\bibfnamefont {J.~M.}\ \bibnamefont
  {Bowen}}\ and\ \bibinfo {author} {\bibfnamefont {J.~W.}\ \bibnamefont {{York,
  Jr.}}},\ }\href@noop {} {\bibfield  {journal} {\bibinfo  {journal} {Phys.\
  Rev.\ D}\ }\textbf {\bibinfo {volume} {21}},\ \bibinfo {pages} {2047}
  (\bibinfo {year} {1980})}\BibitemShut {NoStop}%
\bibitem [{\citenamefont {Brandt}\ and\ \citenamefont
  {Br{\"u}gmann}(1997)}]{Brandt1997}%
  \BibitemOpen
  \bibfield  {author} {\bibinfo {author} {\bibfnamefont {S.}~\bibnamefont
  {Brandt}}\ and\ \bibinfo {author} {\bibfnamefont {B.}~\bibnamefont
  {Br{\"u}gmann}},\ }\href@noop {} {\bibfield  {journal} {\bibinfo  {journal}
  {Phys.\ Rev.\ Lett.}\ }\textbf {\bibinfo {volume} {78}},\ \bibinfo {pages}
  {3606} (\bibinfo {year} {1997})}\BibitemShut {NoStop}%
\bibitem [{\citenamefont {{York, Jr.}}(1980)}]{york80}%
  \BibitemOpen
  \bibfield  {author} {\bibinfo {author} {\bibfnamefont {J.~W.}\ \bibnamefont
  {{York, Jr.}}},\ }in\ \href@noop {} {\emph {\bibinfo {booktitle} {Essays in
  General Relativity}}},\ \bibinfo {editor} {edited by\ \bibinfo {editor}
  {\bibfnamefont {F.~J.}\ \bibnamefont {Tipler}}}\ (\bibinfo  {publisher}
  {Academic},\ \bibinfo {address} {New York},\ \bibinfo {year} {1980})\ pp.\
  \bibinfo {pages} {39--58}\BibitemShut {NoStop}%
\bibitem [{\citenamefont {Garat}\ and\ \citenamefont
  {Price}(2000)}]{GaratPrice:2000}%
  \BibitemOpen
  \bibfield  {author} {\bibinfo {author} {\bibfnamefont {A.}~\bibnamefont
  {Garat}}\ and\ \bibinfo {author} {\bibfnamefont {R.~H.}\ \bibnamefont
  {Price}},\ }\href@noop {} {\bibfield  {journal} {\bibinfo  {journal} {Phys.\
  Rev.\ D}\ }\textbf {\bibinfo {volume} {61}},\ \bibinfo {pages} {124011}
  (\bibinfo {year} {2000})}\BibitemShut {NoStop}%
\bibitem [{\citenamefont {{Valiente Kroon}}(2004)}]{Kroon:2004}%
  \BibitemOpen
  \bibfield  {author} {\bibinfo {author} {\bibfnamefont {J.~A.}\ \bibnamefont
  {{Valiente Kroon}}},\ }\href@noop {} {\bibfield  {journal} {\bibinfo
  {journal} {Phys. Rev. Lett.}\ }\textbf {\bibinfo {volume} {92}},\ \bibinfo
  {pages} {041101} (\bibinfo {year} {2004})}\BibitemShut {NoStop}%
\bibitem [{\citenamefont {Cook}\ and\ \citenamefont {{York,
  Jr.}}(1990)}]{cook90}%
  \BibitemOpen
  \bibfield  {author} {\bibinfo {author} {\bibfnamefont {G.~B.}\ \bibnamefont
  {Cook}}\ and\ \bibinfo {author} {\bibfnamefont {J.~W.}\ \bibnamefont {{York,
  Jr.}}},\ }\href@noop {} {\bibfield  {journal} {\bibinfo  {journal} {Phys.\
  Rev.\ D}\ }\textbf {\bibinfo {volume} {41}},\ \bibinfo {pages} {1077}
  (\bibinfo {year} {1990})}\BibitemShut {NoStop}%
\bibitem [{\citenamefont {Dain}\ \emph {et~al.}(2002)\citenamefont {Dain},
  \citenamefont {Lousto},\ and\ \citenamefont {Takahashi}}]{DainEtAl:2002}%
  \BibitemOpen
  \bibfield  {author} {\bibinfo {author} {\bibfnamefont {S.}~\bibnamefont
  {Dain}}, \bibinfo {author} {\bibfnamefont {C.~O.}\ \bibnamefont {Lousto}}, \
  and\ \bibinfo {author} {\bibfnamefont {R.}~\bibnamefont {Takahashi}},\ }\href
  {\doibase 10.1103/PhysRevD.65.104038} {\bibfield  {journal} {\bibinfo
  {journal} {Phys. Rev. D}\ }\textbf {\bibinfo {volume} {65}},\ \bibinfo
  {pages} {104038} (\bibinfo {year} {2002})}\BibitemShut {NoStop}%
\bibitem [{\citenamefont {Hannam}\ \emph {et~al.}(2009)\citenamefont {Hannam},
  \citenamefont {Husa},\ and\ \citenamefont {Murchadha}}]{HannamEtAl:2009}%
  \BibitemOpen
  \bibfield  {author} {\bibinfo {author} {\bibfnamefont {M.}~\bibnamefont
  {Hannam}}, \bibinfo {author} {\bibfnamefont {S.}~\bibnamefont {Husa}}, \ and\
  \bibinfo {author} {\bibfnamefont {N.~O.}\ \bibnamefont {Murchadha}},\ }\href
  {\doibase 10.1103/PhysRevD.80.124007} {\bibfield  {journal} {\bibinfo
  {journal} {Phys.\ Rev.\ D}\ }\textbf {\bibinfo {volume} {80}},\ \bibinfo
  {pages} {124007} (\bibinfo {year} {2009})}\BibitemShut {NoStop}%
\bibitem [{\citenamefont {Lovelace}\ \emph {et~al.}(2011)\citenamefont
  {Lovelace}, \citenamefont {Scheel},\ and\ \citenamefont {Szil{\'
  a}gyi}}]{Lovelace:2010ne}%
  \BibitemOpen
  \bibfield  {author} {\bibinfo {author} {\bibfnamefont {G.}~\bibnamefont
  {Lovelace}}, \bibinfo {author} {\bibfnamefont {M.~A.}\ \bibnamefont
  {Scheel}}, \ and\ \bibinfo {author} {\bibfnamefont {B.}~\bibnamefont {Szil{\'
  a}gyi}},\ }\href {\doibase 10.1103/PhysRevD.83.024010} {\bibfield  {journal}
  {\bibinfo  {journal} {Phys.\ Rev.\ D}\ }\textbf {\bibinfo {volume} {83}},\
  \bibinfo {pages} {024010} (\bibinfo {year} {2011})},\ \Eprint
  {http://arxiv.org/abs/1010.2777} {arXiv:1010.2777 [gr-qc]} \BibitemShut
  {NoStop}%
\bibitem [{\citenamefont {Lovelace}\ \emph {et~al.}(2012)\citenamefont
  {Lovelace}, \citenamefont {Boyle}, \citenamefont {Scheel},\ and\
  \citenamefont {Szil{\'a}gyi}}]{Lovelace:2011nu}%
  \BibitemOpen
  \bibfield  {author} {\bibinfo {author} {\bibfnamefont {G.}~\bibnamefont
  {Lovelace}}, \bibinfo {author} {\bibfnamefont {M.}~\bibnamefont {Boyle}},
  \bibinfo {author} {\bibfnamefont {M.~A.}\ \bibnamefont {Scheel}}, \ and\
  \bibinfo {author} {\bibfnamefont {B.}~\bibnamefont {Szil{\'a}gyi}},\ }\href
  {\doibase 10.1088/0264-9381/29/4/045003} {\bibfield  {journal} {\bibinfo
  {journal} {Class. Quant. Grav.}\ }\textbf {\bibinfo {volume} {29}},\ \bibinfo
  {pages} {045003} (\bibinfo {year} {2012})},\ \Eprint
  {http://arxiv.org/abs/arXiv:1110.2229} {arXiv:arXiv:1110.2229 [gr-qc]}
  \BibitemShut {NoStop}%
\bibitem [{\citenamefont {{Lovelace}}\ \emph {et~al.}(2013)\citenamefont
  {{Lovelace}}, \citenamefont {{Duez}}, \citenamefont {{Foucart}},
  \citenamefont {{Kidder}}, \citenamefont {{Pfeiffer}}, \citenamefont
  {{Scheel}},\ and\ \citenamefont {{Szil{\'a}gyi}}}]{Lovelace:2013vma}%
  \BibitemOpen
  \bibfield  {author} {\bibinfo {author} {\bibfnamefont {G.}~\bibnamefont
  {{Lovelace}}}, \bibinfo {author} {\bibfnamefont {M.~D.}\ \bibnamefont
  {{Duez}}}, \bibinfo {author} {\bibfnamefont {F.}~\bibnamefont {{Foucart}}},
  \bibinfo {author} {\bibfnamefont {L.~E.}\ \bibnamefont {{Kidder}}}, \bibinfo
  {author} {\bibfnamefont {H.~P.}\ \bibnamefont {{Pfeiffer}}}, \bibinfo
  {author} {\bibfnamefont {M.~A.}\ \bibnamefont {{Scheel}}}, \ and\ \bibinfo
  {author} {\bibfnamefont {B.}~\bibnamefont {{Szil{\'a}gyi}}},\ }\href
  {\doibase 10.1088/0264-9381/30/13/135004} {\bibfield  {journal} {\bibinfo
  {journal} {Class. Quantum Grav.}\ }\textbf {\bibinfo {volume} {30}},\
  \bibinfo {eid} {135004} (\bibinfo {year} {2013})},\ \Eprint
  {http://arxiv.org/abs/1302.6297} {arXiv:1302.6297 [gr-qc]} \BibitemShut
  {NoStop}%
\bibitem [{\citenamefont {Novikov}\ and\ \citenamefont
  {Thorne}(1973)}]{NovikovThorne:1973}%
  \BibitemOpen
  \bibfield  {author} {\bibinfo {author} {\bibfnamefont {I.~D.}\ \bibnamefont
  {Novikov}}\ and\ \bibinfo {author} {\bibfnamefont {K.~S.}\ \bibnamefont
  {Thorne}},\ }\enquote {\bibinfo {title} {Black holes},}\ \ (\bibinfo
  {publisher} {Gordon and Breach},\ \bibinfo {address} {New York},\ \bibinfo
  {year} {1973})\ p.\ \bibinfo {pages} {343}\BibitemShut {NoStop}%
\bibitem [{\citenamefont {Thorne}(1974)}]{Thorne:1974}%
  \BibitemOpen
  \bibfield  {author} {\bibinfo {author} {\bibfnamefont {K.~S.}\ \bibnamefont
  {Thorne}},\ }\href {\doibase 10.1086/152991} {\bibfield  {journal} {\bibinfo
  {journal} {Astrophys.\ J.}\ }\textbf {\bibinfo {volume} {191}},\ \bibinfo
  {pages} {507} (\bibinfo {year} {1974})}\BibitemShut {NoStop}%
\bibitem [{\citenamefont {Hemberger}\ \emph
  {et~al.}(2013{\natexlab{a}})\citenamefont {Hemberger}, \citenamefont
  {Lovelace}, \citenamefont {Loredo}, \citenamefont {Kidder}, \citenamefont
  {Scheel}, \citenamefont {Szil\'agyi}, \citenamefont {Taylor},\ and\
  \citenamefont {Teukolsky}}]{Hemberger:2013hsa}%
  \BibitemOpen
  \bibfield  {author} {\bibinfo {author} {\bibfnamefont {D.~A.}\ \bibnamefont
  {Hemberger}}, \bibinfo {author} {\bibfnamefont {G.}~\bibnamefont {Lovelace}},
  \bibinfo {author} {\bibfnamefont {T.~J.}\ \bibnamefont {Loredo}}, \bibinfo
  {author} {\bibfnamefont {L.~E.}\ \bibnamefont {Kidder}}, \bibinfo {author}
  {\bibfnamefont {M.~A.}\ \bibnamefont {Scheel}}, \bibinfo {author}
  {\bibfnamefont {B.}~\bibnamefont {Szil\'agyi}}, \bibinfo {author}
  {\bibfnamefont {N.~W.}\ \bibnamefont {Taylor}}, \ and\ \bibinfo {author}
  {\bibfnamefont {S.~A.}\ \bibnamefont {Teukolsky}},\ }\href {\doibase
  10.1103/PhysRevD.88.064014} {\bibfield  {journal} {\bibinfo  {journal} {Phys.
  Rev. D}\ }\textbf {\bibinfo {volume} {88}},\ \bibinfo {pages} {064014}
  (\bibinfo {year} {2013}{\natexlab{a}})},\ \Eprint
  {http://arxiv.org/abs/1305.5991} {arXiv:1305.5991 [gr-qc]} \BibitemShut
  {NoStop}%
\bibitem [{\citenamefont {Rezzolla}\ \emph {et~al.}(2008)\citenamefont
  {Rezzolla} \emph {et~al.}}]{Rezzolla:2007xa}%
  \BibitemOpen
  \bibfield  {author} {\bibinfo {author} {\bibfnamefont {L.}~\bibnamefont
  {Rezzolla}} \emph {et~al.},\ }\href@noop {} {\bibfield  {journal} {\bibinfo
  {journal} {Astrophys.\ J.}\ }\textbf {\bibinfo {volume} {679}},\ \bibinfo
  {pages} {1422} (\bibinfo {year} {2008})},\ \Eprint
  {http://arxiv.org/abs/0708.3999} {arXiv:0708.3999 [gr-qc]} \BibitemShut
  {NoStop}%
\bibitem [{\citenamefont {Hannam}\ \emph {et~al.}(2010)\citenamefont {Hannam},
  \citenamefont {Husa}, \citenamefont {Ohme}, \citenamefont {M\"{u}ller},\ and\
  \citenamefont {Br\"{u}gmann}}]{HannamEtAl:2010}%
  \BibitemOpen
  \bibfield  {author} {\bibinfo {author} {\bibfnamefont {M.}~\bibnamefont
  {Hannam}}, \bibinfo {author} {\bibfnamefont {S.}~\bibnamefont {Husa}},
  \bibinfo {author} {\bibfnamefont {F.}~\bibnamefont {Ohme}}, \bibinfo {author}
  {\bibfnamefont {D.}~\bibnamefont {M\"{u}ller}}, \ and\ \bibinfo {author}
  {\bibfnamefont {B.}~\bibnamefont {Br\"{u}gmann}},\ }\href@noop {} {\bibfield
  {journal} {\bibinfo  {journal} {Phys.\ Rev.\ D}\ }\textbf {\bibinfo {volume}
  {82}},\ \bibinfo {pages} {124008} (\bibinfo {year} {2010})},\ \Eprint
  {http://arxiv.org/abs/arXiv:1007.4789} {arXiv:1007.4789} \BibitemShut
  {NoStop}%
\bibitem [{\citenamefont {Marronetti}\ \emph {et~al.}(2008)\citenamefont
  {Marronetti}, \citenamefont {Tichy}, \citenamefont {Br{\"{u}}gmann},
  \citenamefont {Gonz{\'{a}}lez},\ and\ \citenamefont
  {Sperhake}}]{MarronettiEtal:2008}%
  \BibitemOpen
  \bibfield  {author} {\bibinfo {author} {\bibfnamefont {P.}~\bibnamefont
  {Marronetti}}, \bibinfo {author} {\bibfnamefont {W.}~\bibnamefont {Tichy}},
  \bibinfo {author} {\bibfnamefont {B.}~\bibnamefont {Br{\"{u}}gmann}},
  \bibinfo {author} {\bibfnamefont {J.}~\bibnamefont {Gonz{\'{a}}lez}}, \ and\
  \bibinfo {author} {\bibfnamefont {U.}~\bibnamefont {Sperhake}},\ }\href@noop
  {} {\bibfield  {journal} {\bibinfo  {journal} {Phys.\ Rev.\ D}\ }\textbf
  {\bibinfo {volume} {77}},\ \bibinfo {pages} {064010} (\bibinfo {year}
  {2008})}\BibitemShut {NoStop}%
\bibitem [{\citenamefont {Dain}\ \emph {et~al.}(2008)\citenamefont {Dain},
  \citenamefont {Lousto},\ and\ \citenamefont {Zlochower}}]{DainEtAl:2008}%
  \BibitemOpen
  \bibfield  {author} {\bibinfo {author} {\bibfnamefont {S.}~\bibnamefont
  {Dain}}, \bibinfo {author} {\bibfnamefont {C.~O.}\ \bibnamefont {Lousto}}, \
  and\ \bibinfo {author} {\bibfnamefont {Y.}~\bibnamefont {Zlochower}},\
  }\href@noop {} {\bibfield  {journal} {\bibinfo  {journal} {Phys.\ Rev.\ D}\
  }\textbf {\bibinfo {volume} {78}},\ \bibinfo {pages} {024039} (\bibinfo
  {year} {2008})},\ \Eprint {http://arxiv.org/abs/0803.0351v2}
  {arXiv:0803.0351v2 [gr-qc]} \BibitemShut {NoStop}%
\bibitem [{\citenamefont {Ruchlin}\ \emph {et~al.}(2014)\citenamefont
  {Ruchlin}, \citenamefont {Healy}, \citenamefont {Lousto},\ and\ \citenamefont
  {Zlochower}}]{Ruchlin:2014zva}%
  \BibitemOpen
  \bibfield  {author} {\bibinfo {author} {\bibfnamefont {I.}~\bibnamefont
  {Ruchlin}}, \bibinfo {author} {\bibfnamefont {J.}~\bibnamefont {Healy}},
  \bibinfo {author} {\bibfnamefont {C.~O.}\ \bibnamefont {Lousto}}, \ and\
  \bibinfo {author} {\bibfnamefont {Y.}~\bibnamefont {Zlochower}},\ }\href@noop
  {} {\  (\bibinfo {year} {2014})},\ \Eprint {http://arxiv.org/abs/1410.8607}
  {arXiv:1410.8607 [gr-qc]} \BibitemShut {NoStop}%
\bibitem [{SpE()}]{SpECwebsite}%
  \BibitemOpen
  \href@noop {} {}\bibinfo {howpublished}
  {\url{http://www.black-holes.org/SpEC.html}}\BibitemShut {NoStop}%
\bibitem [{\citenamefont {Lovelace}\ \emph {et~al.}(2008)\citenamefont
  {Lovelace}, \citenamefont {Owen}, \citenamefont {Pfeiffer},\ and\
  \citenamefont {Chu}}]{Lovelace2008}%
  \BibitemOpen
  \bibfield  {author} {\bibinfo {author} {\bibfnamefont {G.}~\bibnamefont
  {Lovelace}}, \bibinfo {author} {\bibfnamefont {R.}~\bibnamefont {Owen}},
  \bibinfo {author} {\bibfnamefont {H.~P.}\ \bibnamefont {Pfeiffer}}, \ and\
  \bibinfo {author} {\bibfnamefont {T.}~\bibnamefont {Chu}},\ }\href {\doibase
  10.1103/PhysRevD.78.084017} {\bibfield  {journal} {\bibinfo  {journal}
  {Phys.\ Rev.\ D}\ }\textbf {\bibinfo {volume} {78}},\ \bibinfo {pages}
  {084017} (\bibinfo {year} {2008})}\BibitemShut {NoStop}%
\bibitem [{\citenamefont {Taracchini}\ \emph {et~al.}(2014)\citenamefont
  {Taracchini}, \citenamefont {Buonanno}, \citenamefont {Pan}, \citenamefont
  {Hinderer}, \citenamefont {Boyle}, \citenamefont {Hemberger}, \citenamefont
  {Kidder}, \citenamefont {Lovelace}, \citenamefont {Mroue}, \citenamefont
  {Pfeiffer}, \citenamefont {Scheel}, \citenamefont {Szilagyi},\ and\
  \citenamefont {Zenginoglu}}]{Taracchini:2013rva}%
  \BibitemOpen
  \bibfield  {author} {\bibinfo {author} {\bibfnamefont {A.}~\bibnamefont
  {Taracchini}}, \bibinfo {author} {\bibfnamefont {A.}~\bibnamefont
  {Buonanno}}, \bibinfo {author} {\bibfnamefont {Y.}~\bibnamefont {Pan}},
  \bibinfo {author} {\bibfnamefont {T.}~\bibnamefont {Hinderer}}, \bibinfo
  {author} {\bibfnamefont {M.}~\bibnamefont {Boyle}}, \bibinfo {author}
  {\bibfnamefont {D.~A.}\ \bibnamefont {Hemberger}}, \bibinfo {author}
  {\bibfnamefont {L.~E.}\ \bibnamefont {Kidder}}, \bibinfo {author}
  {\bibfnamefont {G.}~\bibnamefont {Lovelace}}, \bibinfo {author}
  {\bibfnamefont {A.~H.}\ \bibnamefont {Mroue}}, \bibinfo {author}
  {\bibfnamefont {H.~P.}\ \bibnamefont {Pfeiffer}}, \bibinfo {author}
  {\bibfnamefont {M.~A.}\ \bibnamefont {Scheel}}, \bibinfo {author}
  {\bibfnamefont {B.}~\bibnamefont {Szilagyi}}, \ and\ \bibinfo {author}
  {\bibfnamefont {A.}~\bibnamefont {Zenginoglu}},\ }\href@noop {} {\bibfield
  {journal} {\bibinfo  {journal} {Phys.Rev.}\ }\textbf {\bibinfo {volume}
  {D89}},\ \bibinfo {pages} {061502} (\bibinfo {year} {2014})},\ \Eprint
  {http://arxiv.org/abs/1311.2544} {arXiv:1311.2544 [gr-qc]} \BibitemShut
  {NoStop}%
\bibitem [{\citenamefont {Lovelace}\ \emph {et~al.}(2014)\citenamefont
  {Lovelace}, \citenamefont {Scheel}, \citenamefont {Owen}, \citenamefont
  {Giesler}, \citenamefont {Katebi}, \citenamefont {Szil\'{a}gyi},
  \citenamefont {Chu}, \citenamefont {Demos}, \citenamefont {Hemberger},
  \citenamefont {Kidder}, \citenamefont {Pfeiffer},\ and\ \citenamefont
  {Afshari}}]{Lovelace2014}%
  \BibitemOpen
  \bibfield  {author} {\bibinfo {author} {\bibfnamefont {G.}~\bibnamefont
  {Lovelace}}, \bibinfo {author} {\bibfnamefont {M.~A.}\ \bibnamefont
  {Scheel}}, \bibinfo {author} {\bibfnamefont {R.}~\bibnamefont {Owen}},
  \bibinfo {author} {\bibfnamefont {M.}~\bibnamefont {Giesler}}, \bibinfo
  {author} {\bibfnamefont {R.}~\bibnamefont {Katebi}}, \bibinfo {author}
  {\bibfnamefont {B.}~\bibnamefont {Szil\'{a}gyi}}, \bibinfo {author}
  {\bibfnamefont {T.}~\bibnamefont {Chu}}, \bibinfo {author} {\bibfnamefont
  {N.}~\bibnamefont {Demos}}, \bibinfo {author} {\bibfnamefont {D.~A.}\
  \bibnamefont {Hemberger}}, \bibinfo {author} {\bibfnamefont {L.~E.}\
  \bibnamefont {Kidder}}, \bibinfo {author} {\bibfnamefont {H.~P.}\
  \bibnamefont {Pfeiffer}}, \ and\ \bibinfo {author} {\bibfnamefont
  {N.}~\bibnamefont {Afshari}},\ }\href@noop {} {\  (\bibinfo {year} {2014})},\
  \bibinfo {note} {submitted to Class.~Quantum~Grav.},\ \Eprint
  {http://arxiv.org/abs/1411.7297} {arXiv:1411.7297 [gr-qc]} \BibitemShut
  {NoStop}%
\bibitem [{\citenamefont {Pfeiffer}\ \emph {et~al.}(2003)\citenamefont
  {Pfeiffer}, \citenamefont {Kidder}, \citenamefont {Scheel},\ and\
  \citenamefont {Teukolsky}}]{Pfeiffer2003}%
  \BibitemOpen
  \bibfield  {author} {\bibinfo {author} {\bibfnamefont {H.~P.}\ \bibnamefont
  {Pfeiffer}}, \bibinfo {author} {\bibfnamefont {L.~E.}\ \bibnamefont
  {Kidder}}, \bibinfo {author} {\bibfnamefont {M.~A.}\ \bibnamefont {Scheel}},
  \ and\ \bibinfo {author} {\bibfnamefont {S.~A.}\ \bibnamefont {Teukolsky}},\
  }\href {\doibase 10.1016/S0010-4655(02)00847-0} {\bibfield  {journal}
  {\bibinfo  {journal} {Comput.\ Phys.\ Commun.}\ }\textbf {\bibinfo {volume}
  {152}},\ \bibinfo {pages} {253} (\bibinfo {year} {2003})}\BibitemShut
  {NoStop}%
\bibitem [{\citenamefont {{Caudill}}\ \emph {et~al.}(2006)\citenamefont
  {{Caudill}}, \citenamefont {{Cook}}, \citenamefont {{Grigsby}},\ and\
  \citenamefont {{Pfeiffer}}}]{Caudill-etal:2006}%
  \BibitemOpen
  \bibfield  {author} {\bibinfo {author} {\bibfnamefont {M.}~\bibnamefont
  {{Caudill}}}, \bibinfo {author} {\bibfnamefont {G.~B.}\ \bibnamefont
  {{Cook}}}, \bibinfo {author} {\bibfnamefont {J.~D.}\ \bibnamefont
  {{Grigsby}}}, \ and\ \bibinfo {author} {\bibfnamefont {H.~P.}\ \bibnamefont
  {{Pfeiffer}}},\ }\href@noop {} {\bibfield  {journal} {\bibinfo  {journal}
  {Phys.\ Rev.\ D}\ }\textbf {\bibinfo {volume} {74}},\ \bibinfo {pages}
  {064011} (\bibinfo {year} {2006})},\ \Eprint
  {http://arxiv.org/abs/gr-qc/0605053} {gr-qc/0605053} \BibitemShut {NoStop}%
\bibitem [{\citenamefont {York}(1999)}]{York1999}%
  \BibitemOpen
  \bibfield  {author} {\bibinfo {author} {\bibfnamefont {J.~W.}\ \bibnamefont
  {York}},\ }\href {\doibase 10.1103/PhysRevLett.82.1350} {\bibfield  {journal}
  {\bibinfo  {journal} {Phys.\ Rev.\ Lett.}\ }\textbf {\bibinfo {volume}
  {82}},\ \bibinfo {pages} {1350} (\bibinfo {year} {1999})}\BibitemShut
  {NoStop}%
\bibitem [{\citenamefont {Pfeiffer}\ \emph {et~al.}(2007)\citenamefont
  {Pfeiffer}, \citenamefont {Brown}, \citenamefont {Kidder}, \citenamefont
  {Lindblom}, \citenamefont {Lovelace},\ and\ \citenamefont
  {Scheel}}]{Pfeiffer-Brown-etal:2007}%
  \BibitemOpen
  \bibfield  {author} {\bibinfo {author} {\bibfnamefont {H.~P.}\ \bibnamefont
  {Pfeiffer}}, \bibinfo {author} {\bibfnamefont {D.~A.}\ \bibnamefont {Brown}},
  \bibinfo {author} {\bibfnamefont {L.~E.}\ \bibnamefont {Kidder}}, \bibinfo
  {author} {\bibfnamefont {L.}~\bibnamefont {Lindblom}}, \bibinfo {author}
  {\bibfnamefont {G.}~\bibnamefont {Lovelace}}, \ and\ \bibinfo {author}
  {\bibfnamefont {M.~A.}\ \bibnamefont {Scheel}},\ }\href@noop {} {\bibfield
  {journal} {\bibinfo  {journal} {Class.\ Quantum Grav.}\ }\textbf {\bibinfo
  {volume} {24}},\ \bibinfo {pages} {S59} (\bibinfo {year} {2007})},\ \Eprint
  {http://arxiv.org/abs/gr-qc/0702106} {gr-qc/0702106} \BibitemShut {NoStop}%
\bibitem [{\citenamefont {Buonanno}\ \emph {et~al.}(2011)\citenamefont
  {Buonanno}, \citenamefont {Kidder}, \citenamefont {Mrou\'{e}}, \citenamefont
  {Pfeiffer},\ and\ \citenamefont {Taracchini}}]{Buonanno:2010yk}%
  \BibitemOpen
  \bibfield  {author} {\bibinfo {author} {\bibfnamefont {A.}~\bibnamefont
  {Buonanno}}, \bibinfo {author} {\bibfnamefont {L.~E.}\ \bibnamefont
  {Kidder}}, \bibinfo {author} {\bibfnamefont {A.~H.}\ \bibnamefont
  {Mrou\'{e}}}, \bibinfo {author} {\bibfnamefont {H.~P.}\ \bibnamefont
  {Pfeiffer}}, \ and\ \bibinfo {author} {\bibfnamefont {A.}~\bibnamefont
  {Taracchini}},\ }\href {\doibase 10.1103/PhysRevD.83.104034} {\bibfield
  {journal} {\bibinfo  {journal} {Phys.\ Rev.\ D}\ }\textbf {\bibinfo {volume}
  {83}},\ \bibinfo {pages} {104034} (\bibinfo {year} {2011})},\ \Eprint
  {http://arxiv.org/abs/1012.1549} {arXiv:1012.1549 [gr-qc]} \BibitemShut
  {NoStop}%
\bibitem [{\citenamefont {Mrou\'e}\ and\ \citenamefont
  {Pfeiffer}(2012)}]{Mroue:2012kv}%
  \BibitemOpen
  \bibfield  {author} {\bibinfo {author} {\bibfnamefont {A.~H.}\ \bibnamefont
  {Mrou\'e}}\ and\ \bibinfo {author} {\bibfnamefont {H.~P.}\ \bibnamefont
  {Pfeiffer}},\ }\href@noop {} {\  (\bibinfo {year} {2012})},\ \Eprint
  {http://arxiv.org/abs/1210.2958} {arXiv:1210.2958 [gr-qc]} \BibitemShut
  {NoStop}%
\bibitem [{\citenamefont {Friedrich}(1985)}]{Friedrich1985}%
  \BibitemOpen
  \bibfield  {author} {\bibinfo {author} {\bibfnamefont {H.}~\bibnamefont
  {Friedrich}},\ }\href {\doibase 10.1007/BF01217728} {\bibfield  {journal}
  {\bibinfo  {journal} {Commun.\ Math.\ Phys.}\ }\textbf {\bibinfo {volume}
  {100}},\ \bibinfo {pages} {525} (\bibinfo {year} {1985})}\BibitemShut
  {NoStop}%
\bibitem [{\citenamefont {Garfinkle}(2002)}]{Garfinkle2002}%
  \BibitemOpen
  \bibfield  {author} {\bibinfo {author} {\bibfnamefont {D.}~\bibnamefont
  {Garfinkle}},\ }\href@noop {} {\bibfield  {journal} {\bibinfo  {journal}
  {Phys.\ Rev.\ D}\ }\textbf {\bibinfo {volume} {65}},\ \bibinfo {pages}
  {044029} (\bibinfo {year} {2002})}\BibitemShut {NoStop}%
\bibitem [{\citenamefont {Pretorius}(2005{\natexlab{b}})}]{Pretorius2005c}%
  \BibitemOpen
  \bibfield  {author} {\bibinfo {author} {\bibfnamefont {F.}~\bibnamefont
  {Pretorius}},\ }\href {http://stacks.iop.org/0264-9381/22/425} {\bibfield
  {journal} {\bibinfo  {journal} {Class.\ Quantum Grav.}\ }\textbf {\bibinfo
  {volume} {22}},\ \bibinfo {pages} {425} (\bibinfo {year}
  {2005}{\natexlab{b}})}\BibitemShut {NoStop}%
\bibitem [{\citenamefont {Lindblom}\ \emph {et~al.}(2006)\citenamefont
  {Lindblom}, \citenamefont {Scheel}, \citenamefont {Kidder}, \citenamefont
  {Owen},\ and\ \citenamefont {Rinne}}]{Lindblom2006}%
  \BibitemOpen
  \bibfield  {author} {\bibinfo {author} {\bibfnamefont {L.}~\bibnamefont
  {Lindblom}}, \bibinfo {author} {\bibfnamefont {M.~A.}\ \bibnamefont
  {Scheel}}, \bibinfo {author} {\bibfnamefont {L.~E.}\ \bibnamefont {Kidder}},
  \bibinfo {author} {\bibfnamefont {R.}~\bibnamefont {Owen}}, \ and\ \bibinfo
  {author} {\bibfnamefont {O.}~\bibnamefont {Rinne}},\ }\href@noop {}
  {\bibfield  {journal} {\bibinfo  {journal} {Class.\ Quantum Grav.}\ }\textbf
  {\bibinfo {volume} {23}},\ \bibinfo {pages} {S447} (\bibinfo {year}
  {2006})}\BibitemShut {NoStop}%
\bibitem [{\citenamefont {Lindblom}\ and\ \citenamefont
  {Szil\'agyi}(2009)}]{Lindblom2009c}%
  \BibitemOpen
  \bibfield  {author} {\bibinfo {author} {\bibfnamefont {L.}~\bibnamefont
  {Lindblom}}\ and\ \bibinfo {author} {\bibfnamefont {B.}~\bibnamefont
  {Szil\'agyi}},\ }\href@noop {} {\bibfield  {journal} {\bibinfo  {journal}
  {Phys.\ Rev.\ D}\ }\textbf {\bibinfo {volume} {80}},\ \bibinfo {pages}
  {084019} (\bibinfo {year} {2009})},\ \Eprint
  {http://arxiv.org/abs/arXiv:0904.4873} {arXiv:0904.4873} \BibitemShut
  {NoStop}%
\bibitem [{\citenamefont {Choptuik}\ and\ \citenamefont
  {Pretorius}(2010)}]{Choptuik:2009ww}%
  \BibitemOpen
  \bibfield  {author} {\bibinfo {author} {\bibfnamefont {M.~W.}\ \bibnamefont
  {Choptuik}}\ and\ \bibinfo {author} {\bibfnamefont {F.}~\bibnamefont
  {Pretorius}},\ }\href {\doibase 10.1103/PhysRevLett.104.111101} {\bibfield
  {journal} {\bibinfo  {journal} {Phys.\ Rev.\ Lett.}\ }\textbf {\bibinfo
  {volume} {104}},\ \bibinfo {pages} {111101} (\bibinfo {year} {2010})},\
  \Eprint {http://arxiv.org/abs/0908.1780} {arXiv:0908.1780 [gr-qc]}
  \BibitemShut {NoStop}%
\bibitem [{\citenamefont {Szil{\' a}gyi}\ \emph {et~al.}(2009)\citenamefont
  {Szil{\' a}gyi}, \citenamefont {Lindblom},\ and\ \citenamefont
  {Scheel}}]{Szilagyi:2009qz}%
  \BibitemOpen
  \bibfield  {author} {\bibinfo {author} {\bibfnamefont {B.}~\bibnamefont
  {Szil{\' a}gyi}}, \bibinfo {author} {\bibfnamefont {L.}~\bibnamefont
  {Lindblom}}, \ and\ \bibinfo {author} {\bibfnamefont {M.~A.}\ \bibnamefont
  {Scheel}},\ }\href@noop {} {\bibfield  {journal} {\bibinfo  {journal} {Phys.\
  Rev.\ D}\ }\textbf {\bibinfo {volume} {80}},\ \bibinfo {pages} {124010}
  (\bibinfo {year} {2009})},\ \Eprint {http://arxiv.org/abs/0909.3557}
  {arXiv:0909.3557 [gr-qc]} \BibitemShut {NoStop}%
\bibitem [{\citenamefont {Szil{\'a}gyi}(2014)}]{Szilagyi:2014fna}%
  \BibitemOpen
  \bibfield  {author} {\bibinfo {author} {\bibfnamefont {B.}~\bibnamefont
  {Szil{\'a}gyi}},\ }\href {\doibase 10.1142/S0218271814300146} {\bibfield
  {journal} {\bibinfo  {journal} {Int.J.Mod.Phys.}\ }\textbf {\bibinfo {volume}
  {D23}},\ \bibinfo {pages} {1430014} (\bibinfo {year} {2014})},\ \Eprint
  {http://arxiv.org/abs/1405.3693} {arXiv:1405.3693 [gr-qc]} \BibitemShut
  {NoStop}%
\bibitem [{\citenamefont {{M. A. Scheel, M. Boyle, T. Chu, L. E. Kidder, K. D.
  Matthews and H. P. Pfeiffer}}(2009)}]{Scheel2009}%
  \BibitemOpen
  \bibfield  {author} {\bibinfo {author} {\bibnamefont {{M. A. Scheel, M.
  Boyle, T. Chu, L. E. Kidder, K. D. Matthews and H. P. Pfeiffer}}},\
  }\href@noop {} {\bibfield  {journal} {\bibinfo  {journal} {Phys.\ Rev.\ D}\
  }\textbf {\bibinfo {volume} {79}},\ \bibinfo {pages} {024003} (\bibinfo
  {year} {2009})},\ \Eprint {http://arxiv.org/abs/arXiv:gr-qc/0810.1767}
  {arXiv:gr-qc/0810.1767} \BibitemShut {NoStop}%
\bibitem [{\citenamefont {Hemberger}\ \emph
  {et~al.}(2013{\natexlab{b}})\citenamefont {Hemberger}, \citenamefont
  {Scheel}, \citenamefont {Kidder}, \citenamefont {Szil{\'a}gyi}, \citenamefont
  {Lovelace}, \citenamefont {Taylor},\ and\ \citenamefont
  {Teukolsky}}]{Hemberger:2012jz}%
  \BibitemOpen
  \bibfield  {author} {\bibinfo {author} {\bibfnamefont {D.~A.}\ \bibnamefont
  {Hemberger}}, \bibinfo {author} {\bibfnamefont {M.~A.}\ \bibnamefont
  {Scheel}}, \bibinfo {author} {\bibfnamefont {L.~E.}\ \bibnamefont {Kidder}},
  \bibinfo {author} {\bibfnamefont {B.}~\bibnamefont {Szil{\'a}gyi}}, \bibinfo
  {author} {\bibfnamefont {G.}~\bibnamefont {Lovelace}}, \bibinfo {author}
  {\bibfnamefont {N.~W.}\ \bibnamefont {Taylor}}, \ and\ \bibinfo {author}
  {\bibfnamefont {S.~A.}\ \bibnamefont {Teukolsky}},\ }\href {\doibase
  10.1088/0264-9381/30/11/115001} {\bibfield  {journal} {\bibinfo  {journal}
  {Class.\ Quantum Grav.}\ }\textbf {\bibinfo {volume} {30}},\ \bibinfo {pages}
  {115001} (\bibinfo {year} {2013}{\natexlab{b}})},\ \Eprint
  {http://arxiv.org/abs/1211.6079} {arXiv:1211.6079 [gr-qc]} \BibitemShut
  {NoStop}%
\bibitem [{\citenamefont {Ossokine}\ \emph {et~al.}(2013)\citenamefont
  {Ossokine}, \citenamefont {Kidder},\ and\ \citenamefont
  {Pfeiffer}}]{Ossokine:2013zga}%
  \BibitemOpen
  \bibfield  {author} {\bibinfo {author} {\bibfnamefont {S.}~\bibnamefont
  {Ossokine}}, \bibinfo {author} {\bibfnamefont {L.~E.}\ \bibnamefont
  {Kidder}}, \ and\ \bibinfo {author} {\bibfnamefont {H.~P.}\ \bibnamefont
  {Pfeiffer}},\ }\href@noop {} {\bibfield  {journal} {\bibinfo  {journal}
  {arXiv:1304.3067}\ } (\bibinfo {year} {2013})},\ \Eprint
  {http://arxiv.org/abs/1304.3067} {arXiv:1304.3067 [gr-qc]} \BibitemShut
  {NoStop}%
\bibitem [{\citenamefont {Rinne}(2006)}]{Rinne2006}%
  \BibitemOpen
  \bibfield  {author} {\bibinfo {author} {\bibfnamefont {O.}~\bibnamefont
  {Rinne}},\ }\href {http://stacks.iop.org/0264-9381/23/6275} {\bibfield
  {journal} {\bibinfo  {journal} {Class.\ Quantum Grav.}\ }\textbf {\bibinfo
  {volume} {23}},\ \bibinfo {pages} {6275} (\bibinfo {year}
  {2006})}\BibitemShut {NoStop}%
\bibitem [{\citenamefont {Rinne}\ \emph {et~al.}(2007)\citenamefont {Rinne},
  \citenamefont {Lindblom},\ and\ \citenamefont {Scheel}}]{Rinne2007}%
  \BibitemOpen
  \bibfield  {author} {\bibinfo {author} {\bibfnamefont {O.}~\bibnamefont
  {Rinne}}, \bibinfo {author} {\bibfnamefont {L.}~\bibnamefont {Lindblom}}, \
  and\ \bibinfo {author} {\bibfnamefont {M.~A.}\ \bibnamefont {Scheel}},\
  }\href {http://stacks.iop.org/0264-9381/24/4053} {\bibfield  {journal}
  {\bibinfo  {journal} {Class.\ Quantum Grav.}\ }\textbf {\bibinfo {volume}
  {24}},\ \bibinfo {pages} {4053} (\bibinfo {year} {2007})}\BibitemShut
  {NoStop}%
\bibitem [{\citenamefont {Gundlach}(1998)}]{Gundlach1998}%
  \BibitemOpen
  \bibfield  {author} {\bibinfo {author} {\bibfnamefont {C.}~\bibnamefont
  {Gundlach}},\ }\href {\doibase 10.1103/PhysRevD.57.863} {\bibfield  {journal}
  {\bibinfo  {journal} {Phys.\ Rev.\ D}\ }\textbf {\bibinfo {volume} {57}},\
  \bibinfo {pages} {863} (\bibinfo {year} {1998})}\BibitemShut {NoStop}%
\bibitem [{\citenamefont {Cook}\ and\ \citenamefont
  {Whiting}(2007)}]{Cook2007}%
  \BibitemOpen
  \bibfield  {author} {\bibinfo {author} {\bibfnamefont {G.~B.}\ \bibnamefont
  {Cook}}\ and\ \bibinfo {author} {\bibfnamefont {B.~F.}\ \bibnamefont
  {Whiting}},\ }\href {\doibase 10.1103/PhysRevD.76.041501} {\bibfield
  {journal} {\bibinfo  {journal} {Phys.\ Rev.\ D}\ }\textbf {\bibinfo {volume}
  {76}},\ \bibinfo {eid} {041501(R)} (\bibinfo {year} {2007})}\BibitemShut
  {NoStop}%
\bibitem [{\citenamefont {Owen}(2007)}]{OwenThesis}%
  \BibitemOpen
  \bibfield  {author} {\bibinfo {author} {\bibfnamefont {R.}~\bibnamefont
  {Owen}},\ }\emph {\bibinfo {title} {Topics in Numerical Relativity: {T}he
  periodic standing-wave approximation, the stability of constraints in free
  evolution, and the spin of dynamical black holes}},\ \href
  {http://resolver.caltech.edu/CaltechETD:etd-05252007-143511} {Ph.D. thesis},\
  \bibinfo  {school} {California Institute of Technology} (\bibinfo {year}
  {2007})\BibitemShut {NoStop}%
\bibitem [{\citenamefont {Scheel}\ \emph {et~al.}(2006)\citenamefont {Scheel},
  \citenamefont {Pfeiffer}, \citenamefont {Lindblom}, \citenamefont {Kidder},
  \citenamefont {Rinne},\ and\ \citenamefont {Teukolsky}}]{Scheel2006}%
  \BibitemOpen
  \bibfield  {author} {\bibinfo {author} {\bibfnamefont {M.~A.}\ \bibnamefont
  {Scheel}}, \bibinfo {author} {\bibfnamefont {H.~P.}\ \bibnamefont
  {Pfeiffer}}, \bibinfo {author} {\bibfnamefont {L.}~\bibnamefont {Lindblom}},
  \bibinfo {author} {\bibfnamefont {L.~E.}\ \bibnamefont {Kidder}}, \bibinfo
  {author} {\bibfnamefont {O.}~\bibnamefont {Rinne}}, \ and\ \bibinfo {author}
  {\bibfnamefont {S.~A.}\ \bibnamefont {Teukolsky}},\ }\href {\doibase
  10.1103/PhysRevD.74.104006} {\bibfield  {journal} {\bibinfo  {journal}
  {Phys.\ Rev.\ D}\ }\textbf {\bibinfo {volume} {74}},\ \bibinfo {pages}
  {104006} (\bibinfo {year} {2006})}\BibitemShut {NoStop}%
\bibitem [{\citenamefont {Boyle}\ \emph {et~al.}(2007)\citenamefont {Boyle},
  \citenamefont {Brown}, \citenamefont {Kidder}, \citenamefont {Mrou{\'e}},
  \citenamefont {Pfeiffer}, \citenamefont {Scheel}, \citenamefont {Cook},\ and\
  \citenamefont {Teukolsky}}]{Boyle2007}%
  \BibitemOpen
  \bibfield  {author} {\bibinfo {author} {\bibfnamefont {M.}~\bibnamefont
  {Boyle}}, \bibinfo {author} {\bibfnamefont {D.~A.}\ \bibnamefont {Brown}},
  \bibinfo {author} {\bibfnamefont {L.~E.}\ \bibnamefont {Kidder}}, \bibinfo
  {author} {\bibfnamefont {A.~H.}\ \bibnamefont {Mrou{\'e}}}, \bibinfo {author}
  {\bibfnamefont {H.~P.}\ \bibnamefont {Pfeiffer}}, \bibinfo {author}
  {\bibfnamefont {M.~A.}\ \bibnamefont {Scheel}}, \bibinfo {author}
  {\bibfnamefont {G.~B.}\ \bibnamefont {Cook}}, \ and\ \bibinfo {author}
  {\bibfnamefont {S.~A.}\ \bibnamefont {Teukolsky}},\ }\href@noop {} {\bibfield
   {journal} {\bibinfo  {journal} {Phys.\ Rev.\ D}\ }\textbf {\bibinfo {volume}
  {76}},\ \bibinfo {pages} {124038} (\bibinfo {year} {2007})},\ \Eprint
  {http://arxiv.org/abs/0710.0158} {arXiv:0710.0158 [gr-qc]} \BibitemShut
  {NoStop}%
\bibitem [{\citenamefont {Buchman}\ \emph {et~al.}(2012)\citenamefont
  {Buchman}, \citenamefont {Pfeiffer}, \citenamefont {Scheel},\ and\
  \citenamefont {Szil{\' a}gyi}}]{Buchman:2012dw}%
  \BibitemOpen
  \bibfield  {author} {\bibinfo {author} {\bibfnamefont {L.~T.}\ \bibnamefont
  {Buchman}}, \bibinfo {author} {\bibfnamefont {H.~P.}\ \bibnamefont
  {Pfeiffer}}, \bibinfo {author} {\bibfnamefont {M.~A.}\ \bibnamefont
  {Scheel}}, \ and\ \bibinfo {author} {\bibfnamefont {B.}~\bibnamefont {Szil{\'
  a}gyi}},\ }\href@noop {} {\bibfield  {journal} {\bibinfo  {journal} {Phys.\
  Rev.\ D}\ }\textbf {\bibinfo {volume} {86}},\ \bibinfo {pages} {084033}
  (\bibinfo {year} {2012})},\ \Eprint {http://arxiv.org/abs/1206.3015}
  {arXiv:1206.3015 [gr-qc]} \BibitemShut {NoStop}%
\bibitem [{\citenamefont {Chu}\ \emph {et~al.}(2009)\citenamefont {Chu},
  \citenamefont {Pfeiffer},\ and\ \citenamefont {Scheel}}]{Chu2009}%
  \BibitemOpen
  \bibfield  {author} {\bibinfo {author} {\bibfnamefont {T.}~\bibnamefont
  {Chu}}, \bibinfo {author} {\bibfnamefont {H.~P.}\ \bibnamefont {Pfeiffer}}, \
  and\ \bibinfo {author} {\bibfnamefont {M.~A.}\ \bibnamefont {Scheel}},\
  }\href {\doibase 10.1103/PhysRevD.80.124051} {\bibfield  {journal} {\bibinfo
  {journal} {Phys.\ Rev.\ D}\ }\textbf {\bibinfo {volume} {80}},\ \bibinfo
  {pages} {124051} (\bibinfo {year} {2009})},\ \Eprint
  {http://arxiv.org/abs/0909.1313} {arXiv:0909.1313 [gr-qc]} \BibitemShut
  {NoStop}%
\bibitem [{SXS()}]{SXSCatalog}%
  \BibitemOpen
  \href@noop {} {}\bibinfo {howpublished}
  {\url{http://www.black-holes.org/waveforms}}\BibitemShut {NoStop}%
\bibitem [{\citenamefont {Alvi}(2001)}]{Alvi:2001mx}%
  \BibitemOpen
  \bibfield  {author} {\bibinfo {author} {\bibfnamefont {K.}~\bibnamefont
  {Alvi}},\ }\href {http://link.aps.org/doi/10.1103/PhysRevD.64.104020}
  {\bibfield  {journal} {\bibinfo  {journal} {Phys.\ Rev.\ D}\ }\textbf
  {\bibinfo {volume} {64}},\ \bibinfo {pages} {104020} (\bibinfo {year}
  {2001})}\BibitemShut {NoStop}%
\bibitem [{\citenamefont {Owen}\ \emph {et~al.}(2011)\citenamefont {Owen},
  \citenamefont {Brink}, \citenamefont {Chen}, \citenamefont {Kaplan},
  \citenamefont {Lovelace}, \citenamefont {Matthews}, \citenamefont {Nichols},
  \citenamefont {Scheel}, \citenamefont {Zhang}, \citenamefont {Zimmerman},\
  and\ \citenamefont {Thorne}}]{OwenEtAl:2011}%
  \BibitemOpen
  \bibfield  {author} {\bibinfo {author} {\bibfnamefont {R.}~\bibnamefont
  {Owen}}, \bibinfo {author} {\bibfnamefont {J.}~\bibnamefont {Brink}},
  \bibinfo {author} {\bibfnamefont {Y.}~\bibnamefont {Chen}}, \bibinfo {author}
  {\bibfnamefont {J.~D.}\ \bibnamefont {Kaplan}}, \bibinfo {author}
  {\bibfnamefont {G.}~\bibnamefont {Lovelace}}, \bibinfo {author}
  {\bibfnamefont {K.~D.}\ \bibnamefont {Matthews}}, \bibinfo {author}
  {\bibfnamefont {D.~A.}\ \bibnamefont {Nichols}}, \bibinfo {author}
  {\bibfnamefont {M.~A.}\ \bibnamefont {Scheel}}, \bibinfo {author}
  {\bibfnamefont {F.}~\bibnamefont {Zhang}}, \bibinfo {author} {\bibfnamefont
  {A.}~\bibnamefont {Zimmerman}}, \ and\ \bibinfo {author} {\bibfnamefont
  {K.~S.}\ \bibnamefont {Thorne}},\ }\href@noop {} {\bibfield  {journal}
  {\bibinfo  {journal} {Phys.\ Rev.\ Lett.}\ }\textbf {\bibinfo {volume}
  {106}},\ \bibinfo {pages} {151101} (\bibinfo {year} {2011})}\BibitemShut
  {NoStop}%
\bibitem [{\citenamefont {{Chatziioannou}}\ \emph {et~al.}(2013)\citenamefont
  {{Chatziioannou}}, \citenamefont {{Poisson}},\ and\ \citenamefont
  {{Yunes}}}]{Chatziioannou:2013}%
  \BibitemOpen
  \bibfield  {author} {\bibinfo {author} {\bibfnamefont {K.}~\bibnamefont
  {{Chatziioannou}}}, \bibinfo {author} {\bibfnamefont {E.}~\bibnamefont
  {{Poisson}}}, \ and\ \bibinfo {author} {\bibfnamefont {N.}~\bibnamefont
  {{Yunes}}},\ }\href {\doibase 10.1103/PhysRevD.87.044022} {\bibfield
  {journal} {\bibinfo  {journal} {Phys.\ Rev.\ D}\ }\textbf {\bibinfo {volume}
  {87}},\ \bibinfo {eid} {044022} (\bibinfo {year} {2013})},\ \Eprint
  {http://arxiv.org/abs/1211.1686} {arXiv:1211.1686 [gr-qc]} \BibitemShut
  {NoStop}%
\bibitem [{\citenamefont {{Chatziioannou}}\ \emph {et~al.}()\citenamefont
  {{Chatziioannou}}, \citenamefont {{Poisson}},\ and\ \citenamefont
  {{Yunes}}}]{Chatziioannou:2014}%
  \BibitemOpen
  \bibfield  {author} {\bibinfo {author} {\bibfnamefont {K.}~\bibnamefont
  {{Chatziioannou}}}, \bibinfo {author} {\bibfnamefont {E.}~\bibnamefont
  {{Poisson}}}, \ and\ \bibinfo {author} {\bibfnamefont {N.}~\bibnamefont
  {{Yunes}}},\ }\href@noop {} {\ }\bibinfo {note} {In preparation.}\BibitemShut
  {Stop}%
\bibitem [{\citenamefont {Kidder}(1995{\natexlab{a}})}]{kidder95}%
  \BibitemOpen
  \bibfield  {author} {\bibinfo {author} {\bibfnamefont {L.~E.}\ \bibnamefont
  {Kidder}},\ }\href {\doibase 10.1103/PhysRevD.52.821} {\bibfield  {journal}
  {\bibinfo  {journal} {Phys.\ Rev.\ D}\ }\textbf {\bibinfo {volume} {52}},\
  \bibinfo {pages} {821} (\bibinfo {year} {1995}{\natexlab{a}})}\BibitemShut
  {NoStop}%
\bibitem [{\citenamefont {Damour}(2001)}]{Damour01c}%
  \BibitemOpen
  \bibfield  {author} {\bibinfo {author} {\bibfnamefont {T.}~\bibnamefont
  {Damour}},\ }\href {\doibase 10.1103/PhysRevD.64.124013} {\bibfield
  {journal} {\bibinfo  {journal} {Phys.\ Rev.\ D}\ }\textbf {\bibinfo {volume}
  {64}},\ \bibinfo {pages} {124013} (\bibinfo {year} {2001})},\ \Eprint
  {http://arxiv.org/abs/gr-qc/0103018} {arXiv:gr-qc/0103018 [gr-qc]}
  \BibitemShut {NoStop}%
\bibitem [{\citenamefont {Kidder}(1995{\natexlab{b}})}]{Kidder:1995zr}%
  \BibitemOpen
  \bibfield  {author} {\bibinfo {author} {\bibfnamefont {L.~E.}\ \bibnamefont
  {Kidder}},\ }\href {\doibase 10.1103/PhysRevD.52.821} {\bibfield  {journal}
  {\bibinfo  {journal} {Phys. Rev.}\ }\textbf {\bibinfo {volume} {D52}},\
  \bibinfo {pages} {821} (\bibinfo {year} {1995}{\natexlab{b}})},\ \Eprint
  {http://arxiv.org/abs/gr-qc/9506022} {arXiv:gr-qc/9506022} \BibitemShut
  {NoStop}%
\bibitem [{\citenamefont {Campanelli}\ \emph
  {et~al.}(2006{\natexlab{b}})\citenamefont {Campanelli}, \citenamefont
  {Lousto},\ and\ \citenamefont {Zlochower}}]{Campanelli2006c}%
  \BibitemOpen
  \bibfield  {author} {\bibinfo {author} {\bibfnamefont {M.}~\bibnamefont
  {Campanelli}}, \bibinfo {author} {\bibfnamefont {C.~O.}\ \bibnamefont
  {Lousto}}, \ and\ \bibinfo {author} {\bibfnamefont {Y.}~\bibnamefont
  {Zlochower}},\ }\href@noop {} {\bibfield  {journal} {\bibinfo  {journal}
  {Phys.\ Rev.\ D}\ }\textbf {\bibinfo {volume} {74}},\ \bibinfo {pages}
  {041501(R)} (\bibinfo {year} {2006}{\natexlab{b}})},\ \Eprint
  {http://arxiv.org/abs/gr-qc/0604012} {gr-qc/0604012} \BibitemShut {NoStop}%
\bibitem [{\citenamefont {Sarbach}\ and\ \citenamefont
  {Tiglio}(2001)}]{Sarbach2001}%
  \BibitemOpen
  \bibfield  {author} {\bibinfo {author} {\bibfnamefont {O.}~\bibnamefont
  {Sarbach}}\ and\ \bibinfo {author} {\bibfnamefont {M.}~\bibnamefont
  {Tiglio}},\ }\href {\doibase 10.1103/PhysRevD.64.084016} {\bibfield
  {journal} {\bibinfo  {journal} {Phys. Rev. D}\ }\textbf {\bibinfo {volume}
  {64}},\ \bibinfo {pages} {084016} (\bibinfo {year} {2001})}\BibitemShut
  {NoStop}%
\bibitem [{\citenamefont {Rinne}\ \emph {et~al.}(2009)\citenamefont {Rinne},
  \citenamefont {Buchman}, \citenamefont {Scheel},\ and\ \citenamefont
  {Pfeiffer}}]{Rinne2008b}%
  \BibitemOpen
  \bibfield  {author} {\bibinfo {author} {\bibfnamefont {O.}~\bibnamefont
  {Rinne}}, \bibinfo {author} {\bibfnamefont {L.~T.}\ \bibnamefont {Buchman}},
  \bibinfo {author} {\bibfnamefont {M.~A.}\ \bibnamefont {Scheel}}, \ and\
  \bibinfo {author} {\bibfnamefont {H.~P.}\ \bibnamefont {Pfeiffer}},\
  }\href@noop {} {\bibfield  {journal} {\bibinfo  {journal} {Class.\ Quantum
  Grav.}\ }\textbf {\bibinfo {volume} {26}},\ \bibinfo {pages} {075009}
  (\bibinfo {year} {2009})}\BibitemShut {NoStop}%
\bibitem [{\citenamefont {Boyle}(2013)}]{Boyle:2013a}%
  \BibitemOpen
  \bibfield  {author} {\bibinfo {author} {\bibfnamefont {M.}~\bibnamefont
  {Boyle}},\ }\href {\doibase 10.1103/PhysRevD.87.104006} {\bibfield  {journal}
  {\bibinfo  {journal} {Phys.\ Rev.\ D}\ }\textbf {\bibinfo {volume} {87}},\
  \bibinfo {pages} {104006} (\bibinfo {year} {2013})}\BibitemShut {NoStop}%
\bibitem [{\citenamefont {Boyle}\ \emph {et~al.}(2014)\citenamefont {Boyle},
  \citenamefont {Kidder}, \citenamefont {Ossokine},\ and\ \citenamefont
  {Pfeiffer}}]{Boyle:2014}%
  \BibitemOpen
  \bibfield  {author} {\bibinfo {author} {\bibfnamefont {M.}~\bibnamefont
  {Boyle}}, \bibinfo {author} {\bibfnamefont {L.~E.}\ \bibnamefont {Kidder}},
  \bibinfo {author} {\bibfnamefont {S.}~\bibnamefont {Ossokine}}, \ and\
  \bibinfo {author} {\bibfnamefont {H.~P.}\ \bibnamefont {Pfeiffer}},\
  }\href@noop {} {\  (\bibinfo {year} {2014})},\ \bibinfo {note}
  {arXiv:1409.4431},\ \Eprint {http://arxiv.org/abs/1409.4431}
  {arXiv:1409.4431} \BibitemShut {NoStop}%
\bibitem [{\citenamefont {Ossokine}\ \emph {et~al.}(2014)\citenamefont
  {Ossokine}, \citenamefont {Boyle}, \citenamefont {Kidder}, \citenamefont
  {Mrou{\'{e}}}, \citenamefont {Pfeiffer}, \citenamefont {Scheel},\ and\
  \citenamefont {Szil{\'{a}}gyi}}]{OssokineEtAl:2014}%
  \BibitemOpen
  \bibfield  {author} {\bibinfo {author} {\bibfnamefont {S.}~\bibnamefont
  {Ossokine}}, \bibinfo {author} {\bibfnamefont {M.}~\bibnamefont {Boyle}},
  \bibinfo {author} {\bibfnamefont {L.~E.}\ \bibnamefont {Kidder}}, \bibinfo
  {author} {\bibfnamefont {A.~H.}\ \bibnamefont {Mrou{\'{e}}}}, \bibinfo
  {author} {\bibfnamefont {H.~P.}\ \bibnamefont {Pfeiffer}}, \bibinfo {author}
  {\bibfnamefont {M.~A.}\ \bibnamefont {Scheel}}, \ and\ \bibinfo {author}
  {\bibfnamefont {B.}~\bibnamefont {Szil{\'{a}}gyi}},\ }\href@noop {} {\
  (\bibinfo {year} {2014})},\ \bibinfo {note} {in preparation}\BibitemShut
  {NoStop}%
\bibitem [{\citenamefont {Bini}\ and\ \citenamefont
  {Damour}(2013)}]{BiniDamour:2013}%
  \BibitemOpen
  \bibfield  {author} {\bibinfo {author} {\bibfnamefont {D.}~\bibnamefont
  {Bini}}\ and\ \bibinfo {author} {\bibfnamefont {T.}~\bibnamefont {Damour}},\
  }\href {http://arxiv.org/abs/1305.4884} {\enquote {\bibinfo {title}
  {{Analytical determination of the two-body gravitational interaction
  potential at the 4th post-Newtonian approximation}},}\ } (\bibinfo {year}
  {2013}),\ \Eprint {http://arxiv.org/abs/1305.4884} {arXiv:1305.4884 [gr-qc]}
  \BibitemShut {NoStop}%
\bibitem [{\citenamefont {Blanchet}\ \emph {et~al.}(2008)\citenamefont
  {Blanchet}, \citenamefont {Faye}, \citenamefont {Iyer},\ and\ \citenamefont
  {Sinha}}]{BFIS}%
  \BibitemOpen
  \bibfield  {author} {\bibinfo {author} {\bibfnamefont {L.}~\bibnamefont
  {Blanchet}}, \bibinfo {author} {\bibfnamefont {G.}~\bibnamefont {Faye}},
  \bibinfo {author} {\bibfnamefont {B.~R.}\ \bibnamefont {Iyer}}, \ and\
  \bibinfo {author} {\bibfnamefont {S.}~\bibnamefont {Sinha}},\ }\href
  {\doibase 10.1088/0264-9381/25/16/165003} {\bibfield  {journal} {\bibinfo
  {journal} {Class. Quant. Grav.}\ }\textbf {\bibinfo {volume} {25}},\ \bibinfo
  {pages} {165003} (\bibinfo {year} {2008})},\ \Eprint
  {http://arxiv.org/abs/0802.1249} {arXiv:0802.1249 [gr-qc]} \BibitemShut
  {NoStop}%
\bibitem [{\citenamefont {Faye}\ \emph {et~al.}(2012)\citenamefont {Faye},
  \citenamefont {Marsat}, \citenamefont {Blanchet},\ and\ \citenamefont
  {Iyer}}]{FayeEtAl:2012}%
  \BibitemOpen
  \bibfield  {author} {\bibinfo {author} {\bibfnamefont {G.}~\bibnamefont
  {Faye}}, \bibinfo {author} {\bibfnamefont {S.}~\bibnamefont {Marsat}},
  \bibinfo {author} {\bibfnamefont {L.}~\bibnamefont {Blanchet}}, \ and\
  \bibinfo {author} {\bibfnamefont {B.~R.}\ \bibnamefont {Iyer}},\ }\href
  {http://arxiv.org/abs/1204.1043} {\bibfield  {journal} {\bibinfo  {journal}
  {{Class. Quant. Grav.}}\ }\textbf {\bibinfo {volume} {{29}}},\ \bibinfo
  {pages} {175004} (\bibinfo {year} {2012})}\BibitemShut {NoStop}%
\bibitem [{\citenamefont {Faye}\ \emph {et~al.}(2014)\citenamefont {Faye},
  \citenamefont {Blanchet},\ and\ \citenamefont {Iyer}}]{FayeEtAl:2014}%
  \BibitemOpen
  \bibfield  {author} {\bibinfo {author} {\bibfnamefont {G.}~\bibnamefont
  {Faye}}, \bibinfo {author} {\bibfnamefont {L.}~\bibnamefont {Blanchet}}, \
  and\ \bibinfo {author} {\bibfnamefont {B.~R.}\ \bibnamefont {Iyer}},\ }\href
  {http://arxiv.org/abs/1409.3546} {\enquote {\bibinfo {title} {{Non-linear
  multipole interactions and gravitational-wave octupole modes for inspiralling
  compact binaries to third-and-a-half post-Newtonian order}},}\ } (\bibinfo
  {year} {2014}),\ \Eprint {http://arxiv.org/abs/1409.3546} {arXiv:1409.3546
  [gr-qc]} \BibitemShut {NoStop}%
\bibitem [{\citenamefont {Boh\'{e}}\ \emph {et~al.}(2013)\citenamefont
  {Boh\'{e}}, \citenamefont {Marsat}, \citenamefont {Faye},\ and\ \citenamefont
  {Blanchet}}]{BoheEtAl:2012}%
  \BibitemOpen
  \bibfield  {author} {\bibinfo {author} {\bibfnamefont {A.}~\bibnamefont
  {Boh\'{e}}}, \bibinfo {author} {\bibfnamefont {S.}~\bibnamefont {Marsat}},
  \bibinfo {author} {\bibfnamefont {G.}~\bibnamefont {Faye}}, \ and\ \bibinfo
  {author} {\bibfnamefont {L.}~\bibnamefont {Blanchet}},\ }\href
  {http://iopscience.iop.org/0264-9381/30/7/075017} {\bibfield  {journal}
  {\bibinfo  {journal} {{Class. Quant. Grav.}}\ }\textbf {\bibinfo {volume}
  {{30}}},\ \bibinfo {pages} {075017} (\bibinfo {year} {2013})}\BibitemShut
  {NoStop}%
\bibitem [{\citenamefont {Marsat}\ \emph {et~al.}(2013)\citenamefont {Marsat},
  \citenamefont {Blanchet}, \citenamefont {Boh{\'{e}}},\ and\ \citenamefont
  {Faye}}]{MarsatEtAl:2013}%
  \BibitemOpen
  \bibfield  {author} {\bibinfo {author} {\bibfnamefont {S.}~\bibnamefont
  {Marsat}}, \bibinfo {author} {\bibfnamefont {L.}~\bibnamefont {Blanchet}},
  \bibinfo {author} {\bibfnamefont {A.}~\bibnamefont {Boh{\'{e}}}}, \ and\
  \bibinfo {author} {\bibfnamefont {G.}~\bibnamefont {Faye}},\ }\href
  {http://arxiv.org/abs/1312.5375} {\enquote {\bibinfo {title} {{Gravitational
  waves from spinning compact object binaries: New post-Newtonian results}},}\
  } (\bibinfo {year} {2013}),\ \Eprint {http://arxiv.org/abs/1312.5375}
  {arXiv:1312.5375 [gr-qc]} \BibitemShut {NoStop}%
\bibitem [{\citenamefont {Will}\ and\ \citenamefont {Wiseman}(1996)}]{Will96}%
  \BibitemOpen
  \bibfield  {author} {\bibinfo {author} {\bibfnamefont {C.~M.}\ \bibnamefont
  {Will}}\ and\ \bibinfo {author} {\bibfnamefont {A.~G.}\ \bibnamefont
  {Wiseman}},\ }\href {\doibase 10.1103/PhysRevD.54.4813} {\bibfield  {journal}
  {\bibinfo  {journal} {Phys.\ Rev.\ D}\ }\textbf {\bibinfo {volume} {54}},\
  \bibinfo {pages} {4813} (\bibinfo {year} {1996})}\BibitemShut {NoStop}%
\bibitem [{\citenamefont {Arun}\ \emph {et~al.}(2009)\citenamefont {Arun},
  \citenamefont {Buonanno}, \citenamefont {Faye},\ and\ \citenamefont
  {Ochsner}}]{Arun:2009}%
  \BibitemOpen
  \bibfield  {author} {\bibinfo {author} {\bibfnamefont {K.}~\bibnamefont
  {Arun}}, \bibinfo {author} {\bibfnamefont {A.}~\bibnamefont {Buonanno}},
  \bibinfo {author} {\bibfnamefont {G.}~\bibnamefont {Faye}}, \ and\ \bibinfo
  {author} {\bibfnamefont {E.}~\bibnamefont {Ochsner}},\ }\href {\doibase
  10.1103/PhysRevD.79.104023, 10.1103/PhysRevD.84.049901,
  10.1103/PhysRevD.79.104023, 10.1103/PhysRevD.84.049901} {\bibfield  {journal}
  {\bibinfo  {journal} {Phys.\ Rev.\ D}\ }\textbf {\bibinfo {volume} {79}},\
  \bibinfo {pages} {104023} (\bibinfo {year} {2009})},\ \Eprint
  {http://arxiv.org/abs/0810.5336} {arXiv:0810.5336 [gr-qc]} \BibitemShut
  {NoStop}%
\bibitem [{\citenamefont {Buonanno}\ \emph {et~al.}(2013)\citenamefont
  {Buonanno}, \citenamefont {Faye},\ and\ \citenamefont
  {Hinderer}}]{BuonannoFayeHinderer:2013}%
  \BibitemOpen
  \bibfield  {author} {\bibinfo {author} {\bibfnamefont {A.}~\bibnamefont
  {Buonanno}}, \bibinfo {author} {\bibfnamefont {G.}~\bibnamefont {Faye}}, \
  and\ \bibinfo {author} {\bibfnamefont {T.}~\bibnamefont {Hinderer}},\ }\href
  {\doibase 10.1103/PhysRevD.87.044009} {\bibfield  {journal} {\bibinfo
  {journal} {{Phys. Rev. D}}\ }\textbf {\bibinfo {volume} {{87}}},\ \bibinfo
  {pages} {044009} (\bibinfo {year} {2013})}\BibitemShut {NoStop}%
\bibitem [{\citenamefont {Boyle}\ \emph {et~al.}(2008)\citenamefont {Boyle},
  \citenamefont {Buonanno}, \citenamefont {Kidder}, \citenamefont {Mrou\'{e}},
  \citenamefont {Pan} \emph {et~al.}}]{Boyle:2008}%
  \BibitemOpen
  \bibfield  {author} {\bibinfo {author} {\bibfnamefont {M.}~\bibnamefont
  {Boyle}}, \bibinfo {author} {\bibfnamefont {A.}~\bibnamefont {Buonanno}},
  \bibinfo {author} {\bibfnamefont {L.~E.}\ \bibnamefont {Kidder}}, \bibinfo
  {author} {\bibfnamefont {A.~H.}\ \bibnamefont {Mrou\'{e}}}, \bibinfo {author}
  {\bibfnamefont {Y.}~\bibnamefont {Pan}},  \emph {et~al.},\ }\href {\doibase
  10.1103/PhysRevD.78.104020} {\bibfield  {journal} {\bibinfo  {journal}
  {Phys.\ Rev.\ D}\ }\textbf {\bibinfo {volume} {78}},\ \bibinfo {pages}
  {104020} (\bibinfo {year} {2008})},\ \Eprint {http://arxiv.org/abs/0804.4184}
  {arXiv:0804.4184 [gr-qc]} \BibitemShut {NoStop}%
\bibitem [{\citenamefont {{MacDonald}}\ \emph {et~al.}(2011)\citenamefont
  {{MacDonald}}, \citenamefont {Nissanke},\ and\ \citenamefont
  {Pfeiffer}}]{MacDonald:2011ne}%
  \BibitemOpen
  \bibfield  {author} {\bibinfo {author} {\bibfnamefont {I.}~\bibnamefont
  {{MacDonald}}}, \bibinfo {author} {\bibfnamefont {S.}~\bibnamefont
  {Nissanke}}, \ and\ \bibinfo {author} {\bibfnamefont {H.~P.}\ \bibnamefont
  {Pfeiffer}},\ }\href {\doibase 10.1088/0264-9381/28/13/134002} {\bibfield
  {journal} {\bibinfo  {journal} {Class.\ Quantum Grav.}\ }\textbf {\bibinfo
  {volume} {28}},\ \bibinfo {pages} {134002} (\bibinfo {year} {2011})},\
  \Eprint {http://arxiv.org/abs/1102.5128} {arXiv:1102.5128 [gr-qc]}
  \BibitemShut {NoStop}%
\bibitem [{\citenamefont {Ajith}\ \emph {et~al.}(2008)\citenamefont {Ajith},
  \citenamefont {Babak}, \citenamefont {Chen}, \citenamefont {Hewitson},
  \citenamefont {Krishnan} \emph {et~al.}}]{Ajith-Babak-Chen-etal:2007b}%
  \BibitemOpen
  \bibfield  {author} {\bibinfo {author} {\bibfnamefont {P.}~\bibnamefont
  {Ajith}}, \bibinfo {author} {\bibfnamefont {S.}~\bibnamefont {Babak}},
  \bibinfo {author} {\bibfnamefont {Y.}~\bibnamefont {Chen}}, \bibinfo {author}
  {\bibfnamefont {M.}~\bibnamefont {Hewitson}}, \bibinfo {author}
  {\bibfnamefont {B.}~\bibnamefont {Krishnan}},  \emph {et~al.},\ }\href
  {\doibase 10.1103/PhysRevD.77.104017, 10.1103/PhysRevD.79.129901,
  10.1103/PhysRevD.77.104017, 10.1103/PhysRevD.79.129901} {\bibfield  {journal}
  {\bibinfo  {journal} {Phys.\ Rev.\ D}\ }\textbf {\bibinfo {volume} {77}},\
  \bibinfo {pages} {104017} (\bibinfo {year} {2008})},\ \Eprint
  {http://arxiv.org/abs/0710.2335} {arXiv:0710.2335 [gr-qc]} \BibitemShut
  {NoStop}%
\bibitem [{\citenamefont {Loken}\ \emph {et~al.}(2010)\citenamefont {Loken},
  \citenamefont {Gruner}, \citenamefont {Groer}, \citenamefont {Peltier},
  \citenamefont {Bunn}, \citenamefont {Craig}, \citenamefont {Henriques},
  \citenamefont {Dempsey}, \citenamefont {Yu}, \citenamefont {Chen},
  \citenamefont {Dursi}, \citenamefont {Chong}, \citenamefont {Northrup},
  \citenamefont {Pinto}, \citenamefont {Knecht},\ and\ \citenamefont
  {Zon}}]{scinet}%
  \BibitemOpen
  \bibfield  {author} {\bibinfo {author} {\bibfnamefont {C.}~\bibnamefont
  {Loken}}, \bibinfo {author} {\bibfnamefont {D.}~\bibnamefont {Gruner}},
  \bibinfo {author} {\bibfnamefont {L.}~\bibnamefont {Groer}}, \bibinfo
  {author} {\bibfnamefont {R.}~\bibnamefont {Peltier}}, \bibinfo {author}
  {\bibfnamefont {N.}~\bibnamefont {Bunn}}, \bibinfo {author} {\bibfnamefont
  {M.}~\bibnamefont {Craig}}, \bibinfo {author} {\bibfnamefont
  {T.}~\bibnamefont {Henriques}}, \bibinfo {author} {\bibfnamefont
  {J.}~\bibnamefont {Dempsey}}, \bibinfo {author} {\bibfnamefont {C.-H.}\
  \bibnamefont {Yu}}, \bibinfo {author} {\bibfnamefont {J.}~\bibnamefont
  {Chen}}, \bibinfo {author} {\bibfnamefont {L.~J.}\ \bibnamefont {Dursi}},
  \bibinfo {author} {\bibfnamefont {J.}~\bibnamefont {Chong}}, \bibinfo
  {author} {\bibfnamefont {S.}~\bibnamefont {Northrup}}, \bibinfo {author}
  {\bibfnamefont {J.}~\bibnamefont {Pinto}}, \bibinfo {author} {\bibfnamefont
  {N.}~\bibnamefont {Knecht}}, \ and\ \bibinfo {author} {\bibfnamefont {R.~V.}\
  \bibnamefont {Zon}},\ }\href {\doibase 10.1088/1742-6596/256/1/012026}
  {\bibfield  {journal} {\bibinfo  {journal} {J. Phys.: Conf. Ser.}\ }\textbf
  {\bibinfo {volume} {256}},\ \bibinfo {pages} {012026} (\bibinfo {year}
  {2010})}\BibitemShut {NoStop}%
\end{thebibliography}%

\end{document}